%% file: main.tex
\documentclass[useAMS,usenatbib]{mnras}

\input{include_tex/pacchetti}

\input{include_tex/journals}

\input{include_tex/definizioni}

\input{include_tex/additional_def}

\defcitealias{paramita:2018}{B18}
\defcitealias{Gallerani:2010}{G10}

\graphicspath{{figures/RT_plots/AGNconeBHon008SMC/},{figures/RT_plots/AGNsphereBHon008SMC/},{figures/RT_plots/noAGN008SMC/},{figures/RT_plots/AGNconeBHon03SMC/},{figures/RT_plots/AGNsphereBHon03SMC/},{figures/RT_plots/noAGN03SMC/},{figures/RT_plots/noAGN008MW/},{figures/RT_plots/noAGN03MW/},{figures/RT_plots/AGNsphereBHon008MW/},{figures/RT_plots/AGNsphereBHon03MW/},{figures/RT_plots/AGNconeBHon008MW/},{figures/RT_plots/AGNconeBHon03MW/},{figures/comparisons/},{figures/extinction/},{figures/special_sims/}, {figures/individual_SEDs/}, {figures/add_extinction/}}

\date{}
\pagerange{\pageref{firstpage}--\pageref{lastpage}} \pubyear{2020}

\title[IR emission of early galaxies: AGN imprints]{Infrared emission of $z\sim 6$ galaxies: AGN imprints}
\author[F. Di Mascia et al.]{F. Di Mascia$^{1}$\thanks{\href{mailto:fabio.dimascia@sns.it}{fabio.dimascia@sns.it}}, S. Gallerani$^{1}$, C. Behrens$^{2}$, A. Pallottini$^{1}$, S. Carniani$^{1}$,   
\newauthor A. Ferrara$^{1}$, P. Barai$^{3,4}$, F. Vito$^{1}$, T. Zana$^{1}$\\
$^{1}$Scuola Normale Superiore, Piazza dei Cavalieri 7, I-56126 Pisa, Italy\\
$^{2}$Institut f\"{u}r Astrophysik, Georg-August Universit\"{a}t G\"{o}ttingen, Friedrich-Hundt-Platz 1, 37077, G\"{o}ttingen, Germany\\
$^{3}$N\'{u}cleo de Astrof\'{i}sica - Universidade Cidade de S\~ao Paulo\\
$^{4}$Universidade Cruzeiro do Sul, Rua Galv\~ao Bueno 868, S\~ao Paulo, 01506-000, Brasil\\
}

\date{Accepted XXX. Received XXX; in original form XXX}

\begin{document}

\maketitle

\label{firstpage}

\begin{abstract}
We investigate the infrared (IR) emission of high-redshift ($z\sim 6$), highly star-forming (${\SFR>100~\msunyr}$) galaxies, with/without Active Galactic Nuclei (AGN), using a suite of cosmological simulations featuring dust radiative transfer. Synthetic Spectral Energy Distributions (SEDs) are used to quantify the relative contribution of stars/AGN to dust heating. In dusty (${M_{\rm d}\gtrsim 3\times 10^7~M_{\rm \odot}}$) galaxies, $\gtrsim$ 50-90\% of the UV radiation is obscured by dust inhomogeneities on scales ${\gtrsim 100}$~pc. In runs with AGN, a clumpy, warm ($\approx 250$~K) dust component co-exists with a colder ($\approx 60$~K) and more diffuse one, heated by stars. Warm dust provides up to ${50 \%}$ of the total IR luminosity, but only $\lesssim 0.1 \%$ of the total mass content. The AGN boosts the MIR flux by ${10-100 \times}$ with respect to star forming galaxies, without significantly affecting the FIR. Our simulations successfully reproduce the observed SED of bright (${M_{\rm UV}\sim -26}$) ${z\sim 6}$ quasars, and show that these objects are part of complex, dust-rich merging systems, containing multiple sources (accreting BHs and/or star forming galaxies) in agreement with recent HST and ALMA observations. Our results show that the proposed ORIGINS missions will be able to investigate the MIR properties of dusty star forming galaxies and to obtain good quality spectra of bright quasars at $z\sim 6$. Finally, the MIR-to-FIR flux ratio of faint (${M_{\rm UV}\sim -24}$) AGN is ${>10\times}$ higher than for normal star forming galaxies. This implies that combined JWST/ORIGINS/ALMA observations will be crucial to identify faint and/or dust-obscured AGN in the distant Universe. 
\end{abstract}


\section{Introduction}
Gas accretion onto super massive black holes (SMBH, ${M_{\rm BH}\sim 10^{6-10}~\rm M_{\odot}}$) residing in the center of most massive galaxies \citep[$M_\star\sim 10^{9-12}~\rm M_{\odot}$; e.g.][]{kormendy1995,magorrian1998,marconi2004,kormendy2013} turns them into active galactic nuclei (AGN). A large fraction (${\sim 10-50\%}$) of the bolometric luminosity produced by accreting BHs is emitted into optical/ultra-violet (UV) wavelength range \citep{hopkins2007,Lusso:2015,Shen:2020}, adding up to the luminosity produced by  massive OB stars. Thus, restframe optical/UV bands (redshifted in the near-infrared for objects located in the Epoch of Reionization) represent the natural spectral windows for AGN searches.

Over the last decade, thanks to several optical/near infra-red (NIR) surveys, such as the Sloan Digital Sky Survey (SDSS; \citealt{Fan2006SDSS, Jiang2009SDSS}), the UKIDSS Large Area Survey \citep{Venemans2007UKIDSS}, the Canada-France High-z Quasar Survey (CFHQS; \citealt{Willott2007CFHQS}), the VISTA Kilo-Degree Infrared Galaxy Survey (VIKING; \citealt{Venemans2013VIKING, Venemans2015VIKING}), Pan-STARRS1 \citep{Banados2014Pan-STARRS1}, the Very Large Telescope Survey Telescope ATLAS survey \citep{Carnall2015ATLAS}, the Dark Energy Survey (DES; \citealt{Reed2015DES}, and the Subaru High-z Exploration of Low-luminosity Quasars (SHELLQs; \citealt{Kashikawa2015ApJ, Matsuoka2016SHELLQs}), more than 200 quasars have been discovered at the most distant redshifts probed so far \citep[${z\sim 6-8}$,][]{Mortlock2011Natur, Banados2018Natur, Wang2018ApJ, Wang2021ApJ}. 
Follow-up NIR spectroscopical observations of emission lines (e.g. $\MGII$ and $\CIV$) produced by Broad Line Region clouds have confirmed that these sources are powered by ${\sim 10^8-10^{10}~\rm M_{\odot}}$ BHs \citep{Fan2000AJ, Willott2003ApJ, Kurk2007ApJ, Jiang2007AJ, Wu2015Natur}. 
The challenge is to understand how SMBHs have formed in $< 1$~Gyr, namely the age of the Universe at ${z\sim 6}$. Theoretical models of black hole accretion are in fact facing serious difficulties in explaining such a rapid growth \citep[e.g.][]{volonteri2003, tanaka2009, Haiman2013, pacucci2015, lupi2016}, also including the rather uncertain formation mechanism of SMBH seeds \citep{shang2010, schleicher2013, latif2013, ferrara:2014, Latif16}.

The problem is exacerbated by the unsuccessful search for high-$z$ AGN powered by ${\sim 10^{6-7}~\rm M_{\odot}}$ BHs \citep[e.g.][]{xue2011,cowie2020}. Whether these sources are too rare \citep{pezzulli2017}, and/or too faint to be detected by current optical/NIR survey \citep{Willott2010AJ, Jiang2016ApJ, Pacucci2016MNRAS, McGreer2018AJ, Matsuoka2018ApJ, Wang2019ApJ, Kulkarni2019MNRAS}, and/or their optical/UV emission is obscured by dust, remains unclear. This latter hypothesis is supported by at least two observational results: (i) multi-wavelength studies of $\sim 1000$ local AGN show a decrease in the covering factor of the circumnuclear material with increasing accretion rates due to the increase of the dust sublimation radius of the obscuring material with incident luminosity \citep[e.g.][]{ricci2017}; (ii) X-ray observations provide indications that the fraction of obscured AGN increases with redshift \citep[e.g.][]{Vito2014,Vito2018_obscured}, an evidence further supported by studies of Ly$\alpha$ absorption profiles of distant quasars \citep[e.g.][]{Davies:2019}. Both these facts resonate with the expectation that early growth of SMBHs, typically characterized by low accretion rates, is buried in a thick cocoon of dust and gas \citep[e.g.][for a review on this subject]{Hickox2018}.

In this scenario a certain fraction of UV photons are absorbed and/or scattered by dust grains in gas clouds in the host galaxy. By transferring energy and momentum to the surrounding dusty environment, AGN radiation can substantially affect the conditions of the interstellar (ISM) and circumgalactic (CGM) medium of the host galaxy in several ways. UV radiation heats the dust, leading grains to re-emit in the far-infrared. Moreover, radiation pressure on dust grains may drive powerful outflows \citep[e.g.][]{Fabian1999MNRAS, Murray2005ApJ, Wada2016ApJ, venanzi:2020} that push away the gas surrounding the black hole, clean up the line of sight, and prevent further accretion onto the BH \citep{DiMatteo2005Natur,Sijacki2007MNRAS,paramita:2018}. 
In addition to that, it is unclear whether star formation in the host galaxy might be quenched \citep{Schawinski2006Natur, Dubois2010MNRAS, Dubois2013MNRAS, Teyssier2011MNRAS, Schaye2015MNRAS, Weinberger2018MNRAS} or triggered \citep{DeYoung1989ApJ, Silk2005MNRAS, Zubovas2013MNRAS, Zinn2013ApJ, Cresci2015A, Cresci2015ApJ, carniani2016} by AGN-driven outflows.

Signatures of such a complex interplay between AGN/stellar radiation and dust grains remain imprinted in the rest-frame UV-to-FIR spectral energy distribution (SED) of galaxies. Therefore, multi-wavelength SED analysis of galaxies and AGN can be used to infer information on their dust properties (mass, temperature, grain size distribution, composition), to shed light on their star formation and nuclear activities, and to quantify the relative contribution of stars and AGN radiation to dust heating \citep{bongiorno2012,pozzi2012,berta2013,gruppioni2016}.
Telescopes sensitive to Mid-Infrared (MIR, ${5 \lesssim \lambda_{\rm RF} \lesssim 40~\mum}$), like Spitzer \citep{Werner2004ApJS} and Herschel \citep{Pilbratt2010}, and to Far-Infrared (FIR, ${45 \lesssim \lambda_{\rm RF} \lesssim 350~\mum}$) wavelengths (e.g. ALMA, NOEMA) have made possible to study the panchromatic SED of bright (${M_{UV}\lesssim -26}$) quasars at ${z\sim 6}$.

SEDs observations obtained with Herschel and Spitzer in these sources \citep[][]{Leipski2013,Leipski2014ApJ} have been used to disentangle the star formation versus AGN contribution to the total restframe IR emission (TIR, ${8 < \lambda_{\rm RF} < 1000~\mum}$). The result of this study is that star formation may contribute ${25-60\%}$ to the bolometric TIR luminosity, with strong variations from source to source. In particular, \citet{Leipski2014ApJ} performed a multi-component SED analysis on a sample of 69 ${z>5}$ quasars, finding that a clumpy torus model needs to be complemented by an hot (${\sim 1300}$~K) dust component to match the NIR data, and by a cold (${\sim 50}$~K) dust component for the FIR emission. This work shows that, in addition to the standard AGN-heated component, a large variety of dust conditions is required to reproduce the observed SED. Yet these kinds of studies are limited to a small sample of bright sources. Future facilities in the rest-frame MIR, such as the proposed Origins Space Telescope (OST; \citealt{Wiedner2020arXiv}) with a sensitivity $\sim 1000$ higher than its precursors Spitzer and Herschel, will significantly improve our knowledge of dusty galaxies in the Epoch of Reionization.

ALMA and NOEMA observations have provided the opportunity of studying the ISM/CGM properties of bright ${z\sim 6}$ quasar hosts \citep[e.g.][]{CarilliWalter2013,gallerani2017PASA}, by means of rest frame FIR emission lines, as the [CII] line at 158 micron \citep[e.g.][]{Maiolino2005A&A, Walter2009Natur, Wang2013ApJ, Venemans2016ApJ, novak:2019}, CO rotational transitions \citep[e.g.][]{Bertoldi:2003, Walter:2003, Riechers:2009, Gallerani:2014, venemans:2017CO, carniani:2019, li:2020}, and the corresponding dust continuum emission \citep[e.g.][]{bertoldi:2003dust, Venemans2016ApJ, Venemans:2017ApJ, novak:2019}. These observations have shown that these massive galaxies (${M_{\rm dyn}\sim 10^{10}-10^{11}~\msun}$) are characterized by high star formation rates ($SFR\sim 100-1000~\msunyr$), and large amount of molecular gas ($\sim 10^{10}~\msun$) and dust ($\sim 10^8~\msun$), that are typically distributed on galactic scales (${\lesssim 5}$~kpc). In some exceptional cases \citep{maiolino:2012,cicone:2015}, the extension of [CII] emitting gas has been detected up to CGM scales (${\sim 20-30}$~kpc), possibly driven by fast outflowing gas (${v_{\rm out}\gtrsim 1000~\rm km~s^{-1}}$) with extreme mass outflow rate of ${\dot{M}_{\rm out}\sim 1000~M_{\odot}~{\rm yr}^{-1}}$.

These FIR data, combined with X-ray and UV observations, have shown the presence of galaxy/AGN companions in the field of ${z\sim 6}$ quasars. In the X-ray, whereas the fraction of dual AGN can be as high as ${\sim 40-50}$\% out to ${z\sim 4.5}$ \citep[e.g.][]{Koss:2012, Vignali2018, Silverman:2020}, at ${z\sim 6}$, there are only tentative X-ray detections of double systems (e.g. \citealt{Vito:2019_cand,Connor:2019}, but see also \citealt{Connor:2020}.)

The occurrence of UV detected and sub-mm galaxy (SMG) companions is instead more frequent: \citet{Marshall2020ApJ} detected up to nine companions with ${-22\lesssim M_{\rm UV}\lesssim -20}$ in the field of view of six quasars at ${z\sim 6}$ \citep[but see also][]{Mechtley2012ApJ}; \citet{Decarli:2017} reported the [CII] line and $1$~mm continuum ($F_{\rm cont}$) detection of SMGs close to 4 (out of ${\sim 20}$) quasars at ${z\sim 6}$, with ${0.2\lesssim F_{\rm cont} \lesssim 2.0}$~mJy, and reported projected distances between ${\sim 8}$ and ${\sim 60}$~kpc.

ALMA data of ${z\sim 8}$ Lyman Break Galaxies \citep[][]{laporte:2017apj,Bakx:2020} have suggested the presence in these sources of dust hotter than expected \citep[${T\sim 60-90~K}$,][]{Behrens:2018, Arata:2019, Sommovigo:2020}. The origin of warm dust in early galaxies can be traced back to their (i) large SFR surface densities that favour an efficient heating of dust grains \citep{Behrens:2018} and (ii) more compact structure of molecular clouds (MC) that delays their dispersal by stellar feedback, implying that a large fraction (${\sim 40\%}$) of the total UV radiation remains obscured \citep{Sommovigo:2020}. Another possibility concerns the presence of obscured, accreting, massive (${\sim 10^8~M_{\odot}}$) BHs, whose UV luminosity is absorbed by dust located in the ISM of the host (${\lesssim 1}$~kpc) and/or into a central obscurer, closer to the active nuclei (${\sim 1}$~pc), and heated to temperatures as high as 80-500 K, respectively (Orofino et al. submitted). According to this scenario, buried AGN should be searched for among Lyman break galaxies (LBGs) populating the bright-end of their UV luminosity function (${-24<M_{\rm UV}<-22}$), where indeed a large fraction of objects consists of spectroscopically confirmed AGN \citep{ono:2018}. 

Obscured AGN may therefore represent a bridge between LBGs and bright quasars in the galaxy formation process. In this appealing scenario, the following questions arise: {\it (i) If high-$z$ galaxies contain an obscured AGN, does this imply warmer dust temperatures? (ii) Is there a relation between the dust temperature and the BH accretion rate?(iii) What are the most promising spectral ranges and observational strategies to detect obscured AGN?} To answer these questions it is necessary to build up a model that follows the co-evolution of BHs with their host galaxy from their birth up to the formation of SMBHs powering ${z\sim 6}$ quasars, while accounting for AGN and stellar feedback. The final aim is to produce synthetic multi-wavelength SEDs that can be directly compared with the aforementioned observations of ${z\sim 6}$ quasars to validate the underlying galaxy-BH formation model. This can be done by post-processing cosmological hydro-dynamical simulations with dust radiative transfer calculations. 

Several works in the past years made use of radiative transfer simulations to understand the AGN contribution to the total IR emission of a galaxy, mainly focusing on Ultra Luminous Infrared Galaxies (ULIRGs), and late-stage mergers \citep[e.g.][]{Chakrabarti2007ApJ, Chakrabarti2009ApJ, Younger2009MNRAS, Snyder2013ApJ, Roebuck2016ApJ, Blecha2018MNRAS}. However, these studies are limited up to $z\sim 3$ and they rely on hydrodynamical simulations in which the initial conditions of both the dark matter and gas components were set with analytical prescriptions. Recently, \citet{Schneider:2015} have studied the origin of the infrared emission in SDSS J1148+5251, a ${z\sim 6}$ quasar, by applying dust RT calculations to the output of a semi-analytical merger tree code finding that the dust heating by the AGN radiation may contribute up to ${70\%}$ of the total IR luminosity. This is consistent with the results found by \citet{Li2008ApJ} that computed RT calculations on hydrodynamical simulations of luminous quasars to reproduce the SED of SDSS J1148+5251. They also found that the AGN contribution to the IR emission is significant, because dust heating is dominated by the central source during the quasar-phase.

In this work, we investigate the imprints of AGN in the IR emission of ${z\sim 6}$ galaxies by post-processing cosmological hydrodynamic simulations of SMBHs formation \citep[][hereafter {\citetalias{paramita:2018}}]{paramita:2018} with dust RT calculations performed by using the code SKIRT. The \citetalias{paramita:2018} simulations studied the growth of SMBHs (${10^8-10^9~\msun}$ at ${z=6}$) and the impact of different AGN feedback prescriptions on their host galaxies, residing in a ${\sim 10^{12}~\msun}$ dark matter halo. 

The paper is organised as follows: in Section \ref{Numerical_methods} we illustrate both the hydrodynamical simulations (Section \ref{hydro_sim}) and the model adopted for the radiative transfer calculations (Section \ref{RT_sim}). We present our results in Section \ref{sec:results} and we compare them with observations in Section \ref{sec:syn_vs_obs}. We then make predictions for the proposed mission ORIGINS in Section \ref{sec:spica}. Finally we summarise our results in Section \ref{sec:conclusions} along with our conclusions.

\section{Numerical model} \label{Numerical_methods}

We describe the main characteristics of the hydrodyamical simulations adopted in this work in Section \ref{hydro_sim} and we present the Radiative Transfer (RT) post-processing analysis runs performed in Section \ref{RT_sim}, where we also discuss the details of the numerical setup and the assumptions made for the dust properties and emitting sources.

\subsection{Hydrodynamical simulations} \label{hydro_sim}

The hydrodynamical cosmological zoom-in simulations used in this work are described in details in \citetalias{paramita:2018} and we summarise the main points in the following.

{\citetalias{paramita:2018}} use a modified version of the Smooth Particle Hydrodynamics (SPH) N-body code \code{GADGET-3} \citep{Springel:2005} to follow the evolution of a comoving volume of ${(500~{\rm Mpc})^3}$, starting from cosmological initial condition (IC)\footnote{A flat {$\Lambda$}CDM model is assumed with the following cosmological parameters \citep{PlanckCollaboration2016}: ${\Omega_{\rm M,0}= 0.3089}$, ${\Omega_{\rm \Lambda,0}= 0.6911}$, ${\Omega_{\rm B,0}= 0.0486}$, ${H_0 = 67.74~\rm{km~s}^{-1}~{\rm Mpc}^{-1}}$.} generated with \code{music} \citep{hahn:2011} at ${z=100}$ and zooming-in on the most massive dark matter (DM) halo inside the box down to ${z=6}$\footnote{In the low-resolution DM-only simulation, the most massive halo at $z=6$ has a mass of $M_{\rm halo} = 4.4 \times 10^{12}~\msun$ (virial radius $R_{200}=511$~kpc comoving), massive enough to host luminous AGN, as suggested by clustering studies \citep[e.g.][]{allevato2016}.}. The mass resolution is ${m_{\rm DM} = 7.54 \times 10^6~\msun}$ and ${m_{\rm gas} = 1.41 \times 10^6~\msun}$ for DM and gas particles, respectively. For these high-resolution DM and gas particles the gravitational softening length is ${1~ h^{-1}~{\rm kpc}}$ comoving. For the gas, the smoothing length is determined at each time step according to the local density and typically ranges from 300~pc in the ISM ($n\approx 100~\cc$) to 6.5~kpc in the CGM ($n\approx 10^{-2}~\cc$).

The code accounts for radiative heating and cooling according to the tables computed by \citet{Wiersma2009MNRAS}, which also include metal-line cooling. 
Star formation in the ISM is implemented following the multiphase model by \citet{Springel:2003}, adopting a density threshold for star formation of ${n_{SF} = 0.13 \ \cc}$ and a \citet{Chabrier:2003} initial mass function (IMF) in the mass range ${0.1-100~\msun}$. Stellar evolution and chemical enrichment are computed for the eleven element species (H, He, C, Ca, O, N, Ne, Mg, S, Si, Fe) tracked in the simulation, following \citet{Tornatore:2007}. Kinetic feedback from supernovae (SN) is included by relating the wind mass-loss rate ($\dot{M}_{\rm SN}$) with the star formation rate ($\dot{M}_\star$) as $\dot{M}_{\rm SN} = \eta \dot{M}_\star$ and assuming a mass-loading factor $\eta=2$. The wind kinetic energy is set to a fixed fraction $\chi$ of the the SN energy: $\frac{1}{2} \dot{M}_{\rm SN} v^2_{\rm SN} = \chi \epsilon_{\rm SN} \dot{M}_\star$, where $v_{\rm SN}=350$~km~s$^{-1}$ is the wind velocity and $\epsilon_{\rm SN} = 1.1 \times 10^{49}$~erg~$\msun^{-1}$ is the average energy released by a SN for each $\msun$ of stars formed\footnote{In the ISM multiphase model adopted here \citep{Springel:2003}, kicked particles mimicking stellar winds are temporarily hydrodynamically decoupled. This procedure may affect both the properties of the resulting outflows and the structure of the surrounding ISM \citep[e.g.][]{dallavecchia2008}.}.

In the simulation each BH is treated as a collisionless sink particle and the following seeding prescription is used. When a DM halo -- that is not already hosting a BH -- reaches a total mass of ${M_{\rm h} = 10^9~\msun}$, a ${M_{\rm BH} = 10^5~\msun}$ BH is seeded at its gravitational potential minimum location. BHs are allowed to grow by accretion of the surrounding gas or by mergers with other BHs. Gas accretion onto the BH is modelled via the classical Bondi-Hoyle-Littleton accretion rate ${\dot{M}_{\rm Bondi}}$ \citep{Hoyle:1939, Bondi:1944, Bondi:1952} and it is capped at the Eddington rate ${\dot{M}_{\rm Edd}}$. The final BH accretion rate ${\dot{M}_{\rm BH}}$ reads as follows:
\begin{equation}
    \dot{M}_{\rm BH} = {\rm min} (\dot{M}_{\rm Bondi}, \dot{M}_{\rm Edd}).
\end{equation}
To avoid BHs moving from the centre of the halo in which they reside because of numerical spurious effects, we implement BH repositioning or \emph{pinning} \citep[see also e.g.][]{Springel2005MNRAS, Sijacki2007MNRAS, Booth2009MNRAS, Schaye2015MNRAS}: at each time-step BHs are shifted towards the position of minimum gravitational potential within their softening length. During its growth a BH radiates away a fraction of the accreted rest-mass energy, with a bolometric luminosity
\begin{equation}\label{eq:luminosity_bh}
    L_{\rm bol} = \epsilon_{\rm r} \dot{M}_{\rm BH} c^2,
\end{equation}
where $c$ is the speed of light and $\epsilon_{\rm r}$ is the radiative efficiency. \citetalias{paramita:2018} set ${\epsilon_{\rm r} = 0.1}$, a fiducial value for radiatively efficient, geometrically thin, optically thick accretion disks around a Schwarzschild BH \citep{Shakura:1973}. A fraction ${\epsilon_{\rm f} = 0.05}$ of this energy is distributed to the surrounding gas in a kinetic form\footnote{We refer to \citetalias{paramita:2018} for details about the choice of the value for $\epsilon_{\rm f}$ and the numerical implementation of the kinetic feedback.}. 

In this work we consider the following three runs performed by \citetalias{paramita:2018}, starting from the same ICs:
\begin{enumerate}
    \item[$\bullet$]\noAGN{}: control simulation without BHs.
    \item[$\bullet$]\AGNsphere{}: simulation accounting for BH accretion and AGN feedback. The kinetic feedback is distributed according to a spherical geometry.
    \item[$\bullet$]\AGNcone{}: same as the \AGNsphere{} run, but with kinetic feedback distributed inside a bi-cone with an half-opening angle of ${45\degree}$.
\end{enumerate}
In Table \ref{tab:hydro_runs} we report the main physical properties of the zoomed-in halo at ${z=6.3}$ inside a cubic region of $60$ physical kpc size (the virial radius of the most massive halo is ${\approx 60}$~kpc) centred on the halo's centre of mass. This choice allows to have an overview of all the relevant dynamical structures around the central galaxy, i.e. satellites, clumps, filaments, star forming regions. 

\begin{table*}
	\centering
	\begin{tabular*}{0.85\textwidth}{ c c c c c c c c}
	\hline\noalign{\smallskip}
         simulation run      & AGN feedback   & $M_{\rm gas}$ [$\msun$]   &  $M_\star$ [$\msun$]      & $\SFR$ [$\msunyr$]  & $\dot{M}_{\rm BH}$  [$\msunyr$] &$M_{\rm UV}$ [mag]\\
         \hline\hline\noalign{\smallskip}
         \noAGN{}         & no         &   $2.9 \times 10^{11}$    &  $1.2 \times 10^{11}$ & $ 600$              &   -   &  \\
         \AGNsphere{}     & spherical  &   $2.1 \times 10^{11}$    &  $6.5 \times 10^{10}$ & $ 312$              &   $3.1$ &-24.32   \\
         \AGNcone{}       & bi-conical    &   $1.4 \times 10^{11}$    &  $7.0 \times 10^{10}$ & $ 189$              &   $89$  &-27.97  \\
	\hline
	\end{tabular*}
  	\caption{Summary of the hydrodynamic runs of \citetalias{paramita:2018} used in this work. For each run, we indicate the feedback model used in the simulation and the main physical properties of the zoomed-in halo at ${z=6.3}$ within a cubic region of $60$~kpc size (that corresponds to $\sim 50$\% of the virial radius): gas mass ($M_{\rm gas}$), stellar mass ($M_\star$), star formation rate ($\SFR$, averaged over the last $10$~Myr), and the sum of the accretion rate of all the black holes (BHs) in the selected region ($\dot{M}_{\rm BH}$). We further associate to $\dot{M}_{\rm BH}$ an intrinsic UV magnitude $M_{\rm UV}$ (see Appendix \ref{MdotMuvrelation}).
  	\label{tab:hydro_runs}
  	}
\end{table*}

\begin{figure*}
    \centering
    \includegraphics[width=0.95\textwidth]{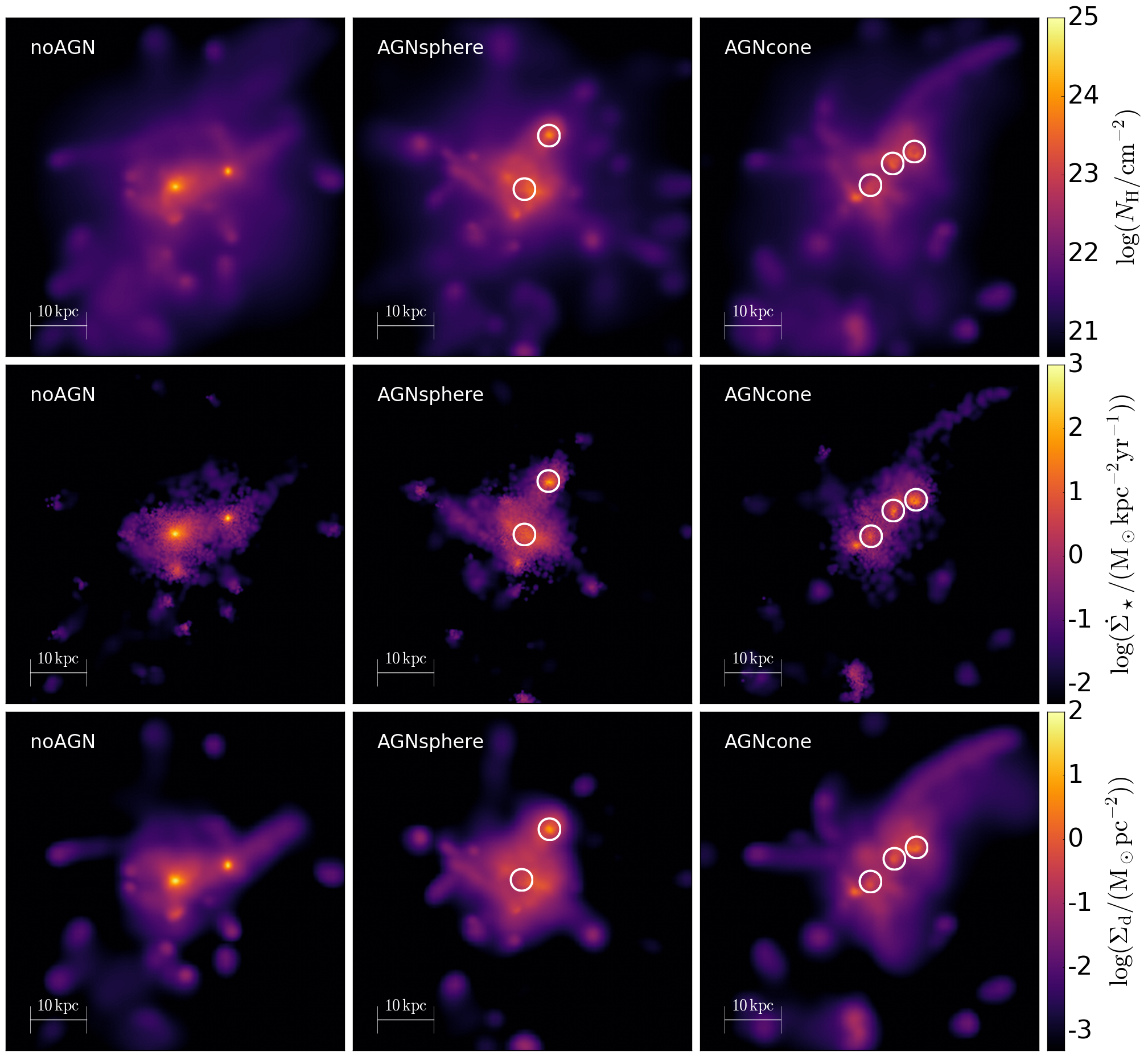}
\caption{Morphology of the most massive halo at ${z=6.3}$ inside a cubic box of $60$~kpc size for the three cosmological simulations of \citetalias{paramita:2018}: \noAGN{} (left column), \AGNsphere{} (middle column) and \AGNcone{} (right column). The top, middle and bottom panel show the hydrogen column density, the star formation rate and the dust surface density (assuming a dust-to-metal ratio ${f_d=0.08}$, see Section \ref{dust_model}), respectively. White empty circles show the location of BHs accreting at ${\dot{M}_{\rm BH}> 1~\msunyr}$.
    \label{fig:runs_hydro}
    }
\end{figure*}

In Fig. \ref{fig:runs_hydro} we show the hydrogen column density (top row) and the star formation rate (middle row) for the zoomed-in halo in the three simulations for a line of sight aligned with the angular momentum of the particles inside the selected region. In the following, this is our reference line of sight.  
From the top row, it can be seen that the central region, corresponding to the main galaxy, is characterised by the highest column density in all the runs. It reaches values of ${N_H \sim 6 \times 10^{24}~\colcm}$ in the \noAGN{} run, whereas it is an order of magnitude lower when AGN feedback is included. This is because kinetic feedback kicks gas away from the accreting BHs. In turn, the decreased gas density quenches the overall SFR density. In fact, star formation rate densities $\Sigma_{\rm SFR}$ as high as ${\Sigma_{\rm SFR}\approx 600~\surfsfr}$ are found in the \noAGN{} run, in sharp contrast with those in the \AGNsphere{} (${\approx 130~\surfsfr}$, characterized by a total BH accretion rate ${\dot{M}_{\rm BH}=3.1~\msunyr}$), and \AGNcone{} (${\approx 50~\surfsfr}$, ${\dot{M}_{\rm BH}=89~\msunyr}$) cases. The same trend is observed also for the total SFR, as reported in Table \ref{tab:hydro_runs}. 

\subsection{Radiative transfer}\label{RT_sim}

We post-process the snapshots at ${z=6.3}$ of the three selected hydrodynamic simulations in {\citetalias{paramita:2018}} by using the publicly available code \code{SKIRT}\footnote{Version 8, \url{http://www.skirt.ugent.be}.} \citep{Baes:2003, Baes:2015, Camps:2015, Camps2016}. \code{SKIRT} solves the continuum radiative transfer problem in a dusty medium with a Monte-Carlo approach, by sampling the SED of the sources with a finite number of photon packets (in the following simply referred to as \emph{photons}). Photons are scattered and/or absorbed by dust grains in the simulation volume according to their properties. Dust grains, after being heated up, thermally re-emit the absorbed energy at IR wavelengths. One of the main advantages of the \code{SKIRT} code is its flexibility: it allows the user to handle input data from different numerical codes (e.g. Adaptive Mesh Refinement and Smooth Particle Hydrodynamic codes), to account for different dust properties (i.e. grain size distribution and composition), to implement different SEDs for the radiating sources (e.g. stars and accreting BHs), to include many physical mechanisms (e.g. dust stochastic heating and self-absorption).

To relate the energy absorbed by dust with its wavelength-dependent emissivity we adopt the dust models described in Section \ref{dust_model}. We describe the SED adopted in different RT runs for stars and accreting BHs in Section \ref{radiation_field}. 

\subsubsection{Dust properties} \label{dust_model}

Dust formation, growth and destruction processes are not tracked in the hydrodynamic simulations considered here. Similarly to other RT works \citep{Behrens:2018,Arata:2019,Liang:2019}, we derive the dust mass distribution by assuming a linear scaling with the gas metallicity\footnote{Throughout this paper the gas metallicity is expressed in solar units, using ${\zsun=0.013}$ as a reference value \citep{Asplund:2009}.} \citep{Draine:2007}, parametrizing the mass fraction of metals locked into dust as:
\begin{equation}
    f_d = M_{\rm d} / M_Z,
\end{equation}
where $M_{\rm d}$ is the dust mass and $M_Z$ is the total mass of all the metals in each gas particle in the hydrodynamical simulation (see Section \ref{hydro_sim}).
The choice of $f_d$ directly affects the total dust content. The RT calculation is sensitive to the $f_d$ value, which is poorly constrained by high-redshift galaxies observations (see \citealt{Wiseman:2017} and references therein) and theoretical models \citep{Nozawa:2015}. In particular, recent theoretical works \citep{Asano:2013, Aoyama:2017} suggest that $f_d$ is constant in the early stages of galaxy evolution and then it grows with metallicity up to the Milky-Way (MW) value of ${f_d = 0.3}$ when/if dust growth becomes important. However, the efficiency of dust growth in the ISM of early galaxies is highly debated \citep{Ferrara:2016b}. In this work, we consider a constant value of $f_d$, and focus our attention on how the dust content of galaxies affects their panchromatic SED.

We adopt two different $f_d$ values for the normalization: i) a MW like value (${f_d=0.3}$); ii) a lower value (${f_d=0.08}$) tuned for hydro-simulations \citep{pallottini:2017althaea, Behrens:2018} to reproduce the observed SED of a ${z\sim 8}$ galaxy \citep{laporte:2017apj}. The dust surface density distribution derived in the ${f_d=0.08}$ case is shown in the bottom row of Figure \ref{fig:runs_hydro}. High dust surface density regions correspond to active star forming regions where gas metal enrichment is more pronounced. Therefore, gas and dust density, and SFR are generally correlated in our simulations, as can be seen in Fig. \ref{fig:runs_hydro}.

The properties of dust as chemical composition and grain size distribution are not known in early (AGN-host) galaxies. The nature and origin of dust at high redshift is in fact a widely debated topic \citep[e.g.][]{Valiante:2009,Stratta:2011,Asano:2013MNRAS,Hirashita:2015,Hirashita:2019}. Some works \citep[][]{Maiolino:2004,Gallerani:2010} have suggested that ${z\gtrsim 4}$ quasars require an extinction curve that is shallower than the Small Magellanic Cloud (SMC), possibly indicating the presence of SN-type dust \citep{Todini:2001,Bianchi:2007}; however, the SMC extinction curve is instead favoured by the analysis of high-z quasars and GRBs performed by other research groups \citep{Zafar2011,Hjorth2013ApJ,Zafar2018MNRASx}.
For the time being, we assume a dust composition and grain size distribution appropriate for the SMC by using the results\footnote{We consider the revised optical properties evaluated in \citet{Draine:2003a,Draine:2003b,Draine:2003c}.} of \citet{Weingartner:2001}. We defer the inclusion of a SN-type extinction curve to a future work.

\subsubsection{Dust implementation in \code{SKIRT}} \label{sec:dust_skirt}

Dust is distributed in the computational domain in an octree grid with a maximum of 8 levels of refinement for high dust density regions, achieving a spatial resolution of ${\approx 230}$~pc in the most refined cells, comparable with the softening length in the hydrodynamic simulation (${\approx 200}$ physical pc at $z=6.3$)\footnote{When distributing the dust content derived from the hydrodynamical simulation into an octree grid, a kernel-based interpolation is required in order to convert the dust content from a particles-based distribution into an octree geometry. This procedure leads to a discrepancy between the total amount of dust carried by the SPH gas particles imported from the hydrodynamical simulation and the effective dust content in the computational domain used for the RT calculation. Therefore, it is important to check that the structure of the dust grid adopted achieves sufficient convergence relative to the overall dust content. We find that the relative difference in the overall dust content is within $0.1\%$, $0.2\%$, $0.4\%$ for \noAGN{}, \AGNsphere{} and \AGNcone{}, respectively.}. We verify in App. \ref{sec:RT_conv} that the number of refinement levels adopted in our fiducial setup is sufficient to achieve converge of the results.  
Adopting an SMC-like dust, the grain size distribution of graphite and silicates is sampled with 5 bins for each component. 
Gas particles hotter than $10^6$~K, are considered dust-free as at these temperatures thermal sputtering is very effective at destroying dust \citep{Draine:1979, Tielens:1994, Hirashita:2015}. This assumption does not affect the main results of our work, as discussed in App. \ref{Dustthermalsputtering}.

Grain temperature and emissivity are evaluated by imposing energy balance between the local radiation field and dust re-emission. By default, when dust emission photons propagate, \code{SKIRT} accounts for the self-absorption by dust, but it does not take this absorption into account when computing the dust temperature, unless the self-absorption flag is turned on. As this effect may be relevant if dust is IR-optically thick, we have enabled a self-consistent evaluation of the dust temperature, iterating the RT calculation for dust absorption and re-emission until the dust IR luminosity converges within $3 \%$. 

We also include non-local thermal equilibrium (NLTE) corrections to dust emission, which include the contribution from small grains that are transiently heated by individual photons. In this case grains of different sizes are no longer at a single equilibrium temperature, but follow a temperature distribution. \citet{Behrens:2018} found in their calculations that stochastic heating affects mostly the MIR portion of the SED (rest-frame wavelength ${\lesssim 80~\mum}$) but it has a minor impact on the FIR and (sub)mm emission. 

We do not include heating from CMB radiation. As discussed in Section \ref{dust_temperature}, only a small fraction of dust grains is at a temperature comparable to $T_{\rm CMB}$. We expect this effect to be negligible, as seen a posteriori from the RT results.

We do not include any subgrid model for dust clumpiness. Recent works (e.g. \citealt{Camps2016, Trayford2017MNRAS, Liang2021MNRAS}) that account for subresolution structures of birth clouds harboring young stars \citep{Jonsson2010MNRAS}, whose typical scales are not resolved by the hydrodynamical simulations, are based on SED templates \citep{Groves2008ApJS} not consistent with our fiducial set up. The stellar emission in the \citet{Groves2008ApJS} template is, in fact, calculated from Starbust99 models \citep{Leitherer1999ApJS}, by assuming a \citet{Kroupa2002} initial mass function, whereas we model stellar emission using the \citet{Bruzual:2003} model (see Section \ref{radiation_field}), based on the \citet{Chabrier:2003} IMF. Moreover, they include PAH molecules in the dust composition that are instead not considered in our work. We notice that \citet{Liang2021MNRAS} find that the \citet{Groves2008ApJS} template mainly affects the IR emission from PAHs which is enhanced up to $50\%$ (see Fig. 23 of their paper). Given that in our model we adopt an SMC dust composition (i.e. no PAHs), we do not expect that the inclusion of a subgrid model that accounts for dust clumpiness would significantly affect the main results of our work.

\subsubsection{Radiation from stars and AGN} \label{radiation_field}

The ultraviolet (UV) radiation field mainly responsible for dust heating is provided by stellar sources and black holes. We describe in the following how the two components are implemented in our model.

Stellar particles in the simulation represent a Single Stellar Population (SSP), i.e. a cluster of stars formed at the same time and with a single metallicity. Given the mass, age and metallicity of the stellar particle imported, \code{SKIRT} builds the individual SEDs according to the \citet{Bruzual:2003} family of stellar synthesis models, placing the sources at the locations of the stellar particles.

Black holes are treated as point source emitters as the typical sizes of the accretion disk and the dusty torus are much smaller (${\lesssim 10}$~pc) than the width of the most refined grid cells (${\approx 230}$~pc, see Section \ref{dust_model}). We implement their emission in \code{SKIRT} adopting a SED as described in Section \ref{sec:AGN_SED}.

The radiation field is sampled using a grid covering the \emph{rest-frame} wavelength range\footnote{The total AGN bolometric luminosity is distributed from the X-ray to the IR according to the SED adopted (see Section \ref{sec:AGN_SED}). The choice of the wavelength range adopted in our simulations affects the fraction of the AGN bolometric luminosity that effectively enters in the calculation (see Fig. \ref{fig:AGN_SED}). For the \emph{fiducial} SED introduced in Sec. \ref{sec:AGN_SED}, this fraction is ${\approx 60 \%}$, whereas it is ${\approx 40 \%}$ for the \emph{UV-steep} SED.} ${[0.1-10^3]~\mum}$.
The choice of the lower limit is quite common for RT simulations in dusty galaxies \citep{Schneider:2015, Behrens:2018} and it is motivated by the fact that codes like \code{SKIRT} typically do not account for the hydrogen absorption of ionising photons (${\lambda < 912~\angstrom}$).
The choice of ${10^3~\mum}$ as the upper limit of the wavelength grid is motivated by the fact that the intrinsic emission from stars and BHs is negligible above this limit. The base wavelength grid is composed of 200 logarithmically spaced bins. 

A total of $10^6$ photon packets per wavelength bin is launched from each source, i.e. stellar particles and BHs\footnote{We verified that the number of packets used is sufficient to achieve numerical convergence by comparing the results with control simulations with $5 \times 10^5$ photon packets per wavelength bin.}. We collect the radiation escaping our computational domain for the 6 lines-of-sight perpendicular to the faces of the cubic computational domain.

\subsubsection{AGN Spectral Energy Distribution} \label{sec:AGN_SED}

\begin{figure*}
	\centering
	\includegraphics[width=0.95\textwidth]{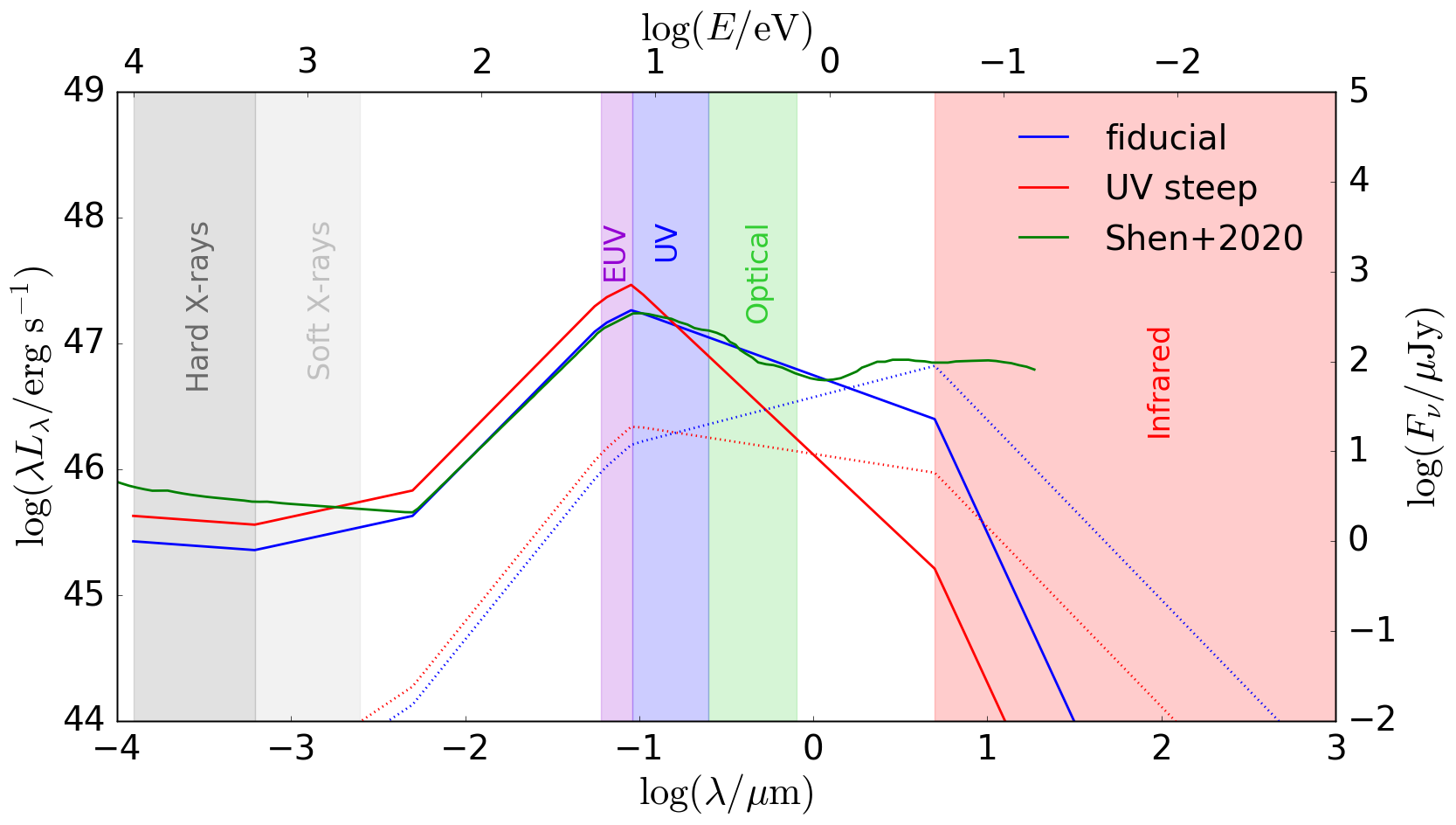}
	\caption{AGN SED for a bolometric luminosity ${L_{\rm bol}=10^{13}~\lsun}$: the \emph{fiducial} SED (${\alpha_{\rm UV} = -1.5}$) is shown with a blue thick line; the \emph{UV-steep} SED (${\alpha_{\rm UV} = -2.3}$) is shown with a red thick line. We plot the SED template derived in \citet{Shen:2020} for comparison with a thick green line, re-scaling it in order to have the same $L_{2500 \angstrom}$ of the fiducial SED. The SEDs differ mainly at wavelength longer than the UV band, with the UV-steep SED dropping faster than the other two. The fiducial SED is in very good agreement with the \citet{Shen:2020} SED up to ${\approx 2~\mum}$, from where the contribution by dust in the torus and in the galaxy included in their IR template begins to dominate the emission in their SED. As a reference for the SED plots in the following, we also plot our two SEDs as the $F_\nu$ (in $\mu Jy$) vs $\lambda$ with dotted lines, keeping the same colour legend.
	\label{fig:AGN_SED}
	}
\end{figure*}

The SED of an AGN is shaped by the numerous physical mechanisms involved in the process of gas accretion onto the BH (see \citealt{Netzer:2015} for a comprehensive review on this topic). AGN SED templates are typically based both on theoretical arguments and observations \citep[e.g.][]{Shakura:1973, VandenBerk:2001, Sazonov:2004, Manti:2016,Shen:2020}, possibly including the dusty torus modelling \citep{Schartmann:2005,Nenkova2008ApJ, Stalevski:2012,Stalevski:2016}. For this work, we adopt a composite power-law for the AGN emission written as:
\begin{equation}\label{AGN_SED_eq2}
     L_\lambda = c_i \ \left(\frac{\lambda}{\mu{\rm m}}\right)^{\alpha_i} \ \left(\frac{L_{\rm bol}}{\lsun}\right) \ \lsun \ {\mum}^{-1},
\end{equation}
where $i$ labels the bands in which we decompose the spectra and the coefficients $c_i$ are determined by imposing the continuity of the function based on the slopes $\alpha_i$. The coefficients $c_i$ and $\alpha_i$ adopted and the relative bands are reported in Table \ref{tab:AGN_SED_coeff} and they are chosen as described in the following.

For the X-ray band, based on the results by \citet{Piconcelli:2005} and \citet{Fiore:1994} in the hard (${2-10}$~keV, ${\alpha_{\rm X,hard} = -1.1 \pm 0.1}$) and soft (${0.5-2}$~keV, ${-0.7 < \alpha_{\rm X,soft} < 0.3}$) band, respectively, we consider ${\alpha_{\rm X,hard} = -1.1}$ and ${\alpha_{\rm X,soft} = -0.7}$.
Consistently with \citet{Shen:2020}, in the wavelength range $50-600~\angstrom$ we use ${\alpha = 0.4}$ (the slope chosen for the soft X-ray band is then adopted up to $50~\angstrom$ for continuity).
For the Extreme UV band (EUV, ${600<\lambda<912~\angstrom}$) we use ${\alpha_{\rm EUV} = -0.3}$ as in \citet{Lusso:2015}. We also note that this value is consistent with the constraints by \citet[][${-1.0 < \alpha_{\rm EUV} < -0.3}$]{Wyithe:2011} based on the analysis of near-zones observed around high redshift quasars.

The analysis of a large sample (4576) of ${z\lesssim 2.2}$ quasars \citep{Richards2003AJ} spectra in the range ${1200 \lesssim \lambda \lesssim 6000~\angstrom}$ has shown that the spectral slopes are distributed in the range (${-2.6 < \alpha < -0.2}$) and peak around ${\alpha=-1.6}$. In the ${912 <\lambda<2500~\angstrom}$ band, \citet{Lusso:2015} have constructed a stacked spectrum of 53 quasars at ${z\sim 2.4}$ finding ${\alpha = -1.39\pm0.01}$. Moreover, \citet{Gallerani:2010} have analysed 33 quasars in the redshift range ${3.9\lesssim z \lesssim 6.4}$ finding that unreddened quasars are characterised by ${\alpha=-1.7\pm0.5}$, whereas reddened quasars prefer steeper slopes (${\alpha < -2.3}$). Finally, from a theoretical point of view, the classical black-body composition for a \citet{Shakura:1973}-disk predicts that ${F_\nu \propto \nu^{1/3}}$, which translates into ${\alpha = -2.3}$. Given the uncertain value of the slope for wavelengths longer than $912~\angstrom$, we consider two possible values for the slope in the range from the UV to NIR: ${\alpha_{\rm UV} = -1.5}$, which is representative of unreddened quasars, and ${\alpha_{\rm UV} = -2.3}$. We will refer to these two models as the \emph{fiducial} and \emph{UV-steep} model, respectively.

At longer wavelengths, the intrinsic AGN emission is expected to follow the Rayleigh-Jeans tail regime ${F_\nu \propto \nu^2}$, which corresponds to ${\alpha_{\rm IR} = -4}$. The transition between the UV slope and the IR one increases with the black hole mass \citep{Shakura:1973,Pringle:1981,Sazonov:2004}. In this work, we adopt a transition wavelength ${\lambda_{\rm trans}= 5~\mum}$. This component represents the IR emission from the accretion disk only. We did not include the emission from the hot dust component from the torus because we cannot resolve the scales ($1-10$~pc) of the torus itself. We discuss how this affects our results in Section \ref{obscured_AGN}. 

\begin{table*}
	\centering
	\begin{tabular*}{0.85\textwidth}{ c c c c c c c }
	\hline\noalign{\smallskip}
            & hard X & soft X & X to EUV & EUV & UV to NIR & NIR to FIR \\
            & $[2-10]$ keV & $[6.2-60]~\angstrom$  & $[50-600]~\angstrom$ & $[600-912]~\angstrom$ & $[0.0912 - 5]~\mum$ & $[5-10^3]~\mum$ \\
    \hline\hline\noalign{\smallskip}
    $c$ (fiducial)   &   2          & 0.042   &   14.133    &     1.972    &  0.111    & 6.225    \\
    $\alpha$ (fiducial)  &   -1.1   & -0.7    &   0.4    &    -0.3     &  -1.5  & -4.0             \\
         \hline
    $c$ (UV-steep)   &   0.003         & 0.066  &   22.499    &     3.140    &  0.026 &  0.402             \\
    $\alpha$ (UV-steep)  &   -1.1   & -0.7    &   0.4    &    -0.3     &  -2.3  & -4.0             \\
	\hline\end{tabular*}
  	\caption{Coefficients of our AGN SEDs models as expressed in eq. \ref{AGN_SED_eq2}. The slopes $\alpha_i$ and the ranges of the piece-wise decomposition were chosen as explained in Section  \ref{sec:AGN_SED}. Imposing the continuity of the function determines the coefficients $c_i$. The SED built in this way is by construction normalised to the bolometric luminosity of the source expressed in $\lsun$ according to eq. \ref{AGN_SED_eq2}.
  	\label{tab:AGN_SED_coeff}
  	}
\end{table*}

The fiducial and UV-steep SEDs adopted in this work are shown in Fig. \ref{fig:AGN_SED} with blue and red lines, respectively. We also report with a green line the SED derived by \citet{Shen:2020}. Our bolometric corrections\footnote{Consistently with \citet{Shen:2020}, we express the UV band luminosity as ${\nu_{1450\angstrom}L_{\nu_{1450\angstrom}}}$, the B band as ${\nu_{4400\angstrom}L_{\nu_{4400\angstrom}}}$, whereas the soft [hard] X-ray luminosity is the integrated luminosity in the 0.5-2 [2-10] keV band.} (reported in Table \ref{tab:bolometric_corr}) are consistent with the ones by \citet{Shen:2020} (reported in the top panel of their Fig. 2), for ${L_{\rm bol} \approx 10^{47}~{\rm erg~s}^{-1}}$. 
We further calculate the ${\alpha_{\rm OX} = 0.384 \log L_\nu(2 {\rm keV}) /L_\nu(2500 \angstrom)}$ index for our SEDs and find that it is in agreement with observations of ${z \sim 6}$ quasars \citep[e.g.][]{Nanni:2017,Gallerani:2017, Vito:2019}.

\begin{table}
	\centering
	\begin{tabular*}{0.475\textwidth}{ c c c c c c c }
	\hline\noalign{\smallskip}
        SED model      & $\frac{L_{\rm bol}}{L_{\rm X,hard}}$  &$\frac{L_{\rm bol}}{L_{\rm X,soft}}$ & $\frac{L_{\rm bol}}{L_{\rm UV}}$   & $\frac{L_{\rm bol}}{L_{\rm B}}$      & $\alpha_{\rm OX}$  \\
        \noalign{\smallskip}\hline\hline\noalign{\smallskip}
         \emph{fiducial}   &   130          & 130   &   3.4    &     6.0     &  -1.65              \\
         \emph{UV-steep}   &   80           & 81    &   3.1    &    13.6     &  -1.51              \\
	\hline
	\end{tabular*}
  	\caption{Bolometric corrections (${L_{\rm bol}/L_{\rm band})}$ and $\alpha_{\rm OX}$ for the \emph{fiducial} (${\alpha_{\rm UV} = 1.5)}$ and \emph{UV-steep} (${\alpha_{\rm UV} = 2.3)}$ AGN SED models adopted in this work. The bands used to compute the bolometric corrections are defined as: hard X-ray $[2-10]$~keV, soft X-ray $[0.5-2]$~keV, UV $[0.1-0.3]~\mum$. $L_{\rm B}$ is defined as ${\lambda L_\lambda}$ at $\lambda=4400~\angstrom$. The two models mostly differ for the luminosity in the B band. For a bolometric luminosity ${L_{\rm bol} = 10^{47}~{\rm erg~s}^{-1}}$, our bolometric corrections are consistent with the observational constraints reported in the top panel of Fig. 2 by \citet{Shen:2020}. 
  	\label{tab:bolometric_corr}
  	}
\end{table}

\section{Results} \label{sec:results}

We perform RT calculations on the three hydrodynamic simulations presented in section \ref{hydro_sim}. For each hydro-simulation we vary the dust to metal ratio from ${f_d = 0.08}$ to ${f_d = 0.3}$; for the \AGNcone{} run we consider both the AGN SEDs described in Section \ref{sec:AGN_SED}. We end up with a total of 8 post-processed runs, as reported in Table \ref{tab:SKIRT_runs_setup}.

\begin{table*}
	\centering
	\begin{tabular*}{0.65\textwidth}{l l c c c c }
		\hline\noalign{\smallskip}
		 RT run name & Hydro run name & Radiation field & AGN SED & $f_d$ \\
		\hline\hline\noalign{\smallskip}
		\emph{noAGN${008}$} & noAGN & stars & & $0.08$ \\
		\emph{noAGN${03}$} & noAGN & stars & & $0.3$ \\
		\emph{AGNsphere$008$} & AGNsphere & stars + BHs & fiducial & $0.08$ \\
		\emph{AGNsphere$03$} & AGNsphere & stars + BHs & fiducial & $0.3$  \\
		\emph{AGNcone$008$} & AGNcone & stars + BHs & fiducial & $0.08$ \\
		\emph{AGNcone$03$} & AGNcone & stars + BHs & fiducial & $0.3$ \\
		\emph{AGNcone$008$UVsteep} & AGNcone & stars + BHs & UV-steep & $0.08$ \\
		\emph{AGNcone$03$UVsteep} & AGNcone & stars + BHs & UV-steep & $0.3$ \\
	\hline\end{tabular*}
  	\caption{\code{SKIRT} post-processing runs performed. The first column labels the RT simulation, the second column indicates the corresponding hydrodynamical run, the third column specifies the radiation field included (e.g. stars with or without black holes), the fourth column specifies the AGN SED used (if black holes are present), and the fifth column contains the dust to metal ratio $f_d$ adopted.
  	  \label{tab:SKIRT_runs_setup}
  	  }
\end{table*}

In this section we present the results obtained through our RT calculations. We first present in Section \ref{results_overview} the morphology of the ultraviolet (UV, ${1000-3000~\angstrom}$) and total infrared (TIR, ${8-1000~\mum}$) emission and discuss how it is affected by the presence of the AGN and total dust content. Then, in Section \ref{dust_temperature} we derive the dust temperature in the different runs. Finally, we discuss in Section \ref{sec:sed} the synthetic SEDs resulting from our calculations.

\subsection{Overview} \label{results_overview}

\begin{figure*}
    \centering
    \includegraphics[width=0.95\textwidth]{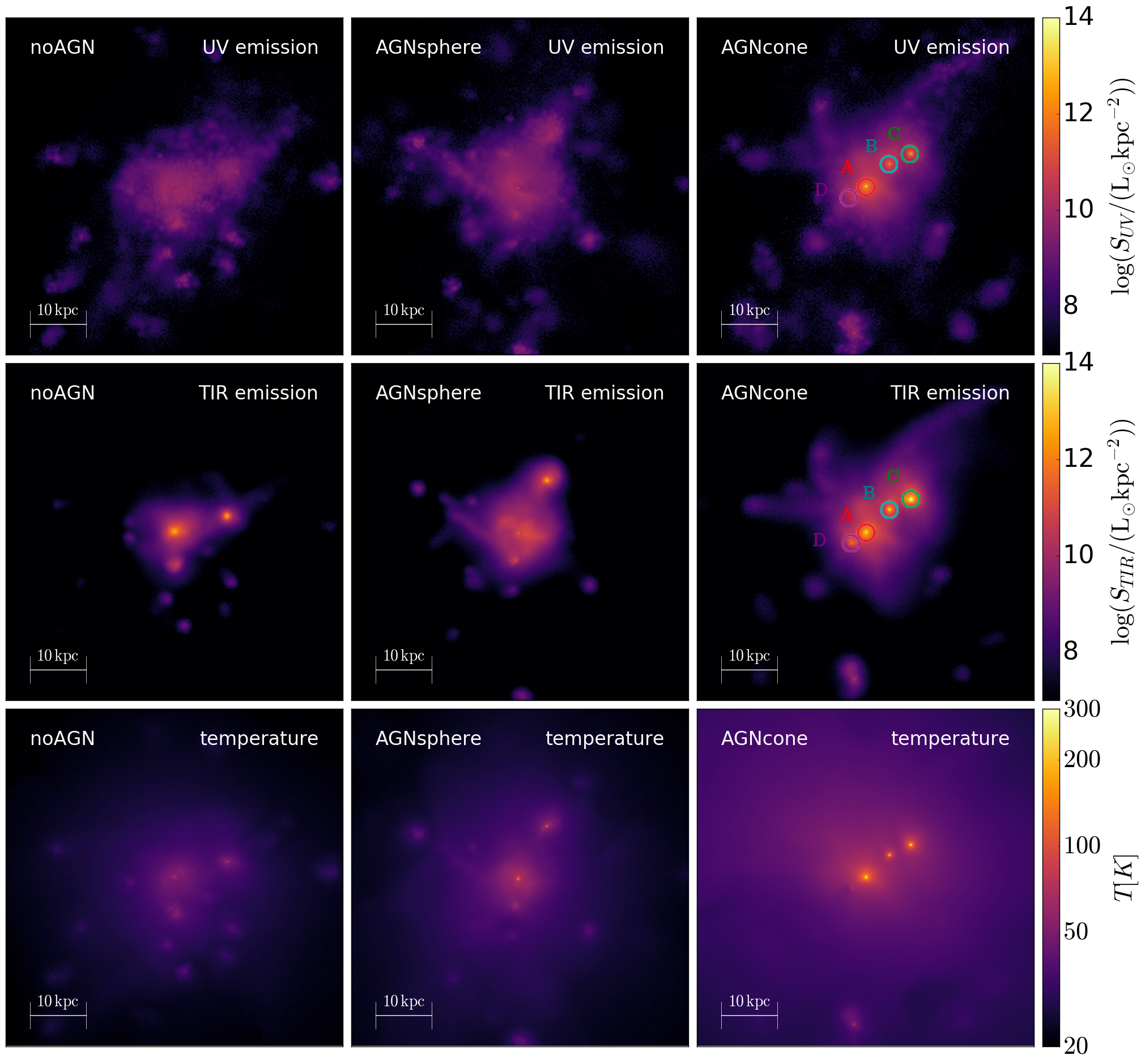}
    \caption{UV (top row), TIR (middle row), luminosity-weighted dust grain temperature (bottom row) maps for the runs with ${f_d=0.08}$. The maps shown are produced with the same line of sight used in Fig. \ref{fig:runs_hydro}. We mark the four most luminous sources in TIR for the \AGNcone{} runs, which will be discussed in more details in Section \ref{sec:individual_sources}. 
    \label{fig:runs_RT_008}
    }
\end{figure*}

\begin{figure*}
    \centering
    \includegraphics[width=0.95\textwidth]{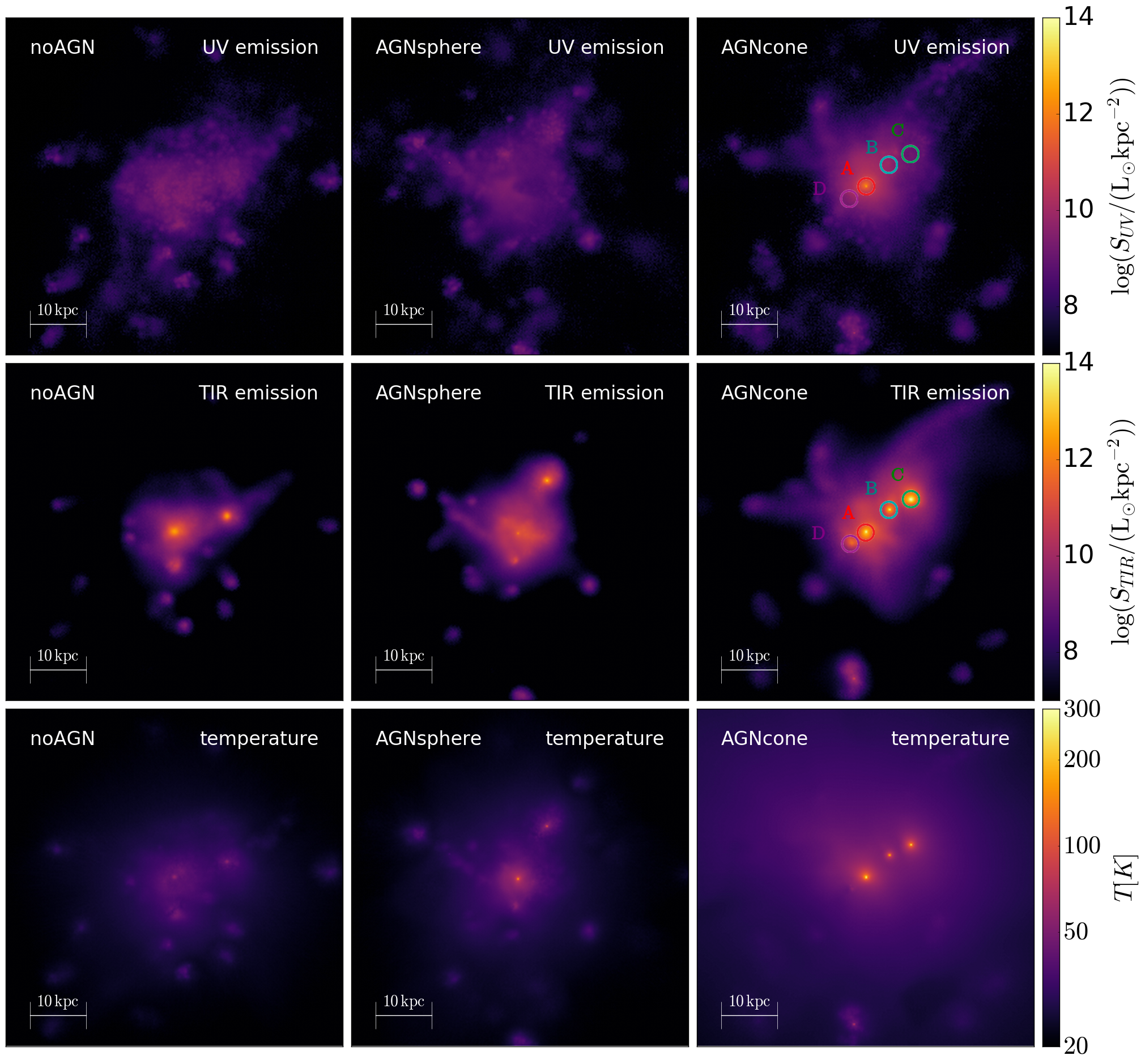}
    \caption{Same as in figure \ref{fig:runs_RT_008} but for for ${f_d=0.3}$. 
    \label{fig:runs_RT_03}
    }
\end{figure*}

In Fig. \ref{fig:runs_RT_008}, we show the UV (top row) and TIR (middle row) emission maps derived for the runs \noAGN{} (left column), \AGNsphere{} (middle column), \AGNcone{} (right column) for ${f_d = 0.08}$. In Fig. \ref{fig:runs_RT_03} we show the same maps but for ${f_d = 0.3}$. We use the same line of sight as in Fig. \ref{fig:runs_hydro}. 

By comparing the TIR maps with the dust surface density (Fig. \ref{fig:runs_hydro}, bottom row) we see that the morphology of the TIR emission matches the dust distribution, as expected. Moreover, the brightest TIR spots in the \noAGN{} (AGN runs) correspond to the locations of the most highly star forming regions (accreting BHs), responsible for the dust grains heating. We discuss in more details the dust temperature in Section  \ref{dust_temperature}.  

For what concerns UV emission, in the \noAGN{} case, its distribution correlates with the star formation surface density (see middle row in Fig.\ref{fig:runs_hydro}); in the AGN runs, the brightest spots are located in correspondence of the AGN positions, identified by white circles. Noticeably, whereas in the \AGNcone{} simulation with ${f_d=0.08}$ three peaks appear in the UV emission map (labelled as A, B, and C in Fig. \ref{fig:runs_RT_008}), corresponding to the AGN positions\footnote{The accretion rate quoted in Table \ref{hydro_sim} for the case \AGNcone{} is in fact the sum of the accretion rates of the most active black holes in the simulations ($\dot{M}_{\rm BH}\approx$ 32, 7 and 50 $\msunyr$, for the sources A, B, and C, respectively).}, in the case ${f_d=0.3}$ only one of them survives to the strong dust obscuration. In Section \ref{sec:individual_sources} we investigate in further details the contribution to the total SED of the different components traced by the UV and TIR maps.

Table \ref{tab:SKIRT_runs_results} reports the UV and TIR luminosities before and after the dust-reprocessing of the radiation. We find that, in the \noAGN{} run ${84-94\%}$ of the total UV emission is extincted by dust if ${f_d=0.08-0.3}$. For comparison, in the \AGNcone{} and \AGNsphere{} runs, the same fraction is ${77-99\%}$ and ${54-95\%}$, respectively. Overall we find that in our simulated dusty galaxies (${M_{\rm d}\gtrsim 3\times 10^7 M_{\rm \odot}}$) a large fraction (${\gtrsim50\%}$) of UV emission is obscured by dust, with some lines of sight characterised by $1\%$ of UV transmission. 

The range reported for the UV reprocessed luminosity in Tab. \ref{tab:SKIRT_runs_results} refers to the variation occurring along different lines of sights: the minimum and maximum values differ by a factor that can be as high as ${\sim 6}$ in the AGN runs. We expect UV luminosity variations along different lines of sight even larger than the ones we find, if UV radiation would intersect dense, compact, dusty, molecular clouds, whose sizes ($\lesssim 100$~pc) and complex internal structure \citep[$\sim 1-10$~pc,][]{Padoan:2011,Padoan:2014,Vallini:2017} are not resolved by of our simulations. 

According to the Unified Model \citep{Urry:1995}, the classification between TypeI (unobscured) and TypeII (obscured) AGN is based on the presence of a dusty, donut-like shaped structure that is responsible for anisotropic obscuration in the circum-nuclear region ($<~10$~pc). Our results show that large UV luminosity variations with viewing angle, in addition to the ones due to the torus, arise from the inhomogeneous distribution of dusty gas surrounding the accreting BH, on ISM scales \citep[${\gtrsim 200}$~pc; see also][]{gilli2014}. 

\begin{table*}
	\centering
	{\def\arraystretch{1.5}
	\begin{tabular}{ l c c c c c c c}
		\hline\noalign{\smallskip}
		RT run & $L_{\rm UV}$            & $L_{\rm TIR}$           & M$_{\rm d}$  & $\langle T_{\rm d}\rangle_L$  & $T^{\rm min/max}_{\rm d}$  & $L_{\rm UV}^{intr}$              & $\tau_{\rm UV}$ \\
		       & [10$^{11}$ $\lsun$] & [10$^{12}$ $\lsun$] & [10$^7$  $\msun$] &      [K]           &   [K]                     & [10$^{12}$ $\lsun$]   &   \\
		\hline\hline\noalign{\smallskip}
		\emph{noAGN${008}$}& $5.7 - 6.7$ & $4.4 - 4.6$  & $9.2$    & $54 \pm 6$   &   $15 - 64$   &  $4.1$           & $1.81 - 1.97$   \\
		\emph{noAGN${03}$}& $3.2 - 3.9$ & $4.7 - 4.9$ & $34$    & $48 \pm 6$   &   $13 - 57$   &  $4.1$            & $2.35 - 2.56$  \\
		\emph{AGNsphere$008$} & $7.0 - 17$ & $3.3 - 3.4$  & $5.1$    & $70 \pm 27$   & $17 - 179$    &$3.7$           & $0.78 - 1.65$  \\
		\emph{AGNsphere$03$}  & $2.8 - 7.9$ & $4.4 - 4.6$ & $19$    & $62 \pm 25$    & $15 - 178$  & $3.7$            & $1.54 - 2.57$ \\
		\emph{AGNcone$008$} & $27 - 90$ & $43 - 50$ & $3.3$    & $208 \pm 78$ &  $22 - 282$ &       $39$      & $1.47 - 2.67$       \\
		\emph{AGNcone$03$}& $5.8 - 32$ & $54 - 71 $  & $13$    & $182 \pm 69$ &  $20 - 272$ &       $39$     & $2.50 - 4.20$ \\
	\hline\end{tabular}
	}
  	\caption{Overview of the main physical properties of the galaxies for the RT runs performed (see Table \ref{tab:hydro_runs}). The table contains: (first column) the name of the run, (second column) the processed UV (integrated in the band ${1000-3000~\angstrom}$) luminosity $L_{\rm UV}$, (third column) the processed total infrared (integrated in the band ${8-1000~\mum}$) luminosity $L_{\rm TIR}$, (fourth column) the total dust mass contained in the simulated region $M_{\rm d}$, (fifth column) the \emph{luminosity-weighted} temperature of the dust grains $\langle T_{\rm d}\rangle_L$, reported as the mean of the PDF within one standard deviation, (sixth column) the minimum and maximum value the dust grains temperature, (seventh column) the intrinsic (i.e, not dust-processed) UV luminosity $L_{\rm UV}^{\rm intr}$, (eighth column) the effective UV optical depth $\tau_{\rm UV}$, estimated as as $e^{-\tau_{\rm UV}} = L_{\rm UV}/L_{\rm UV}^{\rm intr}$. For the dust-processed UV, TIR luminosities and UV optical depth we report the range bracketed by the six line of sights considered for each simulation. 
  	\label{tab:SKIRT_runs_results}
  	  	}
\end{table*}

\subsection{Dust temperature} \label{dust_temperature}

One of the key physical quantities derived from RT calculations is the mass-weighted dust temperature ($\langle T_{\rm d}\rangle_M$). In what follows, we first describe how we compute the luminosity-weighted dust temperature \citep[$\langle T_{\rm d}\rangle_L$, see][]{Behrens:2018,Sommovigo:2020} and compare this value with $\langle T_{\rm d}\rangle_M$; then we discuss how the dust temperature is affected by the total amount of dust, and different types of UV sources (stars vs AGN).

\subsubsection{Luminosity- vs. mass-weighted $T_{\rm d}$} \label{lum_weight_vs_mass_weight_temperature}

To compute $\langle T_{\rm d}\rangle_L$, we assume that each dust cell emits as a grey body\footnote{This approximation holds only for dust cells that are optically thin to IR radiation, although we caveat that a small number of cells in the simulation is actually optically thick.} ${L_{TIR} \propto M_{\rm d} T^{4+\beta_{\rm d}}_{\rm d}}$, where $\beta_{\rm d}$ is the dust emissivity index\footnote{The actual value of $\beta_{\rm d}$ depends on the RT calculation. For example, \citet{Behrens:2018} found a value of $\beta_{\rm d}=1.7$. For computing the luminosity-weighted temperature, we assume $\beta_{\rm d}=2$. This choice does not significantly affect the final results: the estimate of $\langle T_{\rm d}\rangle_L$ varying ${1.5<\beta_{\rm d}<2.5}$ is within 10\% of the value reported in Table 5.}. $\langle T_{\rm d}\rangle_L$, finally depends on the total amount of dust $M_{\rm d}$ in the simulation, determined by our choice of $f_d$. Runs with ${f_d = 0.3}$ are characterised by average dust temperatures $\sim 10$\% lower with respect to the corresponding runs with ${f_d = 0.08}$. This is because the same UV energy is distributed over a larger amount of dust mass.

\begin{figure}
	\centering
	\includegraphics[width=0.45\textwidth]{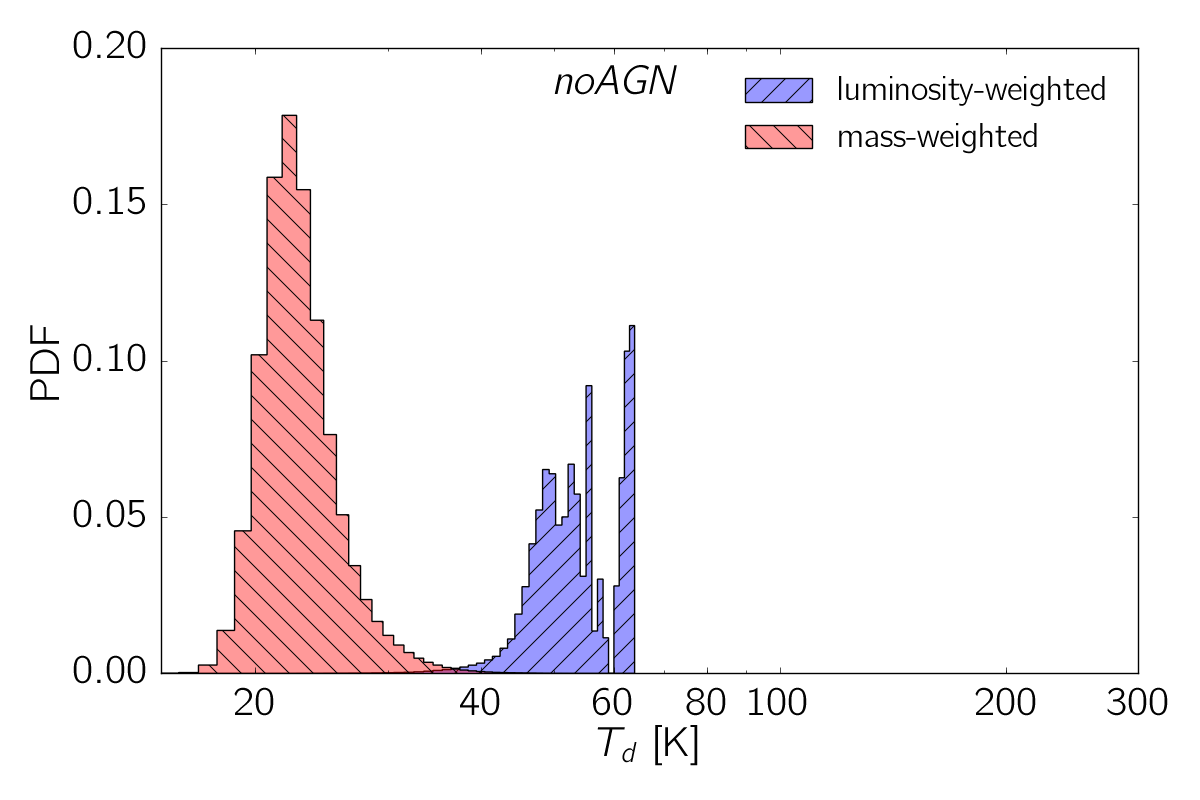}
	\hfill
	\vskip\baselineskip
	\includegraphics[width=0.45\textwidth]{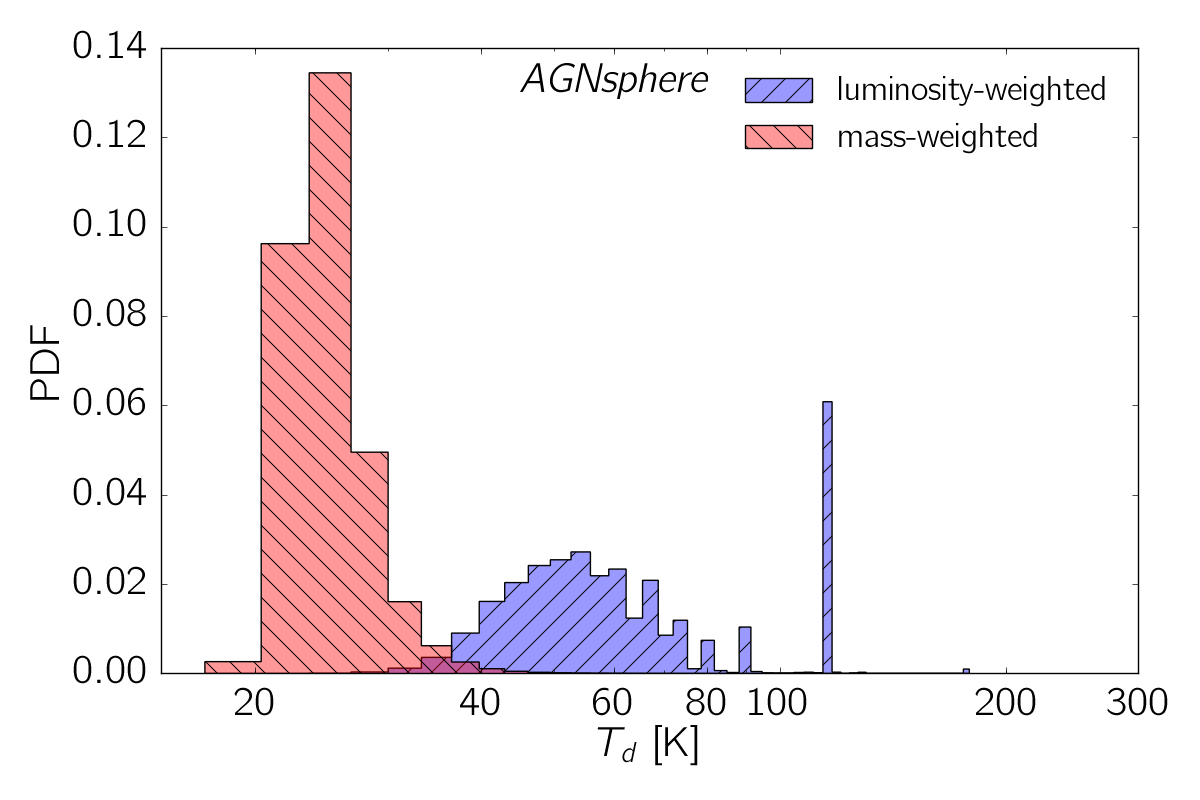}
	\hfill
	\vskip\baselineskip
	\includegraphics[width=0.45\textwidth]{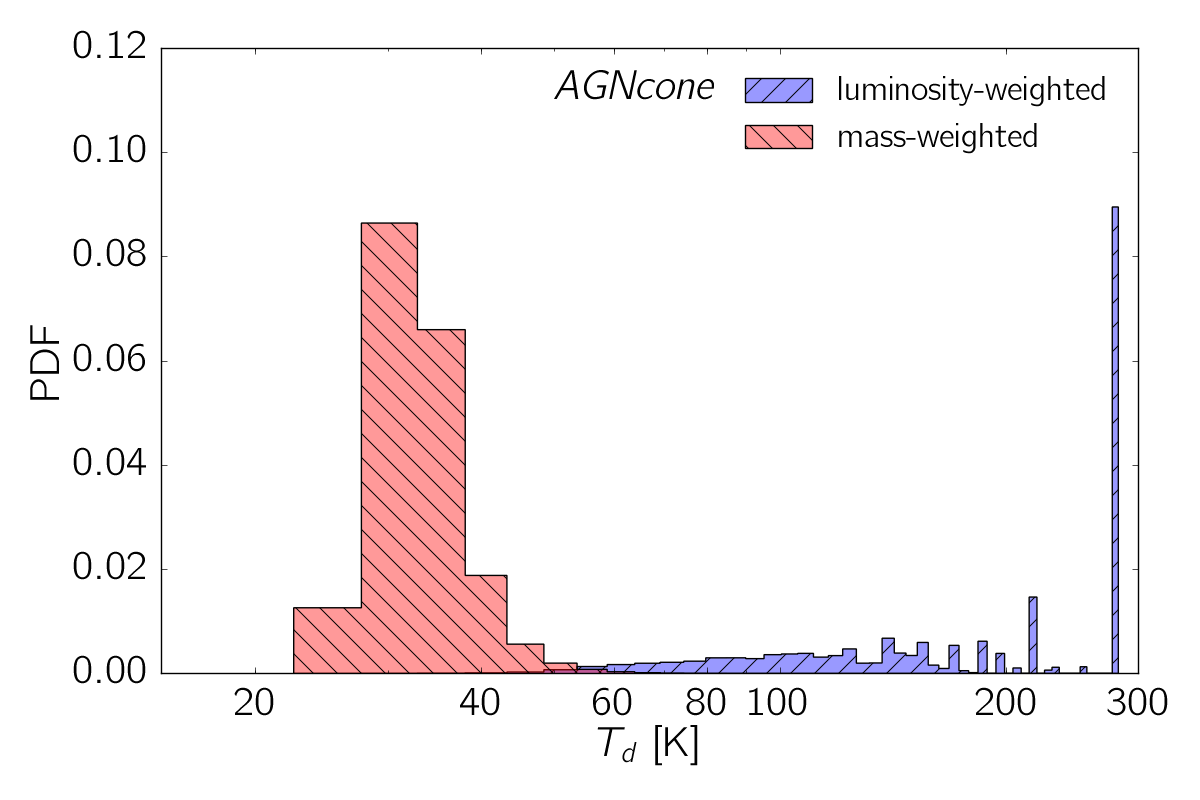}
	\caption{Mass-weighted (red histograms) and luminosity-weighted (blue histograms) dust grains temperature PDF. The panels refer to: (top) \noAGN{}, (middle) \AGNsphere{} and (bottom) \AGNcone{}. As a reference case, we show the results for ${f_d = 0.08}$.
	\label{fig:temp_histograms_008}
	}
\end{figure}

In Fig. \ref{fig:temp_histograms_008} we show the $\langle T_{\rm d}\rangle_L$ PDF (blue histograms), compared with the mass-weighted $\langle T_{\rm d}\rangle_M$ one (red histograms) for the \noAGN{}, \AGNsphere{} and \AGNcone{} simulations with ${f_d=0.08}$, as a reference case. In each run, the PDF of $\langle T_{\rm d}\rangle_M$ peaks at lower dust temperatures with respect to $\langle T_{\rm d}\rangle_L$. The difference between the mass-weighted and luminosity-weighted temperatures is particularly evident in the runs in which AGN radiation is included. In particular, the spikes of the luminosity-weighted histograms correspond to dust cells in the immediate proximity of accreting BHs. This dust component constitutes only a small fraction of the total mass, but it provides a significant contribution to the overall luminosity, as further discussed in the next section.

\subsubsection{Stars and AGN contribution to dust heating} \label{dust_lum_mass_temp}

The brightest TIR spots in Fig. \ref{fig:runs_RT_008} and \ref{fig:runs_RT_03} in the \noAGN{} (AGN runs) correspond to the locations of the most highly star forming regions (accreting BHs). Whereas in the \noAGN{} run the maximum $T_{\rm d}$ value is about $60$~K, in the AGN runs, dust grains reach luminosity-weighted temperatures ${T_{\rm d} \gtrsim 200}$~K close to BHs, and ${T_{\rm d} \approx 60}$~K in the diffuse gas.

We underline that in the \noAGN{} run $\langle T_{\rm d}\rangle_L$ is up to $4$ times lower with respect to the AGN runs despite having a star formation rate $3$ times higher. These results indicate the dominant role played by AGN radiation in the dust heating. This is particularly evident if we compare in more details the run \noAGN{} and \AGNsphere{}. In the \noAGN{} case, ${L_{\rm UV}^{\rm intr}=L_{\rm UV, \rm stars}=4.1\times 10^{12}~\lsun}$; in the \AGNsphere{} case, $L_{\rm UV}^{\rm intr}=L_{\rm UV, stars}+L_{\rm UV,BH}=(2.3+1.4)\times 10^{12}\lsun=3.7\times 10^{12}~\lsun$.

\begin{figure*}
    \centering
    \includegraphics[width=0.95\textwidth]{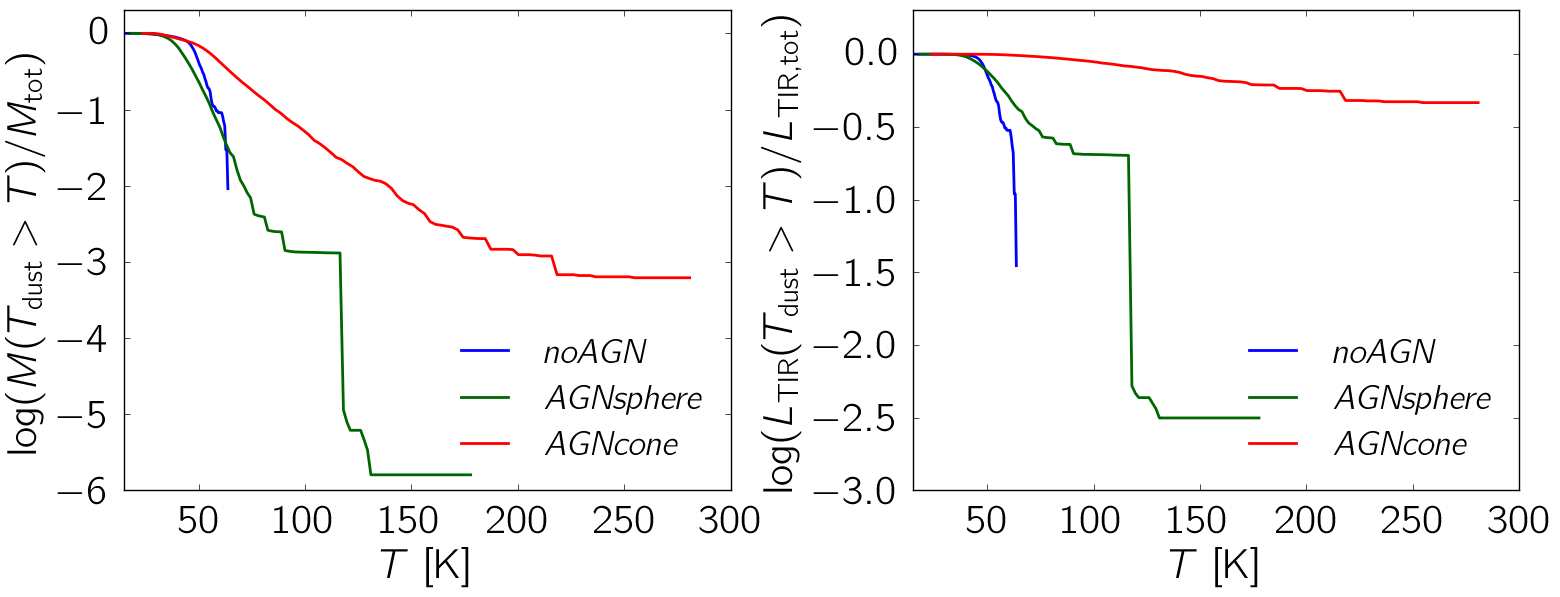}
    \caption{Mass fraction (left) and TIR luminosity fraction (right) of dust with a temperature ${T_{\rm d} > T}$ as a function of the temperature $T$ for the runs \noAGN{} (blue line), \AGNsphere{} (green line), \AGNcone{} (red line). Results for ${f_d=0.08}$ are shown. 
    \label{fig:cumulative_mass_lum}
    }
\end{figure*}

Thus, although the UV budget in the \AGNsphere{} run is mostly provided by stars, and the total UV intrinsic luminosity is comparable to the \noAGN{} case, $T_{\rm d}$ peaks at higher temperature values if BH accretion is present. In Fig. \ref{fig:cumulative_mass_lum}, we compare the fraction of mass (left panel) and TIR luminosity\footnote{The luminosity is computed assuming $\beta_{\rm d}=2$ for consistency with the temperature PDF. The resulting luminosity varying ${1.5<\beta_{\rm d}<2.5}$ differs by $\approx 10 \%$ from the quoted values.} (right panel) from dust with a temperature above a certain threshold for the three runs. In the \noAGN{} run the TIR luminosity is arising from dust with ${T_{\rm d}\lesssim 50}$~K. In the \AGNsphere{} (\AGNcone{}) run $>50$\% of the TIR luminosity is arising from dust with ${T_{\rm d}\gtrsim 70}$~K (${T_{\rm d}\gtrsim 150}$~K); this warm dust only constitutes ~0.1\% of the total dust mass. This confirms that a small mass fraction of warm dust dominates the IR emission, as expected from the scaling ${L_{\rm d} \propto M_{\rm d} T^{4+\beta_{\rm d}}_{\rm d}}$. 

\subsubsection{Spatial extent of FIR emitting regions}

\begin{figure*}
    \centering
    \includegraphics[width=0.95\textwidth]{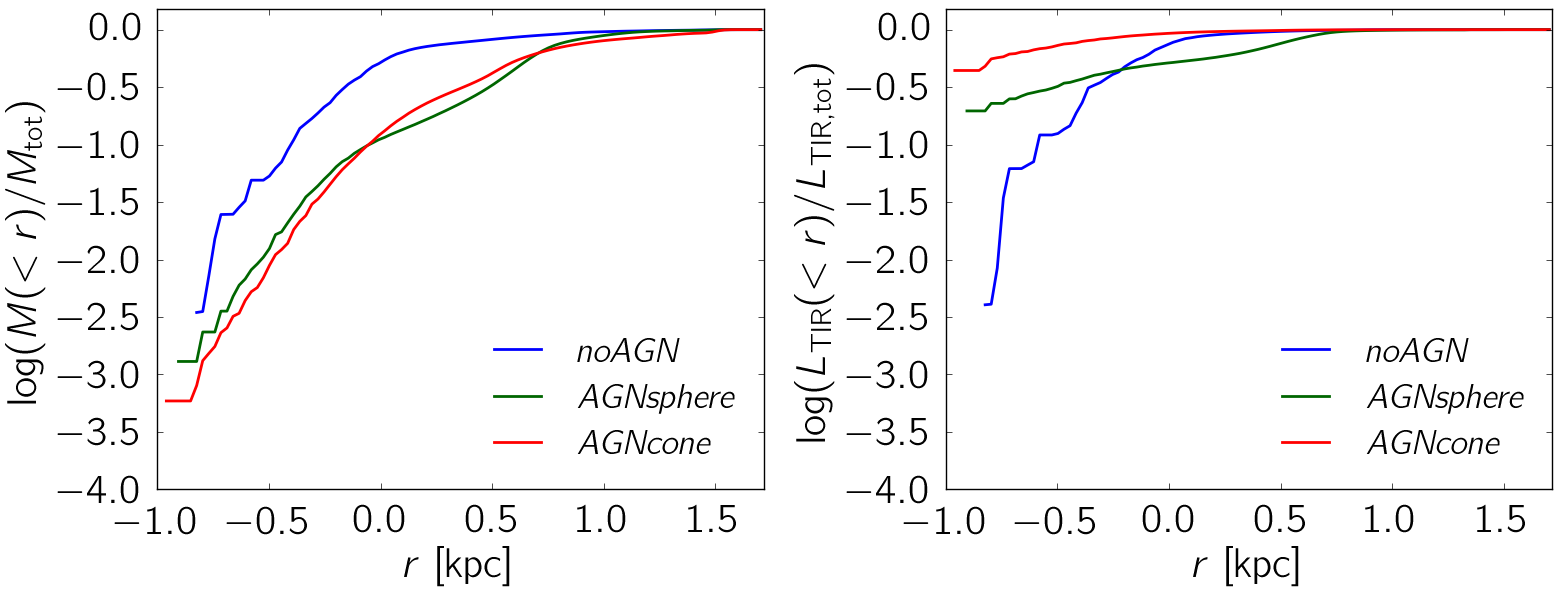}
    \caption{Cumulative mass fraction (left) and TIR luminosity fraction (right) of the dust at a distance $r$ from AGN location or most star forming regions. The lines show the results for \noAGN{} (blue line), \AGNsphere{} (green line) and \AGNcone{} (red line) with ${f_d=0.08}$. 
    \label{fig:mass_lum_distance}
    }
\end{figure*}

Fig. \ref{fig:mass_lum_distance} shows the fraction of dust mass and infrared luminosity as a function of the distance\footnote{Given that there are multiple accreting BHs, we selected the 2 (3) most active ones in the \AGNsphere{} \AGNcone{} run and 2 most accreting star forming regions (the main galaxy and its largest satellite) in the \noAGN{} run. For each cell containing dust in the octree grid we evaluate the distance from each reference source and then we consider the minimum one for this calculation.} from the regions with the highest star formation for the \noAGN{} case and from the BHs with the highest accretion rate for the run \AGNsphere{} and \AGNcone{}.

In the \noAGN{} case, the dust mass within ${r \lesssim 300}$~pc represents ${\sim 0.3 \%}$ of the total dust, and it provides ${\sim 3 \%}$ of the total IR luminosity. In the \AGNsphere{} (\AGNcone{}) case, only $\sim 0.1 \%$ ($\sim 0.06$ \%) of the total dust mass is found at ${r \lesssim 300}$~pc from an accreting BH but it contributes $20 \%$ ($\sim40 \%$) of the total IR luminosity. 

\subsection{Synthetic Spectral Energy Distributions} \label{sec:sed}

\begin{figure*}
    \centering
    \includegraphics[width=0.32\textwidth]{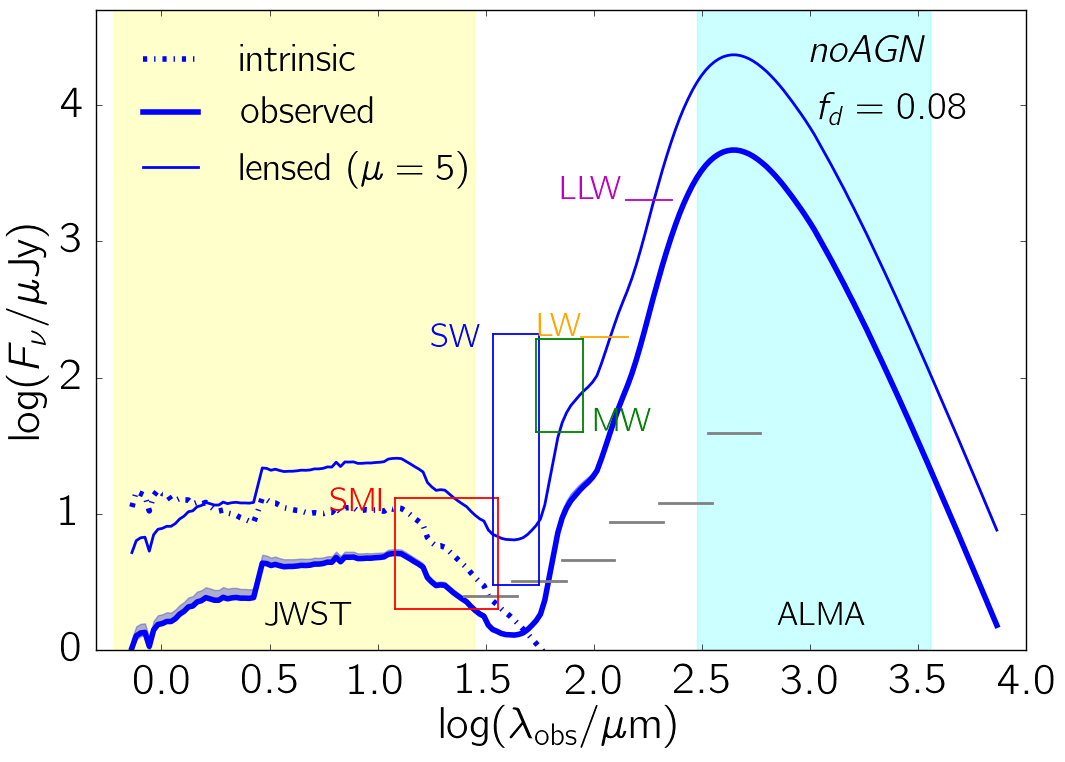}
    \hfill
    \includegraphics[width=0.32\textwidth]{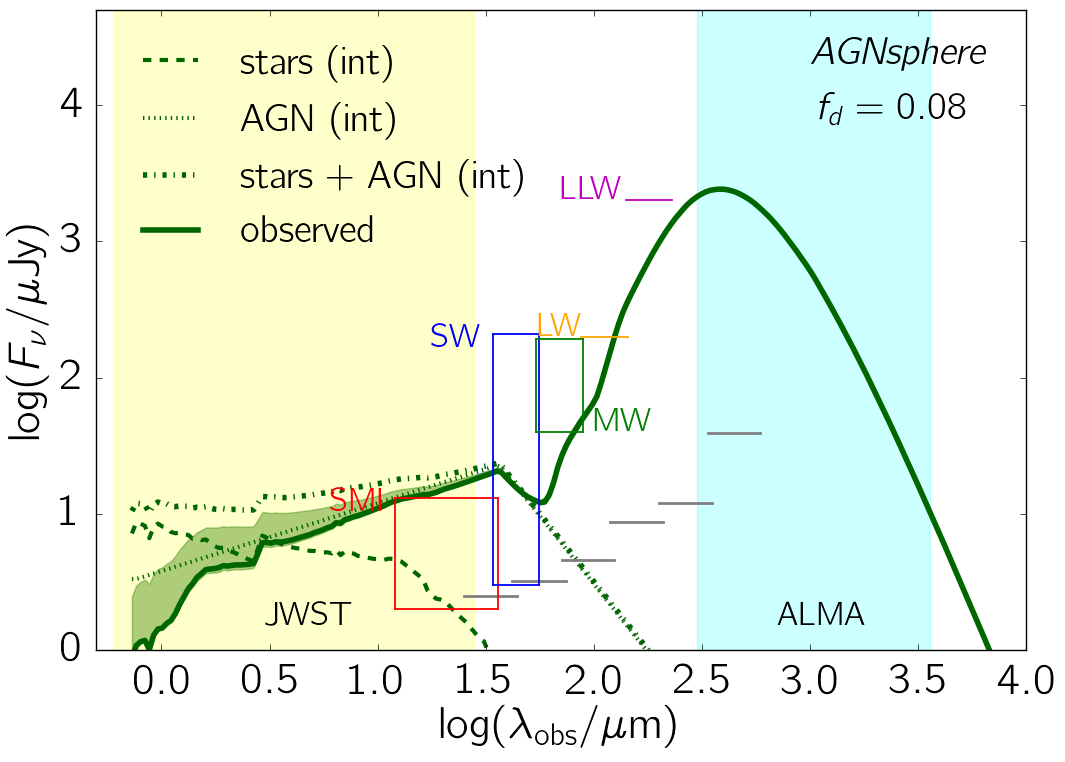}
    \hfill
    \includegraphics[width=0.32\textwidth]{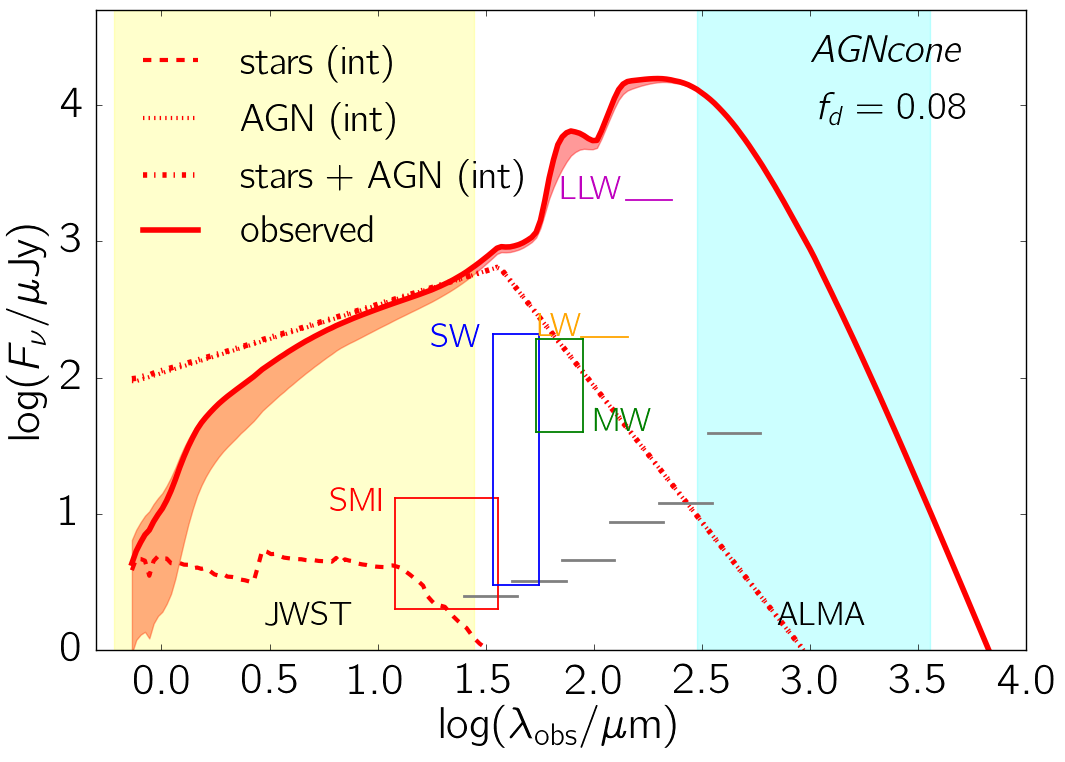}
    \vskip\baselineskip
    \includegraphics[width=0.32\textwidth]{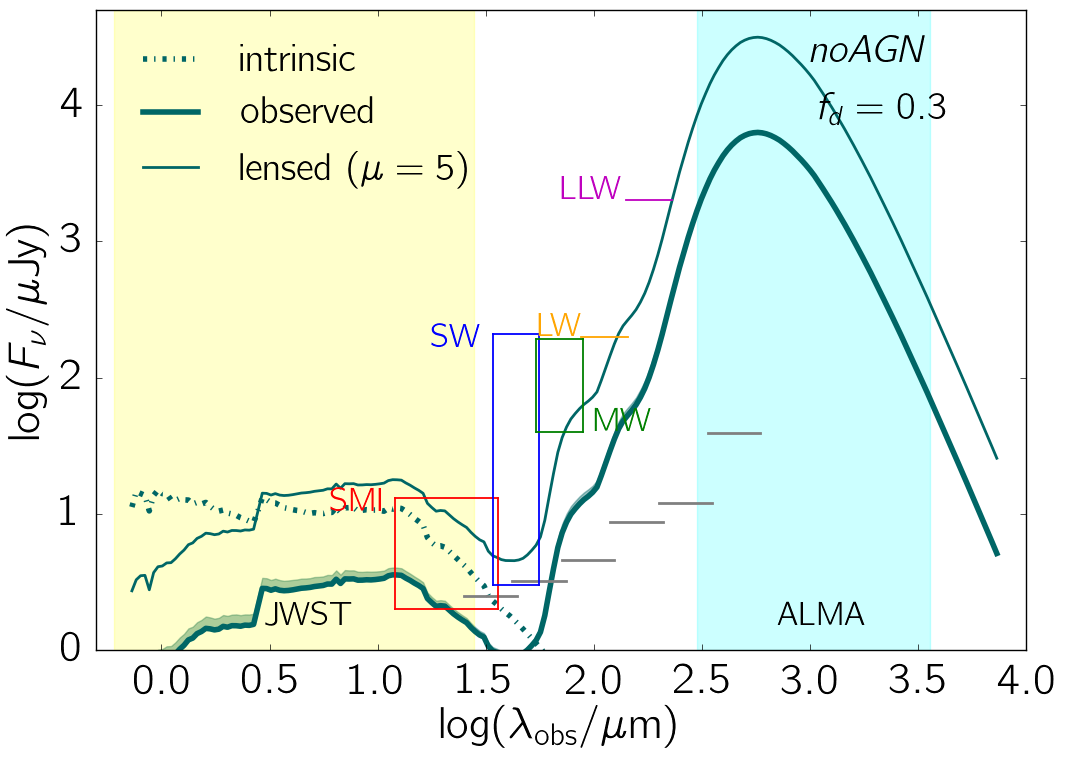}
    \hfill
    \includegraphics[width=0.32\textwidth]{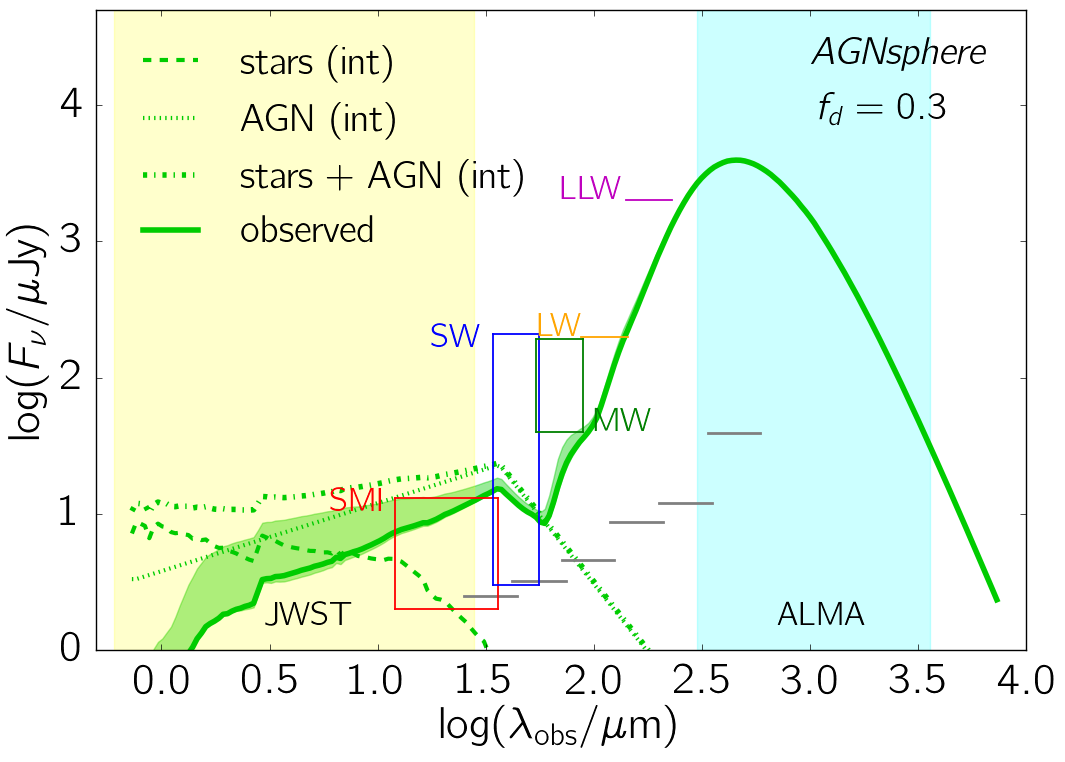}
    \hfill
    \includegraphics[width=0.32\textwidth]{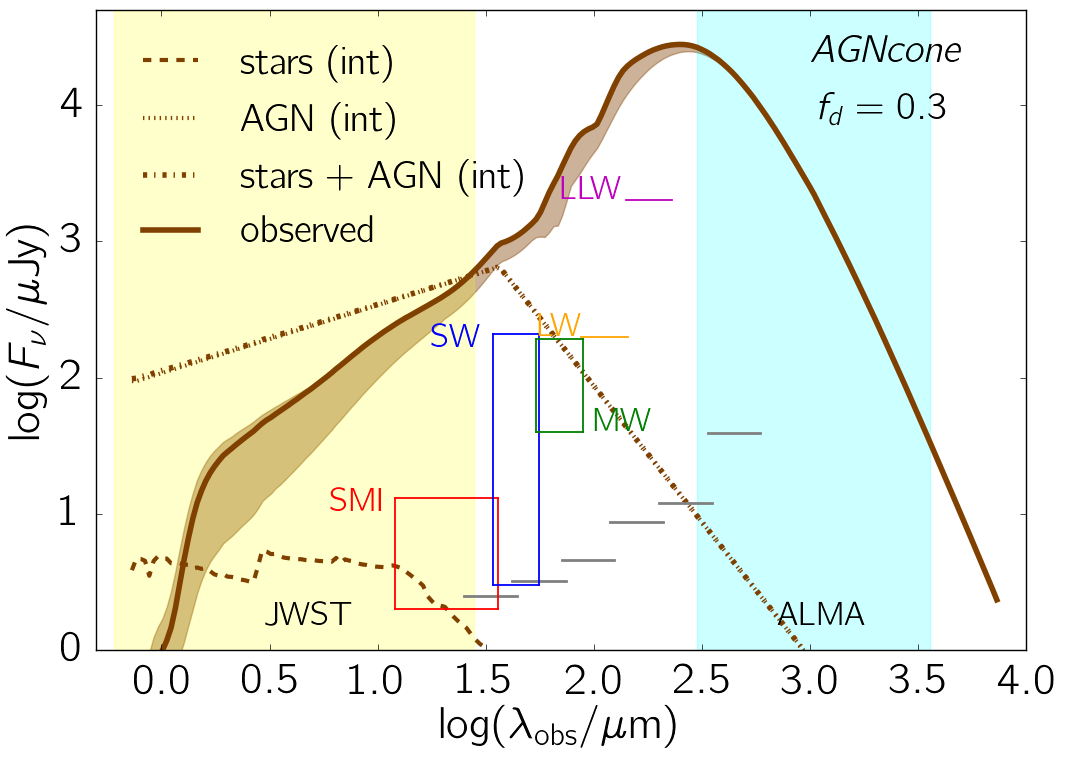}
    \caption{Intrinsic and processed (observed) SED for the first six runs in Table \ref{tab:SKIRT_runs_setup}. The first column refers to \noAGN{}, the second to \AGNsphere{} and the third to \AGNcone{}, whereas the first row to ${f_d=0.08}$, and the second to ${f_d=0.3}$. The solid line shows the observed flux for the reference line of sight, whereas the shaded area brackets the scatter in the observed SED between the six lines of sights used for the computation. The intrinsic flux is also shown with a a dot-dashed line. In the runs with AGN, the individual components are also shown: radiation from stars is denoted with dashed lines, radiation from AGN with dotted lines. In \noAGN{} runs, thin solid lines indicate the flux that would be observed if the galaxy is magnified by a factor ${\mu = 5}$. Sensitivity bands of JWST and ALMA are shown as yellow and cyan shaded regions respectively. The grey lines indicate the sensitivity reached by the ORIGINS telescope at $5\sigma$ in 1 hr of observing time. The colored rectangles and horizontal lines indicate the sensitivity of the two instruments of the SPICA telescope: SMI (${\lambda_{\rm obs}=27~\mum}$, red rectangle), and SAFARI, in photometric mapping mode at short (SW, ${\lambda_{\rm obs}=45~\mum}$, blue rectangle), mid (MW, ${\lambda_{\rm obs}=72~\mum}$, green rectangle), long (LW, ${\lambda_{\rm obs}=115~\mum}$, orange line) and very long wavelengths (LLW, ${\lambda_{\rm obs}=185~\mum}$, violet line). The upper sides of rectangles represent the sensitivity that will be reached by SPICA at $5\sigma$ in 1 hr of observing time. The bottom side of rectangles represents the maximum sensitivity reachable with SPICA, and it is obtained by considering the confusion limit flux at $3\sigma$ (such a high sensitivity can be reached in the case of follow-up observations). If the confusion limit is reached in less than 1 hr, it is shown as a single line.
    \label{fig:SED}
    }
\end{figure*}

Fig. \ref{fig:SED} shows the intrinsic flux density from stars (dashed line) and AGN (dotted line) for the first six runs reported in Table \ref{tab:SKIRT_runs_setup}. The higher value of the flux density from stars in the \noAGN{} run with respect to both AGN runs is due to the negative AGN feedback that in the AGN simulations quenches the star formation rate in the host galaxy (see Section 3.7 of \citetalias{paramita:2018} for an extensive discussion on this topic). This effect is more pronounced in the \AGNcone{} run since it is characterised by a black hole accretion rate that is a factor of $\sim 30$ higher than in \AGNsphere{} (see Table \ref{tab:hydro_runs}). The total intrinsic flux (dotted-dashed line) is comparable between \noAGN{} and \AGNsphere{} (see also Table \ref{tab:SKIRT_runs_results}).

We now analyse the differences between the reprocessed flux density (observed, solid line) resulting from our calculations, focusing on the rest-frame NIR (${1\lesssim \lambda_{\rm RF}\lesssim 5~\mum}$), MIR (${5\lesssim \lambda_{\rm RF}\lesssim 40~\mum}$), and FIR (${40\lesssim \lambda_{\rm RF}\lesssim 350~\mum}$) wavelength ranges.

The intrinsic NIR flux is suppressed by ${\approx 10}$ times in all runs; the highest rest-frame UV attenuation is seen in the \AGNcone{} run, with some (all) lines of sight showing a flux reduced by ${\approx 100}$ times for ${f_d=0.08}$ (${f_d=0.3}$). However, for a fixed dust content, the \AGNcone{} run still provides the highest rest-frame UV flux. In this wavelength range, the SED is nearly constant in the \noAGN{} run whereas it increases toward larger wavelengths in the runs with AGN, as a consequence of the contribution from accretion. The observed optical-NIR flux depends both on the radiation field and dust content.

For what concerns the MIR, at short wavelengths, (${\lambda_{\rm RF}\sim 4-6~\mum}$), the SED is dominated by the almost unattenuated emission from stars and/or AGN; the \AGNcone{} SED is $\sim 30$ times brighter than the \AGNsphere{} one as a consequence of its higher BH activity. At longer wavelengths (${\lambda_{\rm RF}>6~\mum}$), the observed flux arises from heated dust IR emission. The flux density in this wavelength range is the result of the sum of multiple greybodies, each emitting at different temperatures, according to the luminosity-weighted dust temperature PDF discussed in Section \ref{lum_weight_vs_mass_weight_temperature}. The warm dust in AGN runs produces a MIR excess with respect to the \noAGN{} run, and shifts the peak of the emission toward shorter wavelengths: ${\lambda_{\noAGN{}}^{\rm peak} = 59.4~\mum}$, ${\lambda_{\AGNsphere{}}^{\rm peak} = 54.1~\mum}$ and ${\lambda_{\AGNcone{}}^{\rm peak} = 27.0~\mum}$. 

Finally, the Rayleigh-Jeans tail of the FIR emission is mostly sensitive to the total dust content. In fact, by comparing the ${f_d=0.08}$ and ${f_d=0.3}$ cases, we find that the flux at 1 mm scales almost linearly with the dust mass, without a strong dependence on the radiation source.  

To summarise, the SED in the NIR wavelength range depends both on the dust mass (for fixed dust properties) and the type of source (stars and/or AGN); the MIR retains information almost solely on the type of source: the presence of an AGN enhances the flux and shifts the peak of the emission at shorter wavelengths; the flux in the Rayleigh-Jeans tail of the FIR emission mostly depends on the total dust content.

\section{\texorpdfstring{Comparison with \titlelowercase{$z\sim 6$} quasar data}{Comparison with quasar data}} \label{sec:syn_vs_obs}

\begin{figure*}
    \centering
    \includegraphics[width=0.475\textwidth]{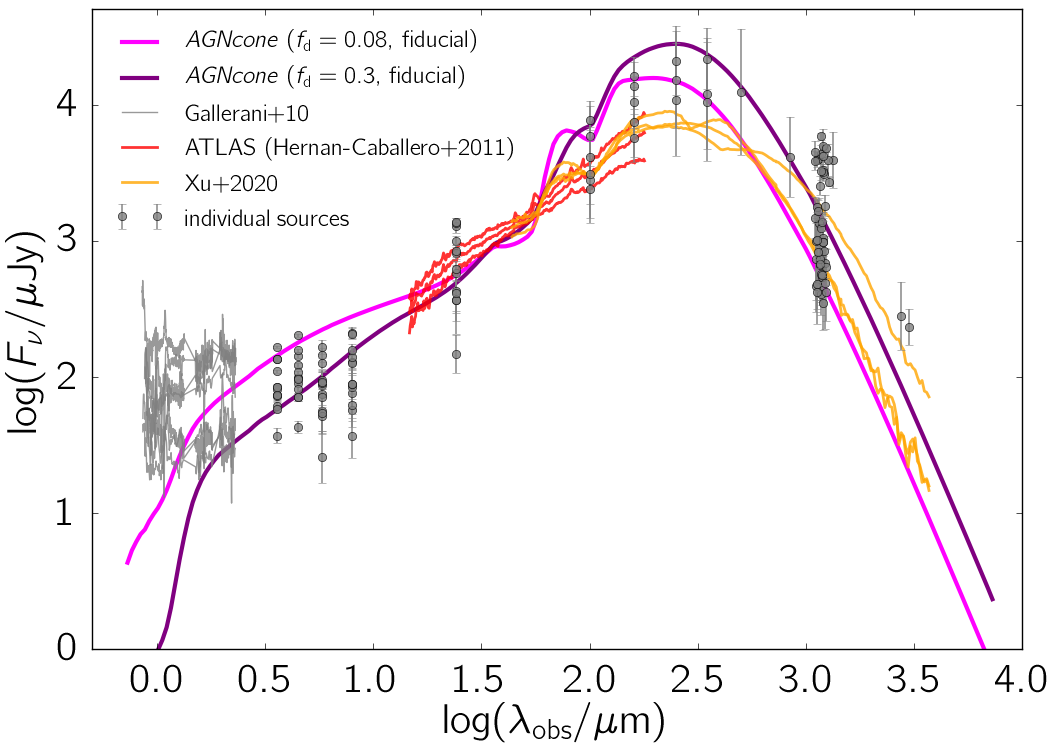}
	\hfill
	\includegraphics[width=0.475\textwidth]{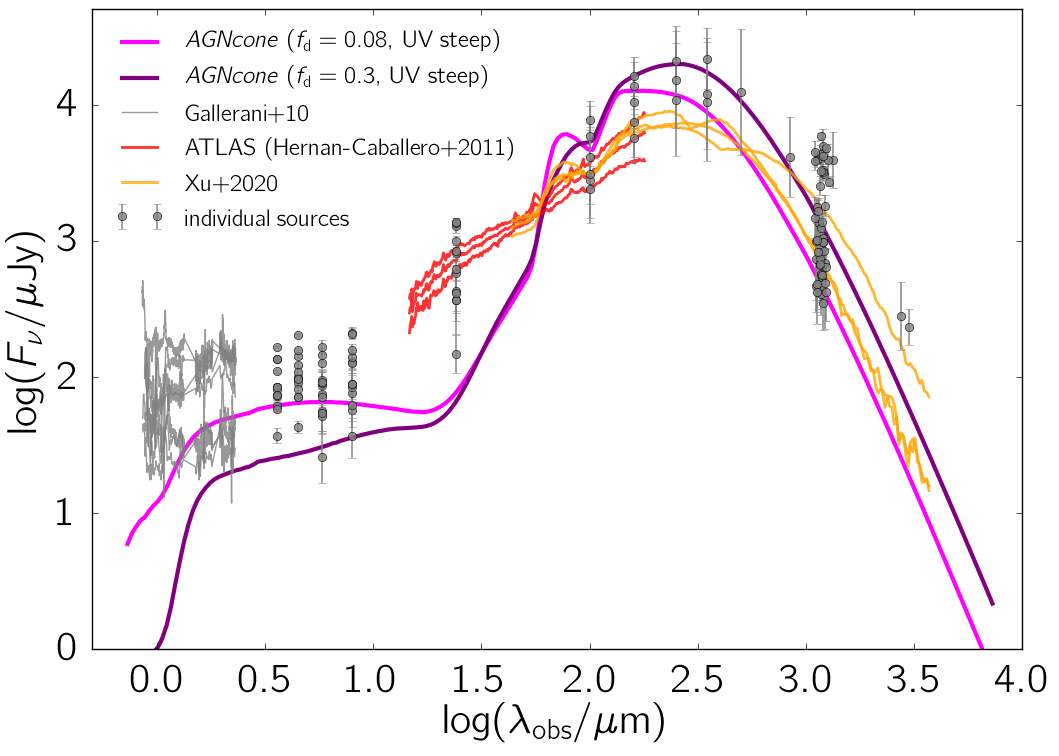}
	\caption{\textit{Left panel}: Comparison between synthetic fiducial SEDs and observations of ${z\sim 6}$ quasars (grey circles, see Table \ref{tab:sources_comparison}). The magenta and violet solid curves denote the models for ${f_d=0.08}$ and ${f_d=0.3}$, respectively. Grey lines represent dust-reddened ${z\gtrsim 6}$ quasar spectra taken with the TNG/GEMINI \citep[${A_{\rm 3000}>0.8}$, see Table 1 in][]{Gallerani:2010}. The spectra are calibrated by using the measures of $\lambda L_\lambda$ at $1450~\angstrom$ provided in Table 1 by \citet{Juarez:2009}. We further show with orange lines the template obtained from the analysis of 125 quasar spectra at ${0.020<z<3.355}$ taken with Spitzer/IRS (the tree lines correspond to the median SED, the 25$^{\rm th}$ and 75$^{\rm th}$ percentiles, in the case of the luminous, ${\log(\lambda L_{5100})>44.55}$, sub-sample by \citet{HernanCaballero2011MNRAS}), and with red lines the IR AGN SED derived by \citet{Xu2020ApJ} from 42 quasars at ${z<0.5}$ (see their Table 3). The spectra of these low redshift sources are reported to get a hint of the spectral slope of AGN in the MIR; no spectroscopy information is available so far in the case of ${z\sim 6}$ quasars. \textit{Right panel}: same as left panel, but for the UV-steep AGN SED models. 
	\label{fig:model_collection}
	}
\end{figure*}

To test the results of our model (SPH simulation post-processed with RT calculations), we compare in Fig. \ref{fig:model_collection} our predictions from the \AGNcone{} run (${M_{\rm UV}=-27.97}$) with multi-wavelength (NIR to FIR) observations of ${z\sim 6}$ bright (${-29\lesssim M_{\rm UV}\lesssim -26}$) quasars (see Table \ref{tab:sources_comparison}). 

\begin{table}
	\centering
	{\def\arraystretch{1.2}
	\begin{tabular}{ l c c c}
		\hline\noalign{\smallskip}
		source & $z$ & $M_{\rm UV}$ & Reference\\
		\hline\hline\noalign{\smallskip}
		J1030+0524 & 6.31& -27.12& [1-2,8]\\ 
		J1048+4637 & 6.23& -27.60& [1-2,8]\\ 
		J1148+5251 & 6.43& -27.85& [1-8]\\ 
		J1306+0356 & 6.03& -26.76& [1-2,8-9]\\ 
		J1602+4228 & 6.07& -26.85& [1-2,8]\\ 
		J1623+3112 & 6.25& -26.71& [1-2,8]\\ 
		J1630+4012 & 6.07& -26.16& [1-2,8]\\
		J0353+0104 & 6.07& -26.56& [8]\\ 
		J0818+1722 & 6.00& -27.44& [8]\\ 
		J0842+1218 & 6.08& -26.85& [8,9]\\ 
		J1137+3549 & 6.01& -27.15& [8]\\ 
		J1250+3130 & 6.13& -27.18& [8]\\ 
		J1427+3312 & 6.12& -26.48& [8]\\ 
		J2054-0005 & 6.04& -26.15& [8]\\ 
		P007+04 & 6.00& -26.58& [9]\\ 
		P009-10 & 6.00& -26.50& [9]\\ 
		J0142-3327 & 6.34& -27.76& [9]\\ 
		P065-26 & 6.19& -27.21& [9]\\ 
		P065-19 & 6.12& -26.57& [9]\\ 
		J0454-4448 & 6.06& -26.41& [9]\\ 
		P159-02 & 6.38& -26.74& [9]\\ 
		J1048-0109 & 6.68& -25.96& [9]\\ 
		J1148+0702 & 6.34& -26.43& [9]\\ 
		J1207+0630 & 6.04& -26.57& [9]\\ 
		P183+05 & 6.44& -26.99& [9]\\ 
		P217-16 & 6.15& -26.89& [9]\\ 
		J1509-1749 & 6.12& -27.09& [9]\\ 
		P231-20 & 6.59& -27.14& [9]\\ 
		P308-21 & 6.23& -26.30& [9]\\ 
		J2211-3206 & 6.34& -26.65& [9]\\ 
		J2318-3113 & 6.44& -26.06& [9]\\ 
		J2318-3029 & 6.15& -26.16& [9]\\ 
		P359-06 & 6.17& -26.74& [9]\\
		J0100+2802 & 6.33& -29.30& [10]\\
  	    P338+29 & 6.66& -26.01& [14]\\
  	    J0305-3150 & 6.61& -26.13& [15]\\
	\hline\end{tabular}
	}
  	\caption{Quasars used for the comparison with the prediction by our model. Columns indicate: (first) source name, (second) redshift, (third) $M_{\rm UV}$ and (fourth) references for the photometric data used in the comparison, according to the legend. [1] \citet{Gallerani:2010}; [2] \citet{Juarez:2009}; [3] \citet{Walter:2003}; [4] \citet{Bertoldi:2003}; [5] \citet{Riechers:2009}; [6] \citet{Gallerani:2014}; [7] \citet{Stefan:2015}; [8] \citet{Leipski2014ApJ}; [9] \citet{Venemans:2018}; [10] \citep{Wang:2016ApJ}; [11] \citep{Venemans:2012ApJ}; [12] \citep{Venemans:2017ApJ}; [13] \citep{Willott:2017ApJ}; [14] \citep{Mazzucchelli:2017ApJ}; [15] \citep{Venemans2016ApJ}. 
  	\label{tab:sources_comparison}
  	  	}
\end{table}

In the NIR, our predicted SEDs are underluminous with respect to the flux of TNG/GEMINI spectra (grey lines in Fig. \ref{fig:model_collection}). This mismatch cannot be solved by decreasing the dust content, since by assuming ${f_d<0.08}$ the synthetic SEDs would become underluminous in the FIR with respect to ALMA data. We instead suggest that a better agreement with observations can be obtained by assuming an extinction curve flatter than the SMC \citep[][Di Mascia et al in preparation]{Gallerani:2010}. 

For what concerns the comparison in the MIR, models with ${\alpha_{\rm UV}^{\rm fid}}$ are in good agreement both with Spitzer/Herschel photometric data and with the slope/shape of templates by \citet{HernanCaballero2011MNRAS} resembling Spitzer/IRS spectra. Vice-versa, models with ${\alpha_{\rm UV}^{\rm steep}}$ are both under-luminous with respect to Spitzer/IRAC observations at ${\lambda_{\rm obs}=24~\mum}$ (namely ${\lambda_{\rm RF}\sim 3~\mum}$ at ${z\sim 6}$) and show a slope in the MIR that does not agree with observed spectra. 

We underline that the model with ${\alpha_{\rm UV}^{\rm steep}}$ can be possibly reconciled with observations if the torus is included. In fact, a dust component with temperature close to sublimation (${\sim 1500}$~K) would enhance the MIR emission exactly at the Spitzer/IRAC wavelengths\footnote{The emission of a greybody at temperature $T_{\rm d}$ and with ${\beta_{\rm d} = 2}$ peaks at ${\lambda_{peak}=(2.9\times 10^3)/T_{\rm d}~\mum}$}. We give a first estimate of the impact of the torus emission on our predicted SEDs in Section \ref{obscured_AGN} (Fig. \ref{fig:SED_torus}) and we defer the inclusion of the torus into our model to a future study.

By comparing our predicted SEDs with FIR observations, we note that both models (${f_d=0.08-0.3}$) provide a reasonable match with FIR data, independently on the assumed UV slope (fiducial vs UV-steep). We find that the models with a larger dust-to-metal ratio ${f_d=0.3}$ are slightly preferred, since in the ${f_d=0.08}$ case we can only explain the less luminous FIR sources. 

Hereafter, we consider as fiducial the model with ${\alpha_{\rm UV}^{\rm fid}}$ and ${f_d=0.3}$. 

\subsection{Multiple merging system} \label{sec:individual_sources}

\begin{figure}
	\centering
	\includegraphics[width=0.475\textwidth]{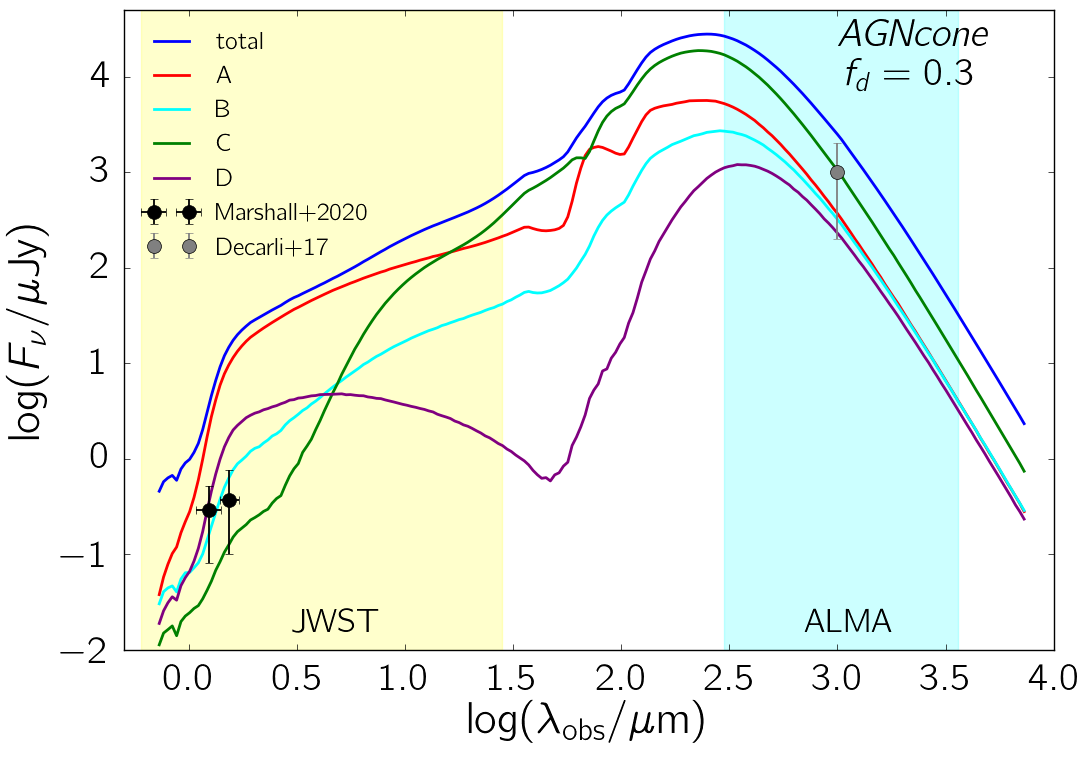}
	\caption{Comparison between the SEDs extracted from sources A, B, C, D in the field of view of the \AGNcone{} run with ${f_d=0.3}$, keeping the same color legend as in Fig. \ref{fig:runs_RT_03}. The total SED is instead plotted with a blue solid line. Black points indicate rest-frame UV limits from deep HST observations by \citet{Marshall2020ApJ}, whereas the grey point FIR fluxes for star-forming companion galaxies around quasars from \citet{Decarli:2017}. 
	\label{fig:AGNcone_individual_seds}
	}
\end{figure}

The most massive halo in the \AGNcone{} run at ${z=6.3}$ hosts a merging system of multiple sources, three of which are AGN (A, B, C) and one is a normal star forming galaxy (D). We show in Fig. \ref{fig:AGNcone_individual_seds} the SEDs extracted from individual sources. In our simulated system, source A is the most luminous UV source, providing ${\sim 70 \%}$ of the total UV flux. However, it does not correspond to the most accreting BH, which is instead powering source C, distant ${\sim 10}$~kpc from A. Despite having the highest intrinsic UV budget, this source is fainter than A in the UV because it is enshrouded by dust: source C is in fact the most luminous IR source of the system and provides ${\sim 70\%-80\%}$ ($\sim40\%$) of the MIR (FIR) flux. The second brightest UV source in our system is source D. 

By comparing our synthetic SEDs with HST and ALMA data\footnote{We do not consider constraints from MIR observations since individual sources cannot be resolved at these wavelengths as a consequence of the poor angular resolution. We further refer to Vito et al. in preparation for a detailed comparison with X-ray observations.} \citep{Marshall2020ApJ, Decarli:2017}, we found that sources A, B and D would be detectable and resolved with HST; for what concerns the FIR, given the angular resolution of current ALMA data (i.e. 1" that corresponds to ${\sim 6}$~kpc at $z=6.3$), it is not possible to disentangle source B and D from A, whereas source C would show up as an SMG companion, even brighter than source A \citep[as in the case of CFHQ J2100-1715 by][]{Decarli:2017}. To summarise, our study shows that, consistently with HST and ALMA observations, bright (${M_{\rm UV}\leq -26}$) ${z\sim 6}$ quasars (e.g. source A in our simulations) are part of complex, dust-rich merging systems, possibly containing highly accreting BHs (e.g. source A, B and C with ${\gtrsim 5~\msunyr}$) and star forming galaxies (e.g. source D). Deeper and higher resolution ALMA data and JWST observations are required to better characterize the properties of galaxy companions in the field of view of ${z\sim 6}$ quasars.

\section{Guiding future MID-IR facilities} \label{sec:spica} 

Given the good agreement between our results and currently available ${z\sim 6}$ quasars observations, we can use our simulations to make predictions for the proposed  Origins Space Telescope (OST\footnote{ OST is a concept study for a $5.9$~m diameter infrared telescope, cryocooled to $4.5$~K, that has been presented to the United States Decadal Survey in 2019 for a possible selection to NASA's large strategic science missions.}; \citealt{Wiedner2020arXiv}). OST covers the wavelength range $2.8-588~\mum$, and is designed to make broad-band imaging (Far-IR Imager Polarimeter, FIP), low resolution ($R \sim 300$) wide-area/deep spectroscopic surveys, and high resolution ($R \sim 40000-300000$) pointed observations (with the Origins Survey Spectrometer, OSS).
We further consider the capability of detecting IR emission from $z \sim 6$ quasars through a $2.5$~m diameter infrared telescope, cryocooled to $8$~K that covers the wavelength range $12-230~\mum$, and is designed to make high-resolution ($R\sim 28000$) in the near-infrared ($12-18~\mum$) and mid-infrared ($30–37~\mum$) broad band mapping, and small field spectroscopic and polarimetric imaging at $100$, $200$ and $350~\mum$. These are the characteristics of the Space Infrared Telescope for Cosmology and Astrophysics (SPICA; e.g. \citealt{Spinoglio:2017, Gruppioni:2017, Egami:2018, Roelfsema2018PASA}), an infrared space mission, initially considered as a candidate for the M5 mission, but cancelled in October 2020 \citep{Clements2020Natur}.

The \noAGN{} case is detectable by ORIGINS in five bands, corresponding to $\approx 6-80~\mum$ rest-frame. ORIGINS would be able to probe the SED of highly star forming galaxies (${\SFR \sim 600~\msunyr}$) at wavelengths shorter than the peak wavelength, which is crucial in order to have a solid determination of the dust temperature \citep{Behrens:2018, Sommovigo:2020}. The \noAGN{} case falls just below the SPICA sensitivity threshold. We thus consider the possibility of observing lensed galaxies with SPICA; the thin solid SED in the \noAGN{} panels in Fig. \ref{fig:SED} accounts for a magnification factor ${\mu \sim 5}$. Our results show that highly star forming galaxies (${\SFR \sim 600~\msunyr}$) without an active AGN will be at the SPICA reach if lensed by a factor ${\mu \gtrsim 5}$.

For what concerns the \AGNsphere{} case, the simulated run corresponds to a faint AGN (${M_{\rm UV} = -23.4}$; X-ray luminosity ${L_{\rm X} \sim 10^{44}~{\rm erg~s}^{-1}}$). This kind of sources is not easily detectable through UV and X-ray observations: (i) less than 20 ${z\sim 6}$ quasars fainter than ${M_{\rm UV} = -23.75}$ have been discovered so far \citep{Matsuoka2018ApJ}; (ii) none ${z \sim 6}$ quasar with ${L_{\rm X} < 4 \times 10^{44}~{\rm erg~s}^{-1}}$ has been detected so far with Chandra \citep{Vito:2019}. Our predictions show that the SED of a faint AGN is instead well above ORIGINS' sensitivities at all wavelengths and also above the sensitivities of two SPICA's bands for all the simulations we performed. This result emphasises the important role that future MIR facilities would have in studying the faint-end of the UV and X-ray luminosity function in ${z\sim 6}$ AGN.

The \AGNcone{} runs show that quasars with ${M_{UV}<-25}$ are very easily detectable both by ORIGINS and SPICA at a signal-to-noise ratio high enough to get good quality spectra even in these very distant sources. We notice that only ${\sim 20}$ quasars have been detected so far with the Spitzer / Herschel telescopes at ${z\gtrsim 6}$ \citep{Leipski2014ApJ,lyu:2016}, and most of them ($>80\%$) are bright (${M_{\rm UV} < -26}$). Quasars fainter than ${M_{UV}=-26}$ have been detected so far at mm wavelengths at $>5\sigma$ only in two ${z\geq 6}$ quasars \citep[J1048-0109 and P167-13 by][]{Venemans:2018}.

Our results show that the ORIGINS telescope will be an extremely powerful instrument for studying the properties of the most distant galaxies and quasars known so far. 

\subsection{Unveiling faint/obscured AGN} \label{obscured_AGN}

Combined ALMA data with follow-up JWST and /or\footnote{The James Webb Space Telescope is planned to fly on October 31, 2021, with 10 years of operation goal. The proposed ORIGINS mission is planned for launch in the early 2030s, so it will ideally continue the work of JWST.} ORIGINS observations will be crucial to discover faint/obscured AGN and to distinguish them from galaxies without an active nuclei. In fact, by comparing the predicted fluxes in ORIGINS band 1 and/or MIRI band at $29~\mum$ ($F_{29\mum}$) with the ones in ALMA band 7 $F_{\rm band 7}$, we find:
\begin{equation*}
    \frac{F_{\rm 29\mum}(\AGNsphere{})/F_{\rm band 7}(\AGNsphere{})}{F_{\rm 29\mum}(\noAGN{})/F_{\rm band 7}(\noAGN{})}\approx 8 - 10,
\end{equation*}
meaning that we expect a a MIR-to-FIR excess of one order of magnitude in the case of a faint AGN host galaxy (\AGNsphere{}) with respect to a star forming galaxy without AGN (\noAGN{}). This result shows that by following up with JWST and/or ORIGINS sources already detected with ALMA it will be possible to discriminate between star forming galaxies and faint/obscured AGN.

\begin{figure*}
	\centering
	\includegraphics[width=0.32\textwidth]{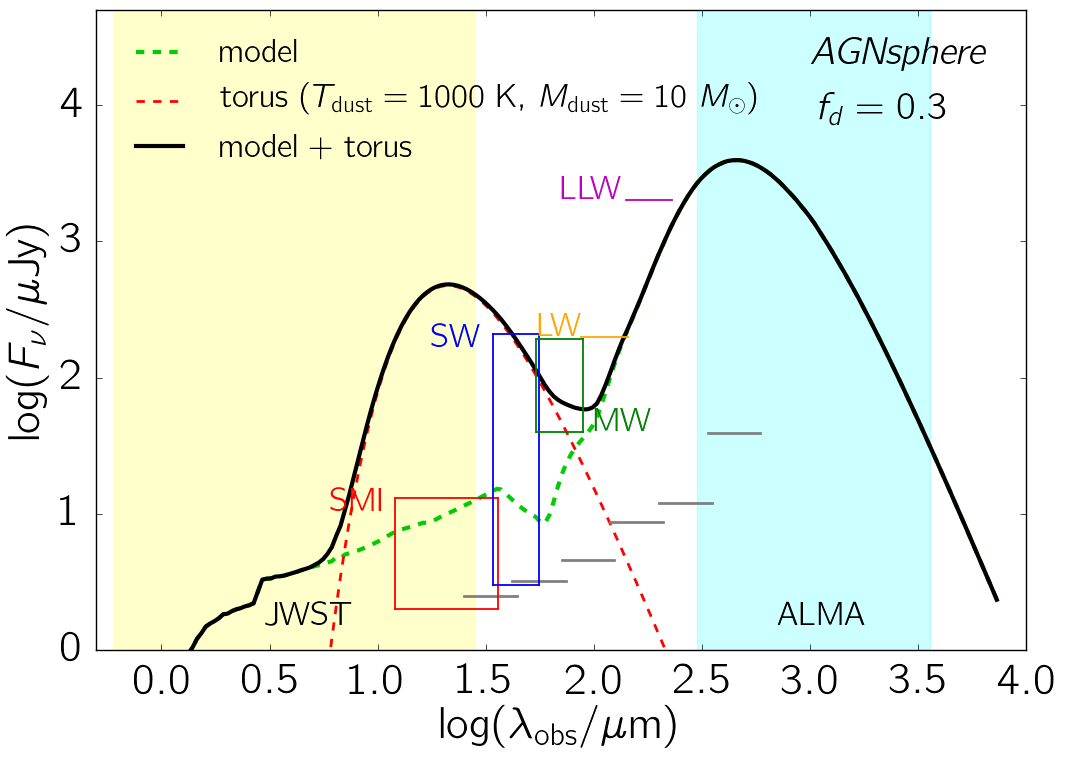}
	\hfill
	\includegraphics[width=0.32\textwidth]{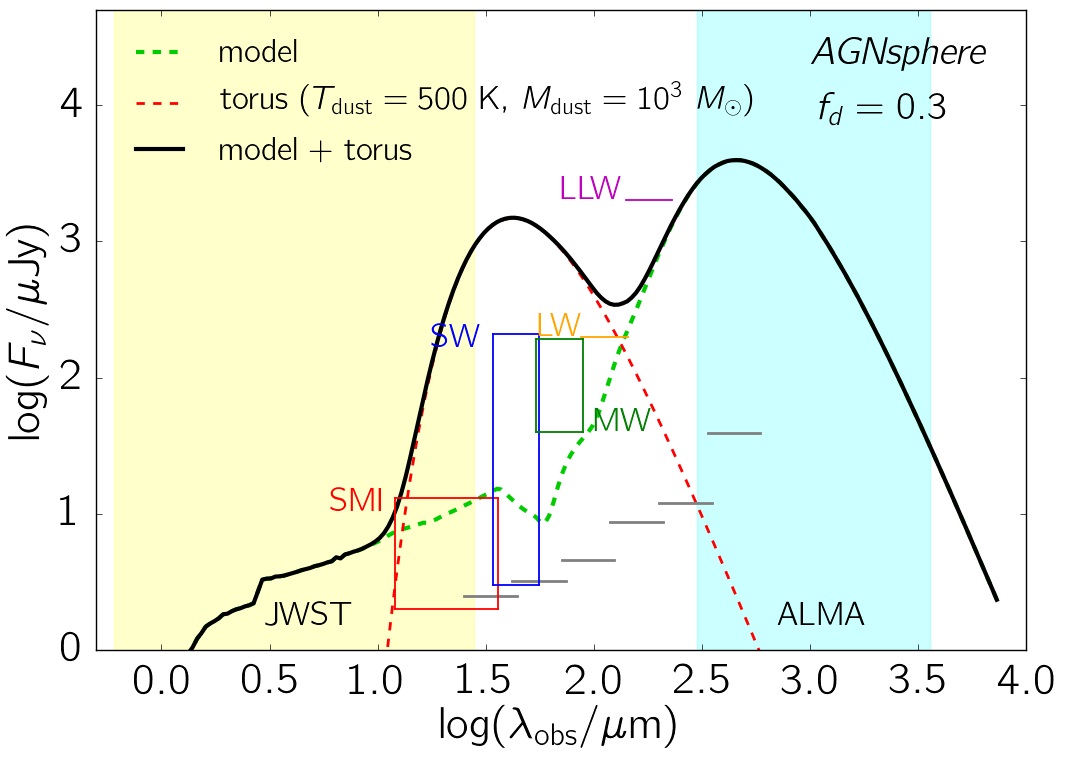}
	\hfill
	\includegraphics[width=0.32\textwidth]{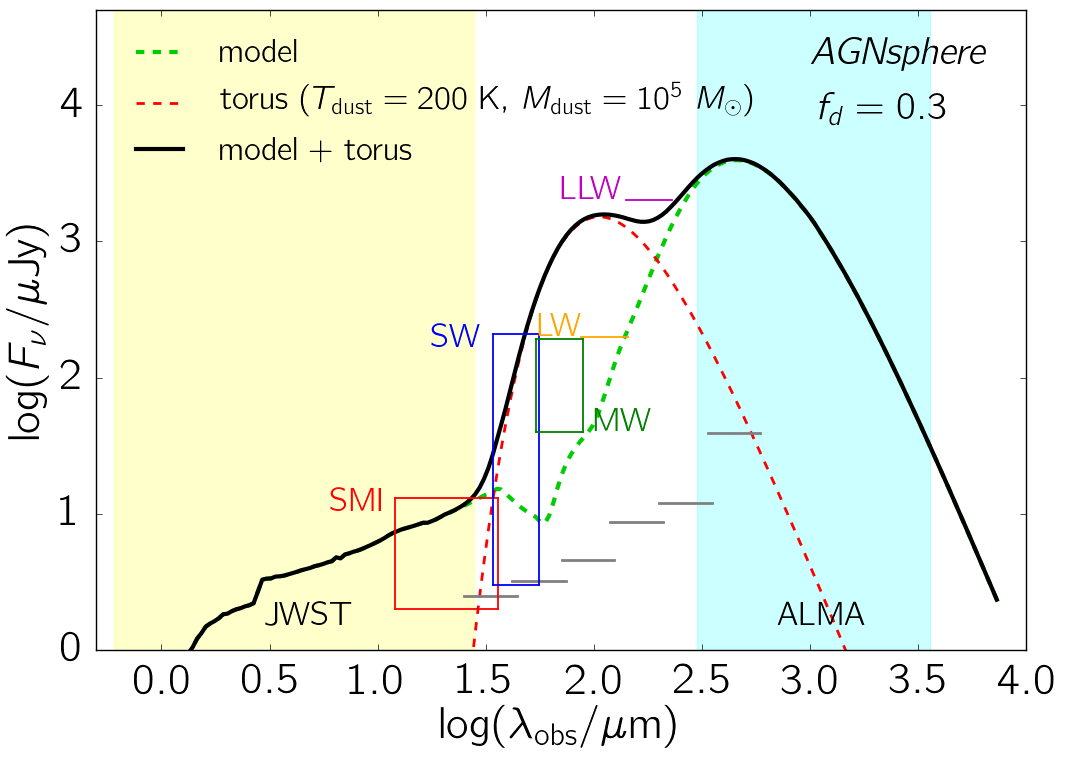}
	\caption{Predicted SEDs with the inclusion of the dusty torus emission to our models for the \AGNsphere{} case with $f_d=0.3$. The green dashed line refers to the original model, the red dashed line to the torus and the black solid line to the sum of the two. The torus emission is modelled as a greybody with $M_{\rm dust}$ and $T_{\rm dust}$ as specified in the panel, and $\beta_{\rm dust} = 2$. 
	\label{fig:SED_torus}
	}
\end{figure*}

We note that, given the limited resolution ($\sim 200$~pc) of the hydrodynamical simulations adopted in this work, we cannot resolve the torus ($\sim 0.1-10$~pc) that is, therefore, not included in our modelling. The presence of a dusty torus surrounding accreting BHs provides an additional source of MIR emission boosting the MIR excess expected in AGN. This can increase both the detectability of faint quasars with a SPICA-like telescope and the possibility of exploiting the synergy between ALMA and MIR facilities to unveil dust-obscured AGN. 
For example, we qualitatively show in Fig. \ref{fig:SED_torus} how our predicted SEDs would change with the inclusion of the emission from the dusty torus. For this comparison we consider the \AGNsphere{} case (${M_{\rm UV} = -23.4}$), since we aim to investigate the ability of MIR telescopes to unveil faint AGN. As a proof of concept, we simply model the torus emission as a single-temperature $T_{\rm dust}$ greybody, with dust mass $M_{\rm dust}$, and $\beta_{\rm dust} = 2$. We consider different models to cover the range in masses ($10^1-10^5~\msun$) and temperatures ($200-1200$~K) constrained by theoretical models \citep{Schartmann:2005, Stalevski:2016} and observations \citep{GarciaBurillo2016, GarciaBurillo2019}.

The MIR emission from the torus brings the SEDs of the faint AGN of our model easily within the reach of a SPICA-like telescope, for a wide range of the torus parameters considered. This further expands the potential of future MIR telescopes in the discovery and the study of faint AGN at high-redshift.

We stress that this is a rough estimate that -- among other things -- neglects the torus geometry, i.e. the fact that UV emission is extinguished along the equatorial plane. Therefore UV photons would escape only towards the polar regions, reducing the amount of dust on $\lesssim 200$~pc scales directly irradiated by the AGN and possibly the IR emission coming from high temperature (${T_{\rm d} \sim 200-300}$~K) regions. 
We plan to include the torus emission in a consistent way in our model in a future work and to further examine its impact on our results and on the potential of future facilities.

\section{Summary and conclusions} \label{sec:conclusions}

In this work, we have considered a suite of zoom-in cosmological hydrodynamic simulations of a massive halo (${\sim 10^{12}~\msun}$) at ${z\sim 6}$ \citep{paramita:2018}. The set of simulations include a control simulation of a highly star forming galaxy (SFR ${\sim 600~\msunyr}$) without BHs (called \noAGN{} run), and two simulations with accreting BHs that account for AGN kinetic feedback distributed according to a spherical (\AGNsphere{} run) and bi-conical (\AGNcone{ run}) geometry. These two different feedback prescriptions result in different SFRs of the host galaxy (${\sim 300}$ and ${\sim 200~\msunyr}$ in \AGNsphere{} and \AGNcone{}, respectively), and AGN activity (${\sim 3}$ and ${\sim 90~\msunyr}$ in \AGNsphere{} and \AGNcone{}, respectively).

We performed dusty radiative transfer calculations of the three runs in post-process by exploiting the code \code{SKIRT} \citep{Baes:2003, Baes:2015, Camps:2015, Camps2016} with the aim of understanding the impact of radiative feedback on the observed spectral energy distributions (SEDs) of ${z\sim 6}$ galaxies. We have considered (i) intrinsic AGN SEDs defined by a composite power-law $F_\lambda \propto \lambda^\alpha$ constrained through observational and theoretical arguments; (ii) SMC dust properties (grain size distribution and composition); (iii) different total dust mass content (parametrized in terms of the dust-to-metal ratios ${f_{\rm d}=0.08}$ and ${f_d=0.3}$), and we have explored how different assumptions affect the observational properties of galaxies in the Epoch of Reionization (EoR). By analyzing the synthetic emission maps and SEDs resulting from our calculations we have found the following results:
\begin{itemize}

    \item In dusty galaxies (${M_{\rm d}\gtrsim 3\times 10^7~\msun}$) a large fraction (${\gtrsim50\%}$) of UV emission is obscured by dust. 

    \item Large UV luminosity variations with viewing angle, can be at least partially due to the inhomogeneous distribution of dusty gas on scales ${\gtrsim 100}$~pc. 
    
    \item Simulations including AGN radiation show the presence of a clumpy, warm (${\approx 200-300}$~K) dust component, in addition to a colder (${\approx 50-70}$~K) and more diffuse dusty medium, heated by stars; warm dust provides up to ${50 \%}$ of the total infrared luminosity, though constituting only a small fraction (${\lesssim 0.1\%}$) of the overall mass content. 

\end{itemize}
We have tested our model by comparing the simulated SEDs with observations of $z{\sim 6}$ bright (${M_{UV}\lesssim -26}$) quasars, the only class of sources for which multi-wavelength observations, ranging from the optical-NIR to the mm, are available so far. For what concerns the intrinsic SEDs, we have considered two variations for the rest-frame UV band: a fiducial value ${\alpha_{\rm UV}^{\rm fid}=-1.7}$ suggested by observations of unreddened quasars \citep{Richards2003AJ}, and a steeper slope ${\alpha_{\rm UV}^{\rm steep}=-2.3}$ supported by observations of reddened quasars \citep{Gallerani:2010} and theoretical arguments \citep{Shakura:1973}. The main findings of this comparison are the following:
\begin{itemize}
    \item We find a good agreement between simulations and both MIR (Spitzer/Herschel) and millimetric (ALMA) data, in the case of ${\alpha_{\rm UV}^{\rm fid}=-1.7}$. In the rest-frame UV, our predicted SEDs are underluminous with respect to data, suggesting peculiar extinction properties \citep[][see also Di Mascia et al. in preparation]{Gallerani:2010}.
    \item The case ${\alpha_{\rm UV}^{\rm steep}=-2.3}$ cannot explain the Spitzer/IRAC flux at ${\lambda_{\rm obs}=24~\mum}$ and show a slope in the MIR that does not agree with Spitzer/IRS spectra. This discrepancy can be possibly alleviated by adding to our model the emission arising from a dusty torus with ${T_{\rm d}\sim 1500}$~K, close to sublimation temperature of graphite and silicate grains \citep{Netzer:2015}. 
    \item Quasars powered by SMBHs are part of complex, dust-rich merging systems, containing both multiple accreting BHs and star forming galaxies that, because of strong dust absorption, are below the detection limit of current deep optical-NIR observations \citep{Mechtley2012ApJ}, but appear as SMG companions, consistently with recent shallow ALMA data \cite{Decarli:2017}. Deeper ALMA and future JWST observations are required to study the environment in which ${z\sim 6}$ quasars form and evolve.
\end{itemize}

Given the good agreement between our results and rest-frame MIR observations, we exploit our simulations to make predictions for the proposed Origins Space Telescope (OST; \citealt{Wiedner2020arXiv}), a possible selection to NASA's large strategic science missions, and for a MIR telescope with the same technical specifications of the Space Infrared Telescope for Cosmology and Astrophysics (SPICA; e.g. \citealt{Spinoglio:2017, Gruppioni:2017, Egami:2018, Roelfsema2018PASA}), an infrared space mission, initially considered as a candidate for the M5 mission, but cancelled in October 2020 \citep{Clements2020Natur}.
We end up with the following conclusions:
\begin{itemize}
    \item Highly star forming galaxies (${\SFR \sim 600~\msunyr}$) without an active AGN will be easily detected by ORIGINS. It will also be able to probe the peak of the dust emission, allowing a solid estimate of the dust temperature in star forming galaxies at high redshift. These galaxies would also be detected by a SPICA-like telescope, if lensed by a factor ${\mu \gtrsim 5}$.
    \item Bright high-$z$ quasars (${M_{UV}<-26}$) are detectable with ORIGINS/SPICA at a signal-to-noise ratio high enough to get high quality spectra even in these very distant sources.
    \item The FIR/MIR flux ratio in star forming galaxies is one order of magnitude higher with respect to AGN hosts, even in the case of low accretion rates (${\dot{M}_{\rm BH}\sim 3~\msunyr}$). By following up with ORIGINS/SPICA galaxies already detected with ALMA it will be possible to unveil faint and/or dust-obscured AGN, whose fraction is expected to be large (${>85 \%}$) at high redshift (e.g. \citealt{Vito2014,Vito2018_obscured}; see also \citealt{Davies:2019}). Our FIR/MIR estimate is quite conservative, because our model does not include the emission from the dusty torus, which is expected to boost the MIR flux by up to two order of magnitudes in ORIGINS/SPICA bands.
\end{itemize}
These results highlight the importance of a new generation of MIR telescopes to understand the properties of dusty galaxies and AGN at the EoR.

\section*{acknowledgements}
FD thanks Alessandro Lupi, Laura Sommovigo and Milena Valentini for helpful discussions and Peter Camps for code support. We acknowledge fruitful discussions with Paola Andreani, Sarah Bosman, Eiichi Egami, Carlotta Gruppioni, Francesca Pozzi, Luigi Spinoglio, Christian Vignali. SG acknowledges support from the ASI-INAF n. 2018-31-HH.0 grant and PRIN-MIUR 2017 (PI Fabrizio Fiore). AF and SC acknowledge support from the ERC Advanced Grant INTERSTELLAR H2020/740120. Any dissemination of results must indicate that it reflects only the author’s view and that the Commission is not responsible for any use that may be made of the information it contains. Generous support from the Carl Friedrich von Siemens-Forschungspreis der Alexander von Humboldt-Stiftung Research Award is kindly acknowledged (AF). We acknowledge usage of the Python programming language \citep{python2,python3}, Astropy \citep{astropy}, Cython \citep{cython}, Matplotlib \citep{matplotlib}, NumPy \citep{numpy}, \code{pynbody} \citep{pynbody}, and SciPy \citep{scipy}.

\section*{Data availability}
Part of the data underlying this article were accessed from the computational resources available to the Cosmology Group at Scuola Normale Superiore, Pisa (IT). The derived data generated in this research will be shared on reasonable request to the corresponding author.

\appendix

\section{Convergence of the dust grid} \label{sec:RT_conv}

\begin{figure}
	\centering
	\includegraphics[width=0.475\textwidth]{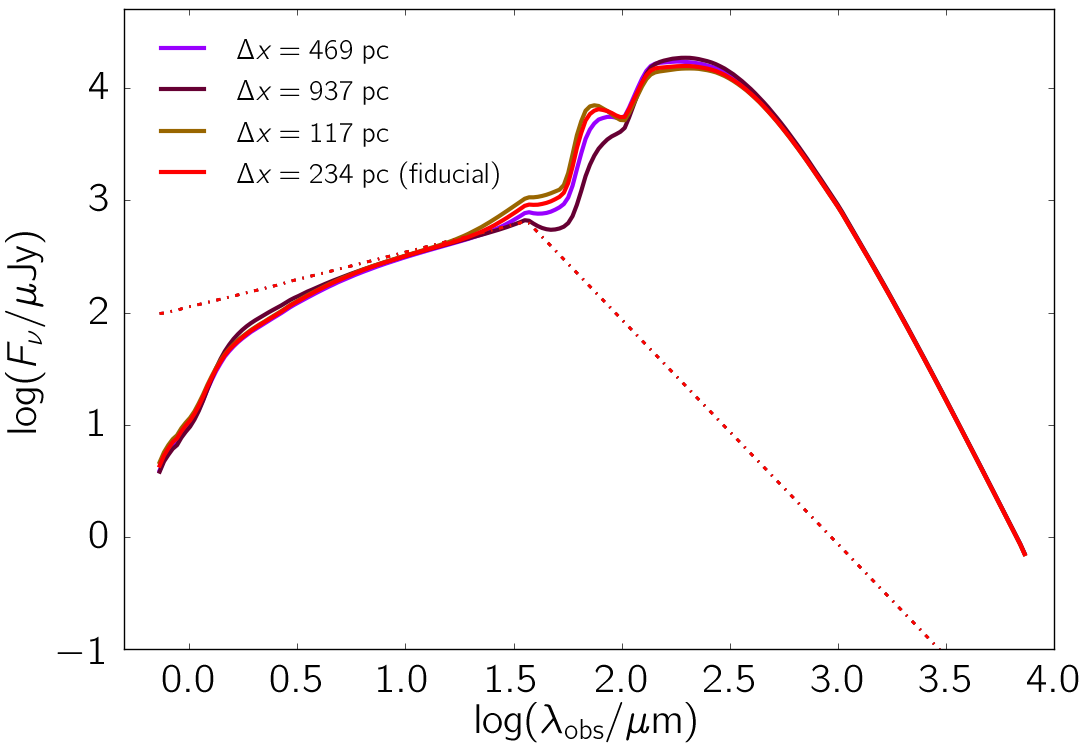}
    \hfill
	\caption{Spectral Energy Distribution of the \AGNcone{} run ($f_d=0.08$, fiducial AGN SED) for different numbers of the maximum refinement levels: 6, 7, 8 (fiducial) and 9, corresponding to 937 pc, 469 pc, 234 pc (fiducial) and 117 pc, respectively.
	\label{fig:AGNcone_spatial_res}
	}
\end{figure}

The dust content derived from the hydrodynamical simulations is distributed in an octree grid with a maximum of 8 level of refinement, achieving a maximum resolution of $\sim 234$ pc, as described in Section \ref{sec:dust_skirt}. This spatial resolution is comparable with the resolution of the hydrodynamical simulations, i.e. $\sim 200$ pc at $z=6$. In this Section, we check if the number of refinement levels adopted in our fiducial setup is sufficient to achieve converge of the results. 
We perform three control simulations, in which the maximum refinement levels are 6, 7 and 9, corresponding to a spatial resolution of 937 pc, 469 pc and 117 pc respectively. In Fig. \ref{fig:AGNcone_spatial_res}, we show the SED plot for the \AGNcone{} run, adopting $f_d=0.08$ and the fiducial AGN SED, for the aforementioned values of the maximum refinement levels. The four SEDs mainly differ in the MIR range ($6-15~\mum$ rest-frame). The MIR emission increases when increasing the number of refinement levels, because dust around AGN, which is heated to the highest temperatures, is better resolved. However, the variation between our fiducial model and the model at the highest resolution is less than $30\%$ in the MIR band, thus we conclude that the spatial resolution of the dust grid adopted in our calculations is sufficient to achieve reasonable numerical convergence.

\section{Dust thermal sputtering} 
\label{Dustthermalsputtering}

\begin{figure}
    \centering
    \includegraphics[width=0.475\textwidth]{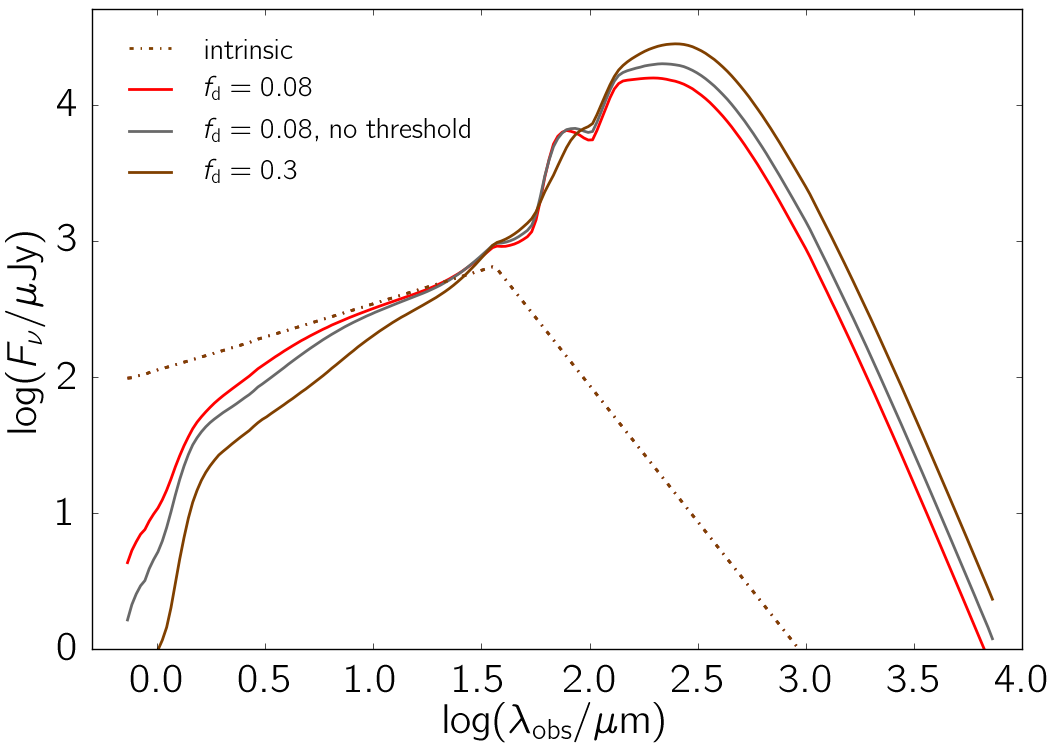}
	\hfill
	\caption{Comparison of the SEDs of the \AGNcone{} run, assuming $f_d=0.08$ (red), $f_d=0.3$ (brown) and $f_d=0.08$ without dust sputtering (grey). 
	\label{fig:impact_noT}
	}
\end{figure}

We have assumed that dust grains with temperature above a given threshold ($T > 10^6$ K) are destroyed by thermal sputtering \citep{Draine:1979, Tielens:1994, Hirashita:2015}, as commonly done in simulations \citep{Liang:2019,ma:2019}. However, this dust destruction process may be inefficient in the proximity of AGN, because of grain charging \citep{Tazaki:2020Drift, Tazaki:2020Charging}. 
To quantify how this assumption affects our results we re-run the \AGNcone{} model with the lower dust content, i.e. $f_d=0.08$ (fifth row in Table \ref{tab:SKIRT_runs_setup}), after removing the threshold on the dust temperature. In this case, the mass of emitting dust is a factor $\sim 2$ higher with respect to the fiducial run ($M_{\rm d}=6 \times 10^7$ $\msun$). In Fig. \ref{fig:impact_noT} we compare the SEDs obtained with $f_d=0.08$ and $f_d=0.3$ (red and brown lines, respectively) with the model in which dust sputtering is ignored (grey line). 
The higher dust mass in the model without dust sputtering increases both the attenuation in the UV and the re-emission in the FIR. The resulting SED lies between the $f_d=0.08$ and $f_d=0.3$ model results, underlining that the temperature threshold adopted does not affect significantly the main results of our work. 

\section{\texorpdfstring{$\dot{M}_{\rm BH}-M_{\rm UV}$ relation}{MBH - MUV relation}} 
\label{MdotMuvrelation}

For a radiation efficiency $\epsilon_r=0.1$, the bolometric luminosity $L_{\rm bol}$ can be related to the BH accretion rate as follows:
\begin{equation}
L_{\rm bol} \approx 1.5 \times 10^{12} \ \left(\frac{\dot{M}_{\rm BH}}{\msunyr}\right) \ \lsun.
\end{equation}
Using the bolometric corrections reported in Table \ref{tab:bolometric_corr}, we can convert the accretion rate into an UV luminosity by multiplying the bolometric luminosity by a factor\footnote{In the case of the fiducial SED, $f_{\rm UV}\approx 0.29$. The results are however very similar in the case of the UV-steep SED.}  ${f_{\rm UV} = L_{\rm UV} / L_{\rm bol}}$.
Then, we adopt the definition of the AB magnitude
\begin{equation*}
m_{\rm AB} = -2.5 \log F_{\nu} - 48.6,
\end{equation*}
where $F_{\nu}$ is in cgs units, and we express $M_{\rm UV}$ in terms of the product $\lambda L_\lambda$:
\begin{equation}
M_{\rm UV} = 89.9 - 2.5 \log\left(\frac{\lambda L_\lambda}{\rm erg \ s^{-1}}\right),
\end{equation}
where\footnote{In this expression we do not include k-corrections for the distance modulus ($\mu$) calculations. At $z=6.3$, the difference between $\mu$ and the k-corrected one $\mu_k$ is $\mu=\mu_k +2.1$.}  $\nu$ and $\lambda L_\lambda$ are evaluated at $\lambda=1450$ $\angstrom$. 
By combining the previous equations we obtain:
\begin{equation}
M_{\rm UV} = -23.1 -2.5 \log_{10}\left(\frac{\dot{M}_{\rm BH}}{\msunyr}\right).
\end{equation}

\bibliographystyle{mnras}
\bibliography{file_bibliography/ref}

\bsp
\label{lastpage}

\end{document}

%% file: include_tex/pacchetti.tex
\usepackage[english]{babel}
\usepackage{amsmath,amssymb}

\usepackage{graphicx}

\usepackage[normalem]{ulem}

\usepackage{verbatim}
\usepackage{color}
\usepackage{multirow}
\usepackage{mathtools}
\usepackage{epstopdf}

\usepackage{url}
\usepackage{xcolor}

\usepackage{gensymb}

\usepackage[usestackEOL]{stackengine}

%% file: include_tex/journals.tex
\def\nat{Nature}

\def\mnras{MNRAS}
\def\apj{ApJ}
\def\apjl{ApJL}
\def\apjs{ApJS}
\def\aap{A\&A}

\def\araa{ARA\&A}
\def\aj{AJ}

\def\pasa{Publ. Astr. Soc. Australia}
\def\pasp{Publ. Astr. Soc. Pac.}

\def\pasj{Pub. Astron. Soc. Japan}

%% file: include_tex/definizioni.tex
\def\be{\begin{equation}} 
\def\ee{\end{equation}} 
\def\ba{\begin{eqnarray}} 
\def\ea{\end{eqnarray}}

\def\cc{\,{\rm {cm^{-3}}}} 
\def\msun{{\Msun}}

\def\CIV{\hbox{C~$\scriptstyle\rm IV\ $}}

\def\gsim{\lower.5ex\hbox{\gtsima}} 
\def\lsim{\lower.5ex\hbox{\ltsima}} \def\gtsima{$\; \buildrel > \over 
\sim \;$} \def\ltsima{$\; \buildrel < \over \sim \;$} \def\prosima{$\; 
\buildrel \propto \over \sim \;$} \def\gsim{\lower.5ex\hbox{\gtsima}} 
\def\lsim{\lower.5ex\hbox{\ltsima}} 
\def\simgt{\lower.5ex\hbox{\gtsima}} 
\def\simlt{\lower.5ex\hbox{\ltsima}} 
\def\simpr{\lower.5ex\hbox{\prosima}} \def\la{\lsim} \def\ga{\gsim} 
  
 \def\gtsima{$\; \buildrel > \over \sim \;$} 
\def\ltsima{$\; \buildrel < \over \sim \;$} 
\def\gsim{\lower.5ex\hbox{\gtsima}} 
\def\lsim{\lower.5ex\hbox{\ltsima}} 
\def\simgt{\lower.5ex\hbox{\gtsima}} 
\def\simlt{\lower.5ex\hbox{\ltsima}} 
\def\simpr{\lower.5ex\hbox{\prosima}} 
\def\la{\lsim} 
\def\ga{\gsim}

\def\msun{\,{\rm \Msun}}

\def\E3{{\cal E}_{\rm g}^{III}}

\def\r12{r_{1/2}} 
\def\x12{x_{1/2}} 
\def\v12{v_{1/2}}

%% file: include_tex/additional_def.tex
\def\MGII{\hbox{Mg~$\scriptstyle\rm II $}} 
\def\CIV{\hbox{C~$\scriptstyle\rm IV $}}

\newcommand\code[1]{\textsc{\MakeLowercase{#1}}}

\def\nh2{n_{\rm H2}}
\def\fh2{f_{\rm H2}}

\def\angstrom{\textrm{A\kern -1.3ex\raisebox{0.6ex}{$^\circ$}}}

\def\mum{\mu{\rm m}}
\def\msun{{\rm M}_{\odot}}
\def\zsun{{\rm Z}_{\odot}}
\def\lsun{{\rm L}_{\odot}}

\def\cc{{\rm cm}^{-3}}
\def\colcm{{\rm cm}^{-2}}
\def\msunyr{\msun\,{\rm yr}^{-1}}

\def\surfsfr{\msun\,{\rm yr}^{-1}\,{\rm kpc}^{-2}}

\def\SFR{{\rm SFR}}

\def\noAGN{\emph{noAGN}}
\def\AGNsphere{\emph{AGNsphere}}
\def\AGNcone{\emph{AGNcone}}

\newcommand\titlelowercase[1]{\texorpdfstring{\lowercase{#1}}{#1}}

%% file: main.bbl
\begin{thebibliography}{}
\makeatletter
\relax
\def\mn@urlcharsother{\let\do\@makeother \do\$\do\&\do\#\do\^\do\_\do\%\do\~}
\def\mn@doi{\begingroup\mn@urlcharsother \@ifnextchar [ {\mn@doi@}
  {\mn@doi@[]}}
\def\mn@doi@[#1]#2{\def\@tempa{#1}\ifx\@tempa\@empty \href
  {http://dx.doi.org/#2} {doi:#2}\else \href {http://dx.doi.org/#2} {#1}\fi
  \endgroup}
\def\mn@eprint#1#2{\mn@eprint@#1:#2::\@nil}
\def\mn@eprint@arXiv#1{\href {http://arxiv.org/abs/#1} {{\tt arXiv:#1}}}
\def\mn@eprint@dblp#1{\href {http://dblp.uni-trier.de/rec/bibtex/#1.xml}
  {dblp:#1}}
\def\mn@eprint@#1:#2:#3:#4\@nil{\def\@tempa {#1}\def\@tempb {#2}\def\@tempc
  {#3}\ifx \@tempc \@empty \let \@tempc \@tempb \let \@tempb \@tempa \fi \ifx
  \@tempb \@empty \def\@tempb {arXiv}\fi \@ifundefined
  {mn@eprint@\@tempb}{\@tempb:\@tempc}{\expandafter \expandafter \csname
  mn@eprint@\@tempb\endcsname \expandafter{\@tempc}}}

\bibitem[\protect\citeauthoryear{{Allevato} et~al.,}{{Allevato}
  et~al.}{2016}]{allevato2016}
{Allevato} V.,  et~al., 2016, \mn@doi [\apj] {10.3847/0004-637X/832/1/70},
  \href {https://ui.adsabs.harvard.edu/abs/2016ApJ...832...70A} {832, 70}

\bibitem[\protect\citeauthoryear{{Aoyama}, {Hou}, {Shimizu}, {Hirashita},
  {Todoroki}, {Choi}  \& {Nagamine}}{{Aoyama} et~al.}{2017}]{Aoyama:2017}
{Aoyama} S.,  {Hou} K.-C.,  {Shimizu} I.,  {Hirashita} H.,  {Todoroki} K.,
  {Choi} J.-H.,   {Nagamine} K.,  2017, \mn@doi [\mnras]
  {10.1093/mnras/stw3061}, \href
  {https://ui.adsabs.harvard.edu/abs/2017MNRAS.466..105A} {466, 105}

\bibitem[\protect\citeauthoryear{{Arata}, {Yajima}, {Nagamine}, {Li}  \&
  {Khochfar}}{{Arata} et~al.}{2019}]{Arata:2019}
{Arata} S.,  {Yajima} H.,  {Nagamine} K.,  {Li} Y.,   {Khochfar} S.,  2019,
  \mn@doi [\mnras] {10.1093/mnras/stz1887}, \href
  {https://ui.adsabs.harvard.edu/abs/2019MNRAS.488.2629A} {488, 2629}

\bibitem[\protect\citeauthoryear{{Asano}, {Takeuchi}, {Hirashita}  \&
  {Inoue}}{{Asano} et~al.}{2013a}]{Asano:2013}
{Asano} R.~S.,  {Takeuchi} T.~T.,  {Hirashita} H.,   {Inoue} A.~K.,  2013a,
  \mn@doi [Earth, Planets, and Space] {10.5047/eps.2012.04.014}, \href
  {https://ui.adsabs.harvard.edu/abs/2013EP&S...65..213A} {65, 213}

\bibitem[\protect\citeauthoryear{{Asano}, {Takeuchi}, {Hirashita}  \&
  {Nozawa}}{{Asano} et~al.}{2013b}]{Asano:2013MNRAS}
{Asano} R.~S.,  {Takeuchi} T.~T.,  {Hirashita} H.,   {Nozawa} T.,  2013b,
  \mn@doi [\mnras] {10.1093/mnras/stt506}, \href
  {https://ui.adsabs.harvard.edu/abs/2013MNRAS.432..637A} {432, 637}

\bibitem[\protect\citeauthoryear{{Asplund}, {Grevesse}, {Sauval}  \&
  {Scott}}{{Asplund} et~al.}{2009}]{Asplund:2009}
{Asplund} M.,  {Grevesse} N.,  {Sauval} A.~J.,   {Scott} P.,  2009, \mn@doi
  [\araa] {10.1146/annurev.astro.46.060407.145222}, \href
  {https://ui.adsabs.harvard.edu/abs/2009ARA&A..47..481A} {47, 481}

\bibitem[\protect\citeauthoryear{{Astropy Collaboration} et~al.,}{{Astropy
  Collaboration} et~al.}{2013}]{astropy}
{Astropy Collaboration} et~al., 2013, \mn@doi [\aap]
  {10.1051/0004-6361/201322068}, \href
  {http://adsabs.harvard.edu/abs/2013A%26A...558A..33A} {558, A33}

\bibitem[\protect\citeauthoryear{{Ba{\~n}ados} et~al.,}{{Ba{\~n}ados}
  et~al.}{2014}]{Banados2014Pan-STARRS1}
{Ba{\~n}ados} E.,  et~al., 2014, \mn@doi [\aj] {10.1088/0004-6256/148/1/14},
  \href {https://ui.adsabs.harvard.edu/abs/2014AJ....148...14B} {148, 14}

\bibitem[\protect\citeauthoryear{{Ba{\~n}ados} et~al.,}{{Ba{\~n}ados}
  et~al.}{2018}]{Banados2018Natur}
{Ba{\~n}ados} E.,  et~al., 2018, \mn@doi [\nat] {10.1038/nature25180}, \href
  {https://ui.adsabs.harvard.edu/abs/2018Natur.553..473B} {553, 473}

\bibitem[\protect\citeauthoryear{{Baes} \& {Camps}}{{Baes} \&
  {Camps}}{2015}]{Baes:2015}
{Baes} M.,  {Camps} P.,  2015, \mn@doi [Astronomy and Computing]
  {10.1016/j.ascom.2015.05.006}, \href
  {https://ui.adsabs.harvard.edu/abs/2015A&C....12...33B} {12, 33}

\bibitem[\protect\citeauthoryear{{Baes} et~al.,}{{Baes}
  et~al.}{2003}]{Baes:2003}
{Baes} M.,  et~al., 2003, \mn@doi [\mnras] {10.1046/j.1365-8711.2003.06770.x},
  \href {https://ui.adsabs.harvard.edu/abs/2003MNRAS.343.1081B} {343, 1081}

\bibitem[\protect\citeauthoryear{{Bakx} et~al.,}{{Bakx}
  et~al.}{2020}]{Bakx:2020}
{Bakx} T. J.~L.~C.,  et~al., 2020, \mn@doi [\mnras] {10.1093/mnras/staa509},
  \href {https://ui.adsabs.harvard.edu/abs/2020MNRAS.493.4294B} {493, 4294}

\bibitem[\protect\citeauthoryear{{Barai}, {Gallerani}, {Pallottini}, {Ferrara},
  {Marconi}, {Cicone}, {Maiolino}  \& {Carniani}}{{Barai}
  et~al.}{2018}]{paramita:2018}
{Barai} P.,  {Gallerani} S.,  {Pallottini} A.,  {Ferrara} A.,  {Marconi} A.,
  {Cicone} C.,  {Maiolino} R.,   {Carniani} S.,  2018, \mn@doi [\mnras]
  {10.1093/mnras/stx2563}, \href
  {http://adsabs.harvard.edu/abs/2018MNRAS.473.4003B} {473, 4003}

\bibitem[\protect\citeauthoryear{Behnel, Bradshaw, Citro, Dalcin, Seljebotn  \&
  Smith}{Behnel et~al.}{2011}]{cython}
Behnel S.,  Bradshaw R.,  Citro C.,  Dalcin L.,  Seljebotn D.,   Smith K.,
  2011, \mn@doi [Computing in Science Engineering] {10.1109/MCSE.2010.118}, 13,
  31

\bibitem[\protect\citeauthoryear{{Behrens}, {Pallottini}, {Ferrara},
  {Gallerani}  \& {Vallini}}{{Behrens} et~al.}{2018}]{Behrens:2018}
{Behrens} C.,  {Pallottini} A.,  {Ferrara} A.,  {Gallerani} S.,   {Vallini} L.,
   2018, \mn@doi [\mnras] {10.1093/mnras/sty552}, \href
  {https://ui.adsabs.harvard.edu/abs/2018MNRAS.477..552B} {477, 552}

\bibitem[\protect\citeauthoryear{{Berta} et~al.,}{{Berta}
  et~al.}{2013}]{berta2013}
{Berta} S.,  et~al., 2013, \mn@doi [\aap] {10.1051/0004-6361/201220859}, \href
  {https://ui.adsabs.harvard.edu/abs/2013A&A...551A.100B} {551, A100}

\bibitem[\protect\citeauthoryear{{Bertoldi}, {Carilli}, {Cox}, {Fan},
  {Strauss}, {Beelen}, {Omont}  \& {Zylka}}{{Bertoldi}
  et~al.}{2003a}]{bertoldi:2003dust}
{Bertoldi} F.,  {Carilli} C.~L.,  {Cox} P.,  {Fan} X.,  {Strauss} M.~A.,
  {Beelen} A.,  {Omont} A.,   {Zylka} R.,  2003a, \mn@doi [\aap]
  {10.1051/0004-6361:20030710}, \href
  {https://ui.adsabs.harvard.edu/abs/2003A&A...406L..55B} {406, L55}

\bibitem[\protect\citeauthoryear{{Bertoldi} et~al.,}{{Bertoldi}
  et~al.}{2003b}]{Bertoldi:2003}
{Bertoldi} F.,  et~al., 2003b, \mn@doi [\aap] {10.1051/0004-6361:20031345},
  \href {https://ui.adsabs.harvard.edu/abs/2003A&A...409L..47B} {409, L47}

\bibitem[\protect\citeauthoryear{{Bianchi} \& {Schneider}}{{Bianchi} \&
  {Schneider}}{2007}]{Bianchi:2007}
{Bianchi} S.,  {Schneider} R.,  2007, \mn@doi [\mnras]
  {10.1111/j.1365-2966.2007.11829.x}, \href
  {https://ui.adsabs.harvard.edu/abs/2007MNRAS.378..973B} {378, 973}

\bibitem[\protect\citeauthoryear{{Blecha}, {Snyder}, {Satyapal}  \&
  {Ellison}}{{Blecha} et~al.}{2018}]{Blecha2018MNRAS}
{Blecha} L.,  {Snyder} G.~F.,  {Satyapal} S.,   {Ellison} S.~L.,  2018, \mn@doi
  [\mnras] {10.1093/mnras/sty1274}, \href
  {https://ui.adsabs.harvard.edu/abs/2018MNRAS.478.3056B} {478, 3056}

\bibitem[\protect\citeauthoryear{{Bondi}}{{Bondi}}{1952}]{Bondi:1952}
{Bondi} H.,  1952, \mn@doi [\mnras] {10.1093/mnras/112.2.195}, \href
  {https://ui.adsabs.harvard.edu/abs/1952MNRAS.112..195B} {112, 195}

\bibitem[\protect\citeauthoryear{{Bondi} \& {Hoyle}}{{Bondi} \&
  {Hoyle}}{1944}]{Bondi:1944}
{Bondi} H.,  {Hoyle} F.,  1944, \mn@doi [\mnras] {10.1093/mnras/104.5.273},
  \href {https://ui.adsabs.harvard.edu/abs/1944MNRAS.104..273B} {104, 273}

\bibitem[\protect\citeauthoryear{{Bongiorno} et~al.,}{{Bongiorno}
  et~al.}{2012}]{bongiorno2012}
{Bongiorno} A.,  et~al., 2012, \mn@doi [\mnras]
  {10.1111/j.1365-2966.2012.22089.x}, \href
  {https://ui.adsabs.harvard.edu/abs/2012MNRAS.427.3103B} {427, 3103}

\bibitem[\protect\citeauthoryear{{Booth} \& {Schaye}}{{Booth} \&
  {Schaye}}{2009}]{Booth2009MNRAS}
{Booth} C.~M.,  {Schaye} J.,  2009, \mn@doi [\mnras]
  {10.1111/j.1365-2966.2009.15043.x}, \href
  {https://ui.adsabs.harvard.edu/abs/2009MNRAS.398...53B} {398, 53}

\bibitem[\protect\citeauthoryear{{Bruzual} \& {Charlot}}{{Bruzual} \&
  {Charlot}}{2003}]{Bruzual:2003}
{Bruzual} G.,  {Charlot} S.,  2003, \mn@doi [\mnras]
  {10.1046/j.1365-8711.2003.06897.x}, \href
  {https://ui.adsabs.harvard.edu/abs/2003MNRAS.344.1000B} {344, 1000}

\bibitem[\protect\citeauthoryear{{Camps} \& {Baes}}{{Camps} \&
  {Baes}}{2015}]{Camps:2015}
{Camps} P.,  {Baes} M.,  2015, \mn@doi [Astronomy and Computing]
  {10.1016/j.ascom.2014.10.004}, \href
  {https://ui.adsabs.harvard.edu/abs/2015A&C.....9...20C} {9, 20}

\bibitem[\protect\citeauthoryear{{Camps}, {Trayford}, {Baes}, {Theuns},
  {Schaller}  \& {Schaye}}{{Camps} et~al.}{2016}]{Camps2016}
{Camps} P.,  {Trayford} J.~W.,  {Baes} M.,  {Theuns} T.,  {Schaller} M.,
  {Schaye} J.,  2016, \mn@doi [\mnras] {10.1093/mnras/stw1735}, \href
  {https://ui.adsabs.harvard.edu/abs/2016MNRAS.462.1057C} {462, 1057}

\bibitem[\protect\citeauthoryear{{Carilli} \& {Walter}}{{Carilli} \&
  {Walter}}{2013}]{CarilliWalter2013}
{Carilli} C.~L.,  {Walter} F.,  2013, \mn@doi [\araa]
  {10.1146/annurev-astro-082812-140953}, \href
  {https://ui.adsabs.harvard.edu/abs/2013ARA&A..51..105C} {51, 105}

\bibitem[\protect\citeauthoryear{{Carnall} et~al.,}{{Carnall}
  et~al.}{2015}]{Carnall2015ATLAS}
{Carnall} A.~C.,  et~al., 2015, \mn@doi [\mnras] {10.1093/mnrasl/slv057}, \href
  {https://ui.adsabs.harvard.edu/abs/2015MNRAS.451L..16C} {451, L16}

\bibitem[\protect\citeauthoryear{{Carniani} et~al.,}{{Carniani}
  et~al.}{2016}]{carniani2016}
{Carniani} S.,  et~al., 2016, \mn@doi [\aap] {10.1051/0004-6361/201528037},
  \href {https://ui.adsabs.harvard.edu/abs/2016A&A...591A..28C} {591, A28}

\bibitem[\protect\citeauthoryear{{Carniani} et~al.,}{{Carniani}
  et~al.}{2019}]{carniani:2019}
{Carniani} S.,  et~al., 2019, \mn@doi [\mnras] {10.1093/mnras/stz2410}, \href
  {https://ui.adsabs.harvard.edu/abs/2019MNRAS.489.3939C} {489, 3939}

\bibitem[\protect\citeauthoryear{{Chabrier}}{{Chabrier}}{2003}]{Chabrier:2003}
{Chabrier} G.,  2003, \mn@doi [\pasp] {10.1086/376392}, \href
  {https://ui.adsabs.harvard.edu/abs/2003PASP..115..763C} {115, 763}

\bibitem[\protect\citeauthoryear{{Chakrabarti} \& {Whitney}}{{Chakrabarti} \&
  {Whitney}}{2009}]{Chakrabarti2009ApJ}
{Chakrabarti} S.,  {Whitney} B.~A.,  2009, \mn@doi [\apj]
  {10.1088/0004-637X/690/2/1432}, \href
  {https://ui.adsabs.harvard.edu/abs/2009ApJ...690.1432C} {690, 1432}

\bibitem[\protect\citeauthoryear{{Chakrabarti}, {Cox}, {Hernquist}, {Hopkins},
  {Robertson}  \& {Di Matteo}}{{Chakrabarti} et~al.}{2007}]{Chakrabarti2007ApJ}
{Chakrabarti} S.,  {Cox} T.~J.,  {Hernquist} L.,  {Hopkins} P.~F.,  {Robertson}
  B.,   {Di Matteo} T.,  2007, \mn@doi [\apj] {10.1086/510113}, \href
  {https://ui.adsabs.harvard.edu/abs/2007ApJ...658..840C} {658, 840}

\bibitem[\protect\citeauthoryear{{Cicone} et~al.,}{{Cicone}
  et~al.}{2015}]{cicone:2015}
{Cicone} C.,  et~al., 2015, \mn@doi [\aap] {10.1051/0004-6361/201424980}, \href
  {https://ui.adsabs.harvard.edu/abs/2015A&A...574A..14C} {574, A14}

\bibitem[\protect\citeauthoryear{{Clements}, {Serjeant}  \& {Jin}}{{Clements}
  et~al.}{2020}]{Clements2020Natur}
{Clements} D.~L.,  {Serjeant} S.,   {Jin} S.,  2020, \mn@doi [\nat]
  {10.1038/d41586-020-03288-z}, \href
  {https://ui.adsabs.harvard.edu/abs/2020Natur.587..548C} {587, 548}

\bibitem[\protect\citeauthoryear{{Connor} et~al.,}{{Connor}
  et~al.}{2019}]{Connor:2019}
{Connor} T.,  et~al., 2019, \mn@doi [\apj] {10.3847/1538-4357/ab5585}, \href
  {https://ui.adsabs.harvard.edu/abs/2019ApJ...887..171C} {887, 171}

\bibitem[\protect\citeauthoryear{{Connor} et~al.,}{{Connor}
  et~al.}{2020}]{Connor:2020}
{Connor} T.,  et~al., 2020, \mn@doi [\apj] {10.3847/1538-4357/abaab9}, \href
  {https://ui.adsabs.harvard.edu/abs/2020ApJ...900..189C} {900, 189}

\bibitem[\protect\citeauthoryear{{Cowie}, {Barger}, {Bauer}  \&
  {Gonz{\'a}lez-L{\'o}pez}}{{Cowie} et~al.}{2020}]{cowie2020}
{Cowie} L.~L.,  {Barger} A.~J.,  {Bauer} F.~E.,   {Gonz{\'a}lez-L{\'o}pez} J.,
  2020, \mn@doi [\apj] {10.3847/1538-4357/ab6aaa}, \href
  {https://ui.adsabs.harvard.edu/abs/2020ApJ...891...69C} {891, 69}

\bibitem[\protect\citeauthoryear{{Cresci} et~al.,}{{Cresci}
  et~al.}{2015a}]{Cresci2015A}
{Cresci} G.,  et~al., 2015a, \mn@doi [\aap] {10.1051/0004-6361/201526581},
  \href {https://ui.adsabs.harvard.edu/abs/2015A&A...582A..63C} {582, A63}

\bibitem[\protect\citeauthoryear{{Cresci} et~al.,}{{Cresci}
  et~al.}{2015b}]{Cresci2015ApJ}
{Cresci} G.,  et~al., 2015b, \mn@doi [\apj] {10.1088/0004-637X/799/1/82}, \href
  {https://ui.adsabs.harvard.edu/abs/2015ApJ...799...82C} {799, 82}

\bibitem[\protect\citeauthoryear{{Dalla Vecchia} \& {Schaye}}{{Dalla Vecchia}
  \& {Schaye}}{2008}]{dallavecchia2008}
{Dalla Vecchia} C.,  {Schaye} J.,  2008, \mn@doi [\mnras]
  {10.1111/j.1365-2966.2008.13322.x}, \href
  {https://ui.adsabs.harvard.edu/abs/2008MNRAS.387.1431D} {387, 1431}

\bibitem[\protect\citeauthoryear{{Davies}, {Hennawi}  \& {Eilers}}{{Davies}
  et~al.}{2019}]{Davies:2019}
{Davies} F.~B.,  {Hennawi} J.~F.,   {Eilers} A.-C.,  2019, \mn@doi [\apjl]
  {10.3847/2041-8213/ab42e3}, \href
  {https://ui.adsabs.harvard.edu/abs/2019ApJ...884L..19D} {884, L19}

\bibitem[\protect\citeauthoryear{{De Young}}{{De Young}}{1989}]{DeYoung1989ApJ}
{De Young} D.~S.,  1989, \mn@doi [\apjl] {10.1086/185484}, \href
  {https://ui.adsabs.harvard.edu/abs/1989ApJ...342L..59D} {342, L59}

\bibitem[\protect\citeauthoryear{{Decarli} et~al.,}{{Decarli}
  et~al.}{2017}]{Decarli:2017}
{Decarli} R.,  et~al., 2017, \mn@doi [\nat] {10.1038/nature22358}, \href
  {https://ui.adsabs.harvard.edu/abs/2017Natur.545..457D} {545, 457}

\bibitem[\protect\citeauthoryear{{Di Matteo}, {Springel}  \& {Hernquist}}{{Di
  Matteo} et~al.}{2005}]{DiMatteo2005Natur}
{Di Matteo} T.,  {Springel} V.,   {Hernquist} L.,  2005, \mn@doi [\nat]
  {10.1038/nature03335}, \href
  {https://ui.adsabs.harvard.edu/abs/2005Natur.433..604D} {433, 604}

\bibitem[\protect\citeauthoryear{{Draine}}{{Draine}}{2003a}]{Draine:2003a}
{Draine} B.~T.,  2003a, \mn@doi [\araa]
  {10.1146/annurev.astro.41.011802.094840}, \href
  {https://ui.adsabs.harvard.edu/abs/2003ARA&A..41..241D} {41, 241}

\bibitem[\protect\citeauthoryear{{Draine}}{{Draine}}{2003b}]{Draine:2003b}
{Draine} B.~T.,  2003b, \mn@doi [\apj] {10.1086/379118}, \href
  {https://ui.adsabs.harvard.edu/abs/2003ApJ...598.1017D} {598, 1017}

\bibitem[\protect\citeauthoryear{{Draine}}{{Draine}}{2003c}]{Draine:2003c}
{Draine} B.~T.,  2003c, \mn@doi [\apj] {10.1086/379123}, \href
  {https://ui.adsabs.harvard.edu/abs/2003ApJ...598.1026D} {598, 1026}

\bibitem[\protect\citeauthoryear{{Draine} \& {Salpeter}}{{Draine} \&
  {Salpeter}}{1979}]{Draine:1979}
{Draine} B.~T.,  {Salpeter} E.~E.,  1979, \mn@doi [\apj] {10.1086/157165},
  \href {https://ui.adsabs.harvard.edu/abs/1979ApJ...231...77D} {231, 77}

\bibitem[\protect\citeauthoryear{{Draine} et~al.,}{{Draine}
  et~al.}{2007}]{Draine:2007}
{Draine} B.~T.,  et~al., 2007, \mn@doi [\apj] {10.1086/518306}, \href
  {https://ui.adsabs.harvard.edu/abs/2007ApJ...663..866D} {663, 866}

\bibitem[\protect\citeauthoryear{{Dubois}, {Devriendt}, {Slyz}  \&
  {Teyssier}}{{Dubois} et~al.}{2010}]{Dubois2010MNRAS}
{Dubois} Y.,  {Devriendt} J.,  {Slyz} A.,   {Teyssier} R.,  2010, \mn@doi
  [\mnras] {10.1111/j.1365-2966.2010.17338.x}, \href
  {https://ui.adsabs.harvard.edu/abs/2010MNRAS.409..985D} {409, 985}

\bibitem[\protect\citeauthoryear{{Dubois}, {Gavazzi}, {Peirani}  \&
  {Silk}}{{Dubois} et~al.}{2013}]{Dubois2013MNRAS}
{Dubois} Y.,  {Gavazzi} R.,  {Peirani} S.,   {Silk} J.,  2013, \mn@doi [\mnras]
  {10.1093/mnras/stt997}, \href
  {https://ui.adsabs.harvard.edu/abs/2013MNRAS.433.3297D} {433, 3297}

\bibitem[\protect\citeauthoryear{{Egami} et~al.,}{{Egami}
  et~al.}{2018}]{Egami:2018}
{Egami} E.,  et~al., 2018, \mn@doi [\pasa] {10.1017/pasa.2018.41}, \href
  {https://ui.adsabs.harvard.edu/abs/2018PASA...35...48E} {35, 48}

\bibitem[\protect\citeauthoryear{{Fabian}}{{Fabian}}{1999}]{Fabian1999MNRAS}
{Fabian} A.~C.,  1999, \mn@doi [\mnras] {10.1046/j.1365-8711.1999.03017.x},
  \href {https://ui.adsabs.harvard.edu/abs/1999MNRAS.308L..39F} {308, L39}

\bibitem[\protect\citeauthoryear{{Fan} et~al.,}{{Fan} et~al.}{2000}]{Fan2000AJ}
{Fan} X.,  et~al., 2000, \mn@doi [\aj] {10.1086/301534}, \href
  {https://ui.adsabs.harvard.edu/abs/2000AJ....120.1167F} {120, 1167}

\bibitem[\protect\citeauthoryear{{Fan} et~al.,}{{Fan}
  et~al.}{2006}]{Fan2006SDSS}
{Fan} X.,  et~al., 2006, \mn@doi [\aj] {10.1086/500296}, \href
  {https://ui.adsabs.harvard.edu/abs/2006AJ....131.1203F} {131, 1203}

\bibitem[\protect\citeauthoryear{{Ferrara}, {Salvadori}, {Yue}  \&
  {Schleicher}}{{Ferrara} et~al.}{2014}]{ferrara:2014}
{Ferrara} A.,  {Salvadori} S.,  {Yue} B.,   {Schleicher} D.,  2014, \mn@doi
  [\mnras] {10.1093/mnras/stu1280}, \href
  {https://ui.adsabs.harvard.edu/abs/2014MNRAS.443.2410F} {443, 2410}

\bibitem[\protect\citeauthoryear{{Ferrara}, {Viti}  \& {Ceccarelli}}{{Ferrara}
  et~al.}{2016}]{Ferrara:2016b}
{Ferrara} A.,  {Viti} S.,   {Ceccarelli} C.,  2016, \mn@doi [\mnras]
  {10.1093/mnrasl/slw165}, \href
  {https://ui.adsabs.harvard.edu/abs/2016MNRAS.463L.112F} {463, L112}

\bibitem[\protect\citeauthoryear{{Fiore}, {Elvis}, {McDowell}, {Siemiginowska}
  \& {Wilkes}}{{Fiore} et~al.}{1994}]{Fiore:1994}
{Fiore} F.,  {Elvis} M.,  {McDowell} J.~C.,  {Siemiginowska} A.,   {Wilkes}
  B.~J.,  1994, \mn@doi [\apj] {10.1086/174504}, \href
  {https://ui.adsabs.harvard.edu/abs/1994ApJ...431..515F} {431, 515}

\bibitem[\protect\citeauthoryear{{Gallerani} et~al.,}{{Gallerani}
  et~al.}{2010}]{Gallerani:2010}
{Gallerani} S.,  et~al., 2010, \mn@doi [\aap] {10.1051/0004-6361/201014721},
  \href {https://ui.adsabs.harvard.edu/abs/2010A&A...523A..85G} {523, A85}

\bibitem[\protect\citeauthoryear{{Gallerani}, {Ferrara}, {Neri}  \&
  {Maiolino}}{{Gallerani} et~al.}{2014}]{Gallerani:2014}
{Gallerani} S.,  {Ferrara} A.,  {Neri} R.,   {Maiolino} R.,  2014, \mn@doi
  [\mnras] {10.1093/mnras/stu2031}, \href
  {https://ui.adsabs.harvard.edu/abs/2014MNRAS.445.2848G} {445, 2848}

\bibitem[\protect\citeauthoryear{{Gallerani}, {Fan}, {Maiolino}  \&
  {Pacucci}}{{Gallerani} et~al.}{2017a}]{gallerani2017PASA}
{Gallerani} S.,  {Fan} X.,  {Maiolino} R.,   {Pacucci} F.,  2017a, \mn@doi
  [\pasa] {10.1017/pasa.2017.14}, \href
  {https://ui.adsabs.harvard.edu/abs/2017PASA...34...22G} {34, e022}

\bibitem[\protect\citeauthoryear{{Gallerani} et~al.,}{{Gallerani}
  et~al.}{2017b}]{Gallerani:2017}
{Gallerani} S.,  et~al., 2017b, \mn@doi [\mnras] {10.1093/mnras/stx363}, \href
  {https://ui.adsabs.harvard.edu/abs/2017MNRAS.467.3590G} {467, 3590}

\bibitem[\protect\citeauthoryear{{Garc{\'\i}a-Burillo}
  et~al.,}{{Garc{\'\i}a-Burillo} et~al.}{2016}]{GarciaBurillo2016}
{Garc{\'\i}a-Burillo} S.,  et~al., 2016, \mn@doi [\apjl]
  {10.3847/2041-8205/823/1/L12}, \href
  {https://ui.adsabs.harvard.edu/abs/2016ApJ...823L..12G} {823, L12}

\bibitem[\protect\citeauthoryear{{Garc{\'\i}a-Burillo}
  et~al.,}{{Garc{\'\i}a-Burillo} et~al.}{2019}]{GarciaBurillo2019}
{Garc{\'\i}a-Burillo} S.,  et~al., 2019, \mn@doi [\aap]
  {10.1051/0004-6361/201936606}, \href
  {https://ui.adsabs.harvard.edu/abs/2019A&A...632A..61G} {632, A61}

\bibitem[\protect\citeauthoryear{{Gilli} et~al.,}{{Gilli}
  et~al.}{2014}]{gilli2014}
{Gilli} R.,  et~al., 2014, \mn@doi [\aap] {10.1051/0004-6361/201322892}, \href
  {https://ui.adsabs.harvard.edu/abs/2014A&A...562A..67G} {562, A67}

\bibitem[\protect\citeauthoryear{{Groves}, {Dopita}, {Sutherland}, {Kewley},
  {Fischera}, {Leitherer}, {Brandl}  \& {van Breugel}}{{Groves}
  et~al.}{2008}]{Groves2008ApJS}
{Groves} B.,  {Dopita} M.~A.,  {Sutherland} R.~S.,  {Kewley} L.~J.,  {Fischera}
  J.,  {Leitherer} C.,  {Brandl} B.,   {van Breugel} W.,  2008, \mn@doi [\apjs]
  {10.1086/528711}, \href
  {https://ui.adsabs.harvard.edu/abs/2008ApJS..176..438G} {176, 438}

\bibitem[\protect\citeauthoryear{{Gruppioni} et~al.,}{{Gruppioni}
  et~al.}{2016}]{gruppioni2016}
{Gruppioni} C.,  et~al., 2016, \mn@doi [\mnras] {10.1093/mnras/stw577}, \href
  {https://ui.adsabs.harvard.edu/abs/2016MNRAS.458.4297G} {458, 4297}

\bibitem[\protect\citeauthoryear{{Gruppioni} et~al.,}{{Gruppioni}
  et~al.}{2017}]{Gruppioni:2017}
{Gruppioni} C.,  et~al., 2017, \mn@doi [\pasa] {10.1017/pasa.2017.49}, \href
  {https://ui.adsabs.harvard.edu/abs/2017PASA...34...55G} {34, e055}

\bibitem[\protect\citeauthoryear{{Hahn} \& {Abel}}{{Hahn} \&
  {Abel}}{2011}]{hahn:2011}
{Hahn} O.,  {Abel} T.,  2011, \mn@doi [\mnras]
  {10.1111/j.1365-2966.2011.18820.x}, \href
  {https://ui.adsabs.harvard.edu/abs/2011MNRAS.415.2101H} {415, 2101}

\bibitem[\protect\citeauthoryear{{Haiman}}{{Haiman}}{2013}]{Haiman2013}
{Haiman} Z.,  2013, {The Formation of the First Massive Black Holes}.
Astrophysics and Space Science Library, p.~293

\bibitem[\protect\citeauthoryear{{Hern{\'a}n-Caballero} \&
  {Hatziminaoglou}}{{Hern{\'a}n-Caballero} \&
  {Hatziminaoglou}}{2011}]{HernanCaballero2011MNRAS}
{Hern{\'a}n-Caballero} A.,  {Hatziminaoglou} E.,  2011, \mn@doi [\mnras]
  {10.1111/j.1365-2966.2011.18413.x}, \href
  {https://ui.adsabs.harvard.edu/abs/2011MNRAS.414..500H} {414, 500}

\bibitem[\protect\citeauthoryear{{Hickox} \& {Alexander}}{{Hickox} \&
  {Alexander}}{2018}]{Hickox2018}
{Hickox} R.~C.,  {Alexander} D.~M.,  2018, \mn@doi [\araa]
  {10.1146/annurev-astro-081817-051803}, \href
  {https://ui.adsabs.harvard.edu/abs/2018ARA&A..56..625H} {56, 625}

\bibitem[\protect\citeauthoryear{{Hirashita} \& {Aoyama}}{{Hirashita} \&
  {Aoyama}}{2019}]{Hirashita:2019}
{Hirashita} H.,  {Aoyama} S.,  2019, \mn@doi [\mnras] {10.1093/mnras/sty2838},
  \href {https://ui.adsabs.harvard.edu/abs/2019MNRAS.482.2555H} {482, 2555}

\bibitem[\protect\citeauthoryear{{Hirashita}, {Nozawa}, {Villaume}  \&
  {Srinivasan}}{{Hirashita} et~al.}{2015}]{Hirashita:2015}
{Hirashita} H.,  {Nozawa} T.,  {Villaume} A.,   {Srinivasan} S.,  2015, \mn@doi
  [\mnras] {10.1093/mnras/stv2095}, \href
  {https://ui.adsabs.harvard.edu/abs/2015MNRAS.454.1620H} {454, 1620}

\bibitem[\protect\citeauthoryear{{Hjorth}, {Vreeswijk}, {Gall}  \&
  {Watson}}{{Hjorth} et~al.}{2013}]{Hjorth2013ApJ}
{Hjorth} J.,  {Vreeswijk} P.~M.,  {Gall} C.,   {Watson} D.,  2013, \mn@doi
  [\apj] {10.1088/0004-637X/768/2/173}, \href
  {https://ui.adsabs.harvard.edu/abs/2013ApJ...768..173H} {768, 173}

\bibitem[\protect\citeauthoryear{{Hopkins}, {Richards}  \&
  {Hernquist}}{{Hopkins} et~al.}{2007}]{hopkins2007}
{Hopkins} P.~F.,  {Richards} G.~T.,   {Hernquist} L.,  2007, \mn@doi [\apj]
  {10.1086/509629}, \href
  {https://ui.adsabs.harvard.edu/abs/2007ApJ...654..731H} {654, 731}

\bibitem[\protect\citeauthoryear{{Hoyle} \& {Lyttleton}}{{Hoyle} \&
  {Lyttleton}}{1939}]{Hoyle:1939}
{Hoyle} F.,  {Lyttleton} R.~A.,  1939, \mn@doi [Proceedings of the Cambridge
  Philosophical Society] {10.1017/S0305004100021150}, \href
  {https://ui.adsabs.harvard.edu/abs/1939PCPS...35..405H} {35, 405}

\bibitem[\protect\citeauthoryear{Hunter}{Hunter}{2007}]{matplotlib}
Hunter J.~D.,  2007, \mn@doi [Computing in Science Engineering]
  {10.1109/MCSE.2007.55}, 9, 90

\bibitem[\protect\citeauthoryear{{Jiang}, {Fan}, {Vestergaard}, {Kurk},
  {Walter}, {Kelly}  \& {Strauss}}{{Jiang} et~al.}{2007}]{Jiang2007AJ}
{Jiang} L.,  {Fan} X.,  {Vestergaard} M.,  {Kurk} J.~D.,  {Walter} F.,  {Kelly}
  B.~C.,   {Strauss} M.~A.,  2007, \mn@doi [\aj] {10.1086/520811}, \href
  {https://ui.adsabs.harvard.edu/abs/2007AJ....134.1150J} {134, 1150}

\bibitem[\protect\citeauthoryear{{Jiang} et~al.,}{{Jiang}
  et~al.}{2009}]{Jiang2009SDSS}
{Jiang} L.,  et~al., 2009, \mn@doi [\aj] {10.1088/0004-6256/138/1/305}, \href
  {https://ui.adsabs.harvard.edu/abs/2009AJ....138..305J} {138, 305}

\bibitem[\protect\citeauthoryear{{Jiang} et~al.,}{{Jiang}
  et~al.}{2016}]{Jiang2016ApJ}
{Jiang} L.,  et~al., 2016, \mn@doi [\apj] {10.3847/1538-4357/833/2/222}, \href
  {https://ui.adsabs.harvard.edu/abs/2016ApJ...833..222J} {833, 222}

\bibitem[\protect\citeauthoryear{{Jonsson}, {Groves}  \& {Cox}}{{Jonsson}
  et~al.}{2010}]{Jonsson2010MNRAS}
{Jonsson} P.,  {Groves} B.~A.,   {Cox} T.~J.,  2010, \mn@doi [\mnras]
  {10.1111/j.1365-2966.2009.16087.x}, \href
  {https://ui.adsabs.harvard.edu/abs/2010MNRAS.403...17J} {403, 17}

\bibitem[\protect\citeauthoryear{{Juarez}, {Maiolino}, {Mujica}, {Pedani},
  {Marinoni}, {Nagao}, {Marconi}  \& {Oliva}}{{Juarez}
  et~al.}{2009}]{Juarez:2009}
{Juarez} Y.,  {Maiolino} R.,  {Mujica} R.,  {Pedani} M.,  {Marinoni} S.,
  {Nagao} T.,  {Marconi} A.,   {Oliva} E.,  2009, \mn@doi [\aap]
  {10.1051/0004-6361:200811415}, \href
  {https://ui.adsabs.harvard.edu/abs/2009A&A...494L..25J} {494, L25}

\bibitem[\protect\citeauthoryear{{Kashikawa} et~al.,}{{Kashikawa}
  et~al.}{2015}]{Kashikawa2015ApJ}
{Kashikawa} N.,  et~al., 2015, \mn@doi [\apj] {10.1088/0004-637X/798/1/28},
  \href {https://ui.adsabs.harvard.edu/abs/2015ApJ...798...28K} {798, 28}

\bibitem[\protect\citeauthoryear{{Kormendy} \& {Ho}}{{Kormendy} \&
  {Ho}}{2013}]{kormendy2013}
{Kormendy} J.,  {Ho} L.~C.,  2013, \mn@doi [\araa]
  {10.1146/annurev-astro-082708-101811}, \href
  {https://ui.adsabs.harvard.edu/abs/2013ARA&A..51..511K} {51, 511}

\bibitem[\protect\citeauthoryear{{Kormendy} \& {Richstone}}{{Kormendy} \&
  {Richstone}}{1995}]{kormendy1995}
{Kormendy} J.,  {Richstone} D.,  1995, \mn@doi [\araa]
  {10.1146/annurev.aa.33.090195.003053}, \href
  {https://ui.adsabs.harvard.edu/abs/1995ARA&A..33..581K} {33, 581}

\bibitem[\protect\citeauthoryear{{Koss}, {Mushotzky}, {Treister}, {Veilleux},
  {Vasudevan}  \& {Trippe}}{{Koss} et~al.}{2012}]{Koss:2012}
{Koss} M.,  {Mushotzky} R.,  {Treister} E.,  {Veilleux} S.,  {Vasudevan} R.,
  {Trippe} M.,  2012, \mn@doi [\apjl] {10.1088/2041-8205/746/2/L22}, \href
  {https://ui.adsabs.harvard.edu/abs/2012ApJ...746L..22K} {746, L22}

\bibitem[\protect\citeauthoryear{{Kroupa}}{{Kroupa}}{2002}]{Kroupa2002}
{Kroupa} P.,  2002, \mn@doi [Science] {10.1126/science.1067524}, \href
  {https://ui.adsabs.harvard.edu/abs/2002Sci...295...82K} {295, 82}

\bibitem[\protect\citeauthoryear{{Kulkarni}, {Worseck}  \&
  {Hennawi}}{{Kulkarni} et~al.}{2019}]{Kulkarni2019MNRAS}
{Kulkarni} G.,  {Worseck} G.,   {Hennawi} J.~F.,  2019, \mn@doi [\mnras]
  {10.1093/mnras/stz1493}, \href
  {https://ui.adsabs.harvard.edu/abs/2019MNRAS.488.1035K} {488, 1035}

\bibitem[\protect\citeauthoryear{{Kurk} et~al.,}{{Kurk}
  et~al.}{2007}]{Kurk2007ApJ}
{Kurk} J.~D.,  et~al., 2007, \mn@doi [\apj] {10.1086/521596}, \href
  {https://ui.adsabs.harvard.edu/abs/2007ApJ...669...32K} {669, 32}

\bibitem[\protect\citeauthoryear{{Laporte} et~al.,}{{Laporte}
  et~al.}{2017}]{laporte:2017apj}
{Laporte} N.,  et~al., 2017, \mn@doi [\apjl] {10.3847/2041-8213/aa62aa}, \href
  {http://adsabs.harvard.edu/abs/2017ApJ...837L..21L} {837, L21}

\bibitem[\protect\citeauthoryear{{Latif} \& {Ferrara}}{{Latif} \&
  {Ferrara}}{2016}]{Latif16}
{Latif} M.~A.,  {Ferrara} A.,  2016, \mn@doi [\pasa] {10.1017/pasa.2016.41},
  \href {https://ui.adsabs.harvard.edu/abs/2016PASA...33...51L} {33, e051}

\bibitem[\protect\citeauthoryear{{Latif}, {Schleicher}, {Schmidt}  \&
  {Niemeyer}}{{Latif} et~al.}{2013}]{latif2013}
{Latif} M.~A.,  {Schleicher} D.~R.~G.,  {Schmidt} W.,   {Niemeyer} J.,  2013,
  \mn@doi [\mnras] {10.1093/mnras/stt834}, \href
  {https://ui.adsabs.harvard.edu/abs/2013MNRAS.433.1607L} {433, 1607}

\bibitem[\protect\citeauthoryear{{Leipski} et~al.,}{{Leipski}
  et~al.}{2013}]{Leipski2013}
{Leipski} C.,  et~al., 2013, \mn@doi [\apj] {10.1088/0004-637X/772/2/103},
  \href {https://ui.adsabs.harvard.edu/abs/2013ApJ...772..103L} {772, 103}

\bibitem[\protect\citeauthoryear{{Leipski} et~al.,}{{Leipski}
  et~al.}{2014}]{Leipski2014ApJ}
{Leipski} C.,  et~al., 2014, \mn@doi [\apj] {10.1088/0004-637X/785/2/154},
  \href {https://ui.adsabs.harvard.edu/abs/2014ApJ...785..154L} {785, 154}

\bibitem[\protect\citeauthoryear{{Leitherer} et~al.,}{{Leitherer}
  et~al.}{1999}]{Leitherer1999ApJS}
{Leitherer} C.,  et~al., 1999, \mn@doi [\apjs] {10.1086/313233}, \href
  {https://ui.adsabs.harvard.edu/abs/1999ApJS..123....3L} {123, 3}

\bibitem[\protect\citeauthoryear{{Li} et~al.,}{{Li} et~al.}{2008}]{Li2008ApJ}
{Li} Y.,  et~al., 2008, \mn@doi [\apj] {10.1086/529364}, \href
  {https://ui.adsabs.harvard.edu/abs/2008ApJ...678...41L} {678, 41}

\bibitem[\protect\citeauthoryear{{Li} et~al.,}{{Li} et~al.}{2020}]{li:2020}
{Li} J.,  et~al., 2020, \mn@doi [\apj] {10.3847/1538-4357/ab65fa}, \href
  {https://ui.adsabs.harvard.edu/abs/2020ApJ...889..162L} {889, 162}

\bibitem[\protect\citeauthoryear{{Liang} et~al.,}{{Liang}
  et~al.}{2019}]{Liang:2019}
{Liang} L.,  et~al., 2019, \mn@doi [\mnras] {10.1093/mnras/stz2134}, \href
  {https://ui.adsabs.harvard.edu/abs/2019MNRAS.489.1397L} {489, 1397}

\bibitem[\protect\citeauthoryear{{Liang}, {Feldmann}, {Hayward}, {Narayanan},
  {{\c{C}}atmabacak}, {Keres}, {Faucher-Gigu{\'e}re}  \& {Hopkins}}{{Liang}
  et~al.}{2021}]{Liang2021MNRAS}
{Liang} L.,  {Feldmann} R.,  {Hayward} C.~C.,  {Narayanan} D.,
  {{\c{C}}atmabacak} O.,  {Keres} D.,  {Faucher-Gigu{\'e}re} C.-A.,   {Hopkins}
  P.~F.,  2021, \mn@doi [\mnras] {10.1093/mnras/stab096}, \href
  {https://ui.adsabs.harvard.edu/abs/2021MNRAS.tmp..122L} {}

\bibitem[\protect\citeauthoryear{{Lupi}, {Haardt}, {Dotti}, {Fiacconi}, {Mayer}
   \& {Madau}}{{Lupi} et~al.}{2016}]{lupi2016}
{Lupi} A.,  {Haardt} F.,  {Dotti} M.,  {Fiacconi} D.,  {Mayer} L.,   {Madau}
  P.,  2016, \mn@doi [\mnras] {10.1093/mnras/stv2877}, \href
  {https://ui.adsabs.harvard.edu/abs/2016MNRAS.456.2993L} {456, 2993}

\bibitem[\protect\citeauthoryear{{Lusso}, {Worseck}, {Hennawi}, {Prochaska},
  {Vignali}, {Stern}  \& {O'Meara}}{{Lusso} et~al.}{2015}]{Lusso:2015}
{Lusso} E.,  {Worseck} G.,  {Hennawi} J.~F.,  {Prochaska} J.~X.,  {Vignali} C.,
   {Stern} J.,   {O'Meara} J.~M.,  2015, \mn@doi [\mnras]
  {10.1093/mnras/stv516}, \href
  {https://ui.adsabs.harvard.edu/abs/2015MNRAS.449.4204L} {449, 4204}

\bibitem[\protect\citeauthoryear{{Lyu}, {Rieke}  \& {Alberts}}{{Lyu}
  et~al.}{2016}]{lyu:2016}
{Lyu} J.,  {Rieke} G.~H.,   {Alberts} S.,  2016, \mn@doi [\apj]
  {10.3847/0004-637X/816/2/85}, \href
  {https://ui.adsabs.harvard.edu/abs/2016ApJ...816...85L} {816, 85}

\bibitem[\protect\citeauthoryear{{Ma} et~al.,}{{Ma} et~al.}{2019}]{ma:2019}
{Ma} X.,  et~al., 2019, \mn@doi [\mnras] {10.1093/mnras/stz1324}, \href
  {https://ui.adsabs.harvard.edu/abs/2019MNRAS.487.1844M} {487, 1844}

\bibitem[\protect\citeauthoryear{{Magorrian} et~al.,}{{Magorrian}
  et~al.}{1998}]{magorrian1998}
{Magorrian} J.,  et~al., 1998, \mn@doi [\aj] {10.1086/300353}, \href
  {https://ui.adsabs.harvard.edu/abs/1998AJ....115.2285M} {115, 2285}

\bibitem[\protect\citeauthoryear{{Maiolino}, {Schneider}, {Oliva}, {Bianchi},
  {Ferrara}, {Mannucci}, {Pedani}  \& {Roca Sogorb}}{{Maiolino}
  et~al.}{2004}]{Maiolino:2004}
{Maiolino} R.,  {Schneider} R.,  {Oliva} E.,  {Bianchi} S.,  {Ferrara} A.,
  {Mannucci} F.,  {Pedani} M.,   {Roca Sogorb} M.,  2004, \mn@doi [\nat]
  {10.1038/nature02930}, \href
  {https://ui.adsabs.harvard.edu/abs/2004Natur.431..533M} {431, 533}

\bibitem[\protect\citeauthoryear{{Maiolino} et~al.,}{{Maiolino}
  et~al.}{2005}]{Maiolino2005A&A}
{Maiolino} R.,  et~al., 2005, \mn@doi [\aap] {10.1051/0004-6361:200500165},
  \href {https://ui.adsabs.harvard.edu/abs/2005A&A...440L..51M} {440, L51}

\bibitem[\protect\citeauthoryear{{Maiolino} et~al.,}{{Maiolino}
  et~al.}{2012}]{maiolino:2012}
{Maiolino} R.,  et~al., 2012, \mn@doi [\mnras]
  {10.1111/j.1745-3933.2012.01303.x}, \href
  {https://ui.adsabs.harvard.edu/abs/2012MNRAS.425L..66M} {425, L66}

\bibitem[\protect\citeauthoryear{{Manti}, {Gallerani}, {Ferrara}, {Feruglio},
  {Graziani}  \& {Bernardi}}{{Manti} et~al.}{2016}]{Manti:2016}
{Manti} S.,  {Gallerani} S.,  {Ferrara} A.,  {Feruglio} C.,  {Graziani} L.,
  {Bernardi} G.,  2016, \mn@doi [\mnras] {10.1093/mnras/stv2635}, \href
  {https://ui.adsabs.harvard.edu/abs/2016MNRAS.456...98M} {456, 98}

\bibitem[\protect\citeauthoryear{{Marconi}, {Risaliti}, {Gilli}, {Hunt},
  {Maiolino}  \& {Salvati}}{{Marconi} et~al.}{2004}]{marconi2004}
{Marconi} A.,  {Risaliti} G.,  {Gilli} R.,  {Hunt} L.~K.,  {Maiolino} R.,
  {Salvati} M.,  2004, \mn@doi [\mnras] {10.1111/j.1365-2966.2004.07765.x},
  \href {https://ui.adsabs.harvard.edu/abs/2004MNRAS.351..169M} {351, 169}

\bibitem[\protect\citeauthoryear{{Marshall} et~al.,}{{Marshall}
  et~al.}{2020}]{Marshall2020ApJ}
{Marshall} M.~A.,  et~al., 2020, \mn@doi [\apj] {10.3847/1538-4357/abaa4c},
  \href {https://ui.adsabs.harvard.edu/abs/2020ApJ...900...21M} {900, 21}

\bibitem[\protect\citeauthoryear{{Matsuoka} et~al.,}{{Matsuoka}
  et~al.}{2016}]{Matsuoka2016SHELLQs}
{Matsuoka} Y.,  et~al., 2016, \mn@doi [\apj] {10.3847/0004-637X/828/1/26},
  \href {https://ui.adsabs.harvard.edu/abs/2016ApJ...828...26M} {828, 26}

\bibitem[\protect\citeauthoryear{{Matsuoka} et~al.,}{{Matsuoka}
  et~al.}{2018}]{Matsuoka2018ApJ}
{Matsuoka} Y.,  et~al., 2018, \mn@doi [\apj] {10.3847/1538-4357/aaee7a}, \href
  {https://ui.adsabs.harvard.edu/abs/2018ApJ...869..150M} {869, 150}

\bibitem[\protect\citeauthoryear{{Mazzucchelli} et~al.,}{{Mazzucchelli}
  et~al.}{2017}]{Mazzucchelli:2017ApJ}
{Mazzucchelli} C.,  et~al., 2017, \mn@doi [\apj] {10.3847/1538-4357/aa9185},
  \href {https://ui.adsabs.harvard.edu/abs/2017ApJ...849...91M} {849, 91}

\bibitem[\protect\citeauthoryear{{McGreer}, {Fan}, {Jiang}  \& {Cai}}{{McGreer}
  et~al.}{2018}]{McGreer2018AJ}
{McGreer} I.~D.,  {Fan} X.,  {Jiang} L.,   {Cai} Z.,  2018, \mn@doi [\aj]
  {10.3847/1538-3881/aaaab4}, \href
  {https://ui.adsabs.harvard.edu/abs/2018AJ....155..131M} {155, 131}

\bibitem[\protect\citeauthoryear{{Mechtley} et~al.,}{{Mechtley}
  et~al.}{2012}]{Mechtley2012ApJ}
{Mechtley} M.,  et~al., 2012, \mn@doi [\apjl] {10.1088/2041-8205/756/2/L38},
  \href {https://ui.adsabs.harvard.edu/abs/2012ApJ...756L..38M} {756, L38}

\bibitem[\protect\citeauthoryear{{Mortlock} et~al.,}{{Mortlock}
  et~al.}{2011}]{Mortlock2011Natur}
{Mortlock} D.~J.,  et~al., 2011, \mn@doi [\nat] {10.1038/nature10159}, \href
  {https://ui.adsabs.harvard.edu/abs/2011Natur.474..616M} {474, 616}

\bibitem[\protect\citeauthoryear{{Murray}, {Quataert}  \& {Thompson}}{{Murray}
  et~al.}{2005}]{Murray2005ApJ}
{Murray} N.,  {Quataert} E.,   {Thompson} T.~A.,  2005, \mn@doi [\apj]
  {10.1086/426067}, \href
  {https://ui.adsabs.harvard.edu/abs/2005ApJ...618..569M} {618, 569}

\bibitem[\protect\citeauthoryear{{Nanni}, {Vignali}, {Gilli}, {Moretti}  \&
  {Brand t}}{{Nanni} et~al.}{2017}]{Nanni:2017}
{Nanni} R.,  {Vignali} C.,  {Gilli} R.,  {Moretti} A.,   {Brand t} W.~N.,
  2017, \mn@doi [\aap] {10.1051/0004-6361/201730484}, \href
  {https://ui.adsabs.harvard.edu/abs/2017A&A...603A.128N} {603, A128}

\bibitem[\protect\citeauthoryear{{Nenkova}, {Sirocky}, {Ivezi{\'c}}  \&
  {Elitzur}}{{Nenkova} et~al.}{2008}]{Nenkova2008ApJ}
{Nenkova} M.,  {Sirocky} M.~M.,  {Ivezi{\'c}} Z.,   {Elitzur} M.,  2008,
  \mn@doi [\apj] {10.1086/590482}, \href
  {https://ui.adsabs.harvard.edu/abs/2008ApJ...685..147N} {685, 147}

\bibitem[\protect\citeauthoryear{{Netzer}}{{Netzer}}{2015}]{Netzer:2015}
{Netzer} H.,  2015, \mn@doi [Annual Review of Astronomy and Astrophysics]
  {10.1146/annurev-astro-082214-122302}, \href
  {https://ui.adsabs.harvard.edu/abs/2015ARA&A..53..365N} {53, 365}

\bibitem[\protect\citeauthoryear{{Novak} et~al.,}{{Novak}
  et~al.}{2019}]{novak:2019}
{Novak} M.,  et~al., 2019, \mn@doi [\apj] {10.3847/1538-4357/ab2beb}, \href
  {https://ui.adsabs.harvard.edu/abs/2019ApJ...881...63N} {881, 63}

\bibitem[\protect\citeauthoryear{{Nozawa}, {Asano}, {Hirashita}  \&
  {Takeuchi}}{{Nozawa} et~al.}{2015}]{Nozawa:2015}
{Nozawa} T.,  {Asano} R.~S.,  {Hirashita} H.,   {Takeuchi} T.~T.,  2015,
  \mn@doi [\mnras] {10.1093/mnrasl/slu175}, \href
  {https://ui.adsabs.harvard.edu/abs/2015MNRAS.447L..16N} {447, L16}

\bibitem[\protect\citeauthoryear{{Ono} et~al.,}{{Ono} et~al.}{2018}]{ono:2018}
{Ono} Y.,  et~al., 2018, \mn@doi [\pasj] {10.1093/pasj/psx103}, \href
  {https://ui.adsabs.harvard.edu/abs/2018PASJ...70S..10O} {70, S10}

\bibitem[\protect\citeauthoryear{{Pacucci}, {Volonteri}  \&
  {Ferrara}}{{Pacucci} et~al.}{2015}]{pacucci2015}
{Pacucci} F.,  {Volonteri} M.,   {Ferrara} A.,  2015, \mn@doi [\mnras]
  {10.1093/mnras/stv1465}, \href
  {https://ui.adsabs.harvard.edu/abs/2015MNRAS.452.1922P} {452, 1922}

\bibitem[\protect\citeauthoryear{{Pacucci}, {Ferrara}, {Grazian}, {Fiore},
  {Giallongo}  \& {Puccetti}}{{Pacucci} et~al.}{2016}]{Pacucci2016MNRAS}
{Pacucci} F.,  {Ferrara} A.,  {Grazian} A.,  {Fiore} F.,  {Giallongo} E.,
  {Puccetti} S.,  2016, \mn@doi [\mnras] {10.1093/mnras/stw725}, \href
  {https://ui.adsabs.harvard.edu/abs/2016MNRAS.459.1432P} {459, 1432}

\bibitem[\protect\citeauthoryear{{Padoan} \& {Nordlund}}{{Padoan} \&
  {Nordlund}}{2011}]{Padoan:2011}
{Padoan} P.,  {Nordlund} a.,  2011, \mn@doi [\apj]
  {10.1088/0004-637X/730/1/40}, \href
  {https://ui.adsabs.harvard.edu/abs/2011ApJ...730...40P} {730, 40}

\bibitem[\protect\citeauthoryear{{Padoan}, {Federrath}, {Chabrier}, {Evans},
  {Johnstone}, {J{\o}rgensen}, {McKee}  \& {Nordlund}}{{Padoan}
  et~al.}{2014}]{Padoan:2014}
{Padoan} P.,  {Federrath} C.,  {Chabrier} G.,  {Evans} N.~J. I.,  {Johnstone}
  D.,  {J{\o}rgensen} J.~K.,  {McKee} C.~F.,   {Nordlund} a.,  2014, in
  Protostars and Planets VI. p.~77

\bibitem[\protect\citeauthoryear{{Pallottini}, {Ferrara}, {Bovino}, {Vallini},
  {Gallerani}, {Maiolino}  \& {Salvadori}}{{Pallottini}
  et~al.}{2017}]{pallottini:2017althaea}
{Pallottini} A.,  {Ferrara} A.,  {Bovino} S.,  {Vallini} L.,  {Gallerani} S.,
  {Maiolino} R.,   {Salvadori} S.,  2017, \mn@doi [\mnras]
  {10.1093/mnras/stx1792}, \href
  {http://adsabs.harvard.edu/abs/2017MNRAS.471.4128P} {471, 4128}

\bibitem[\protect\citeauthoryear{{Pezzulli}, {Valiante}, {Orofino},
  {Schneider}, {Gallerani}  \& {Sbarrato}}{{Pezzulli}
  et~al.}{2017}]{pezzulli2017}
{Pezzulli} E.,  {Valiante} R.,  {Orofino} M.~C.,  {Schneider} R.,  {Gallerani}
  S.,   {Sbarrato} T.,  2017, \mn@doi [\mnras] {10.1093/mnras/stw3243}, \href
  {https://ui.adsabs.harvard.edu/abs/2017MNRAS.466.2131P} {466, 2131}

\bibitem[\protect\citeauthoryear{{Piconcelli}, {Jimenez-Bail{\'o}n},
  {Guainazzi}, {Schartel}, {Rodr{\'\i}guez-Pascual}  \&
  {Santos-Lle{\'o}}}{{Piconcelli} et~al.}{2005}]{Piconcelli:2005}
{Piconcelli} E.,  {Jimenez-Bail{\'o}n} E.,  {Guainazzi} M.,  {Schartel} N.,
  {Rodr{\'\i}guez-Pascual} P.~M.,   {Santos-Lle{\'o}} M.,  2005, \mn@doi [\aap]
  {10.1051/0004-6361:20041621}, \href
  {https://ui.adsabs.harvard.edu/abs/2005A&A...432...15P} {432, 15}

\bibitem[\protect\citeauthoryear{{Pilbratt} et~al.,}{{Pilbratt}
  et~al.}{2010}]{Pilbratt2010}
{Pilbratt} G.~L.,  et~al., 2010, \mn@doi [\aap] {10.1051/0004-6361/201014759},
  \href {https://ui.adsabs.harvard.edu/abs/2010A&A...518L...1P} {518, L1}

\bibitem[\protect\citeauthoryear{{Planck Collaboration} et~al.,}{{Planck
  Collaboration} et~al.}{2016}]{PlanckCollaboration2016}
{Planck Collaboration} et~al., 2016, \mn@doi [\aap]
  {10.1051/0004-6361/201525830}, \href
  {https://ui.adsabs.harvard.edu/abs/2016A&A...594A..13P} {594, A13}

\bibitem[\protect\citeauthoryear{{Pontzen}, {Rovskar}, {Stinson}, {Woods},
  {Reed}, {Coles}  \& {Quinn}}{{Pontzen} et~al.}{2013}]{pynbody}
{Pontzen} A.,  {Rovskar} R.,  {Stinson} G.~S.,  {Woods} R.,  {Reed} D.~M.,
  {Coles} J.,   {Quinn} T.~R.,  2013, {pynbody: Astrophysics Simulation
  Analysis for Python}

\bibitem[\protect\citeauthoryear{{Pozzi} et~al.,}{{Pozzi}
  et~al.}{2012}]{pozzi2012}
{Pozzi} F.,  et~al., 2012, \mn@doi [\mnras] {10.1111/j.1365-2966.2012.21015.x},
  \href {https://ui.adsabs.harvard.edu/abs/2012MNRAS.423.1909P} {423, 1909}

\bibitem[\protect\citeauthoryear{{Pringle}}{{Pringle}}{1981}]{Pringle:1981}
{Pringle} J.~E.,  1981, \mn@doi [\araa] {10.1146/annurev.aa.19.090181.001033},
  \href {https://ui.adsabs.harvard.edu/abs/1981ARA&A..19..137P} {19, 137}

\bibitem[\protect\citeauthoryear{{Reed} et~al.,}{{Reed}
  et~al.}{2015}]{Reed2015DES}
{Reed} S.~L.,  et~al., 2015, \mn@doi [\mnras] {10.1093/mnras/stv2031}, \href
  {https://ui.adsabs.harvard.edu/abs/2015MNRAS.454.3952R} {454, 3952}

\bibitem[\protect\citeauthoryear{{Ricci} et~al.,}{{Ricci}
  et~al.}{2017}]{ricci2017}
{Ricci} C.,  et~al., 2017, \mn@doi [\nat] {10.1038/nature23906}, \href
  {https://ui.adsabs.harvard.edu/abs/2017Natur.549..488R} {549, 488}

\bibitem[\protect\citeauthoryear{{Richards} et~al.,}{{Richards}
  et~al.}{2003}]{Richards2003AJ}
{Richards} G.~T.,  et~al., 2003, \mn@doi [\aj] {10.1086/377014}, 126, 1131

\bibitem[\protect\citeauthoryear{{Riechers} et~al.,}{{Riechers}
  et~al.}{2009}]{Riechers:2009}
{Riechers} D.~A.,  et~al., 2009, \mn@doi [\apj] {10.1088/0004-637X/703/2/1338},
  \href {https://ui.adsabs.harvard.edu/abs/2009ApJ...703.1338R} {703, 1338}

\bibitem[\protect\citeauthoryear{{Roebuck}, {Sajina}, {Hayward}, {Pope},
  {Kirkpatrick}, {Hernquist}  \& {Yan}}{{Roebuck}
  et~al.}{2016}]{Roebuck2016ApJ}
{Roebuck} E.,  {Sajina} A.,  {Hayward} C.~C.,  {Pope} A.,  {Kirkpatrick} A.,
  {Hernquist} L.,   {Yan} L.,  2016, \mn@doi [\apj]
  {10.3847/1538-4357/833/1/60}, \href
  {https://ui.adsabs.harvard.edu/abs/2016ApJ...833...60R} {833, 60}

\bibitem[\protect\citeauthoryear{{Roelfsema} et~al.,}{{Roelfsema}
  et~al.}{2018}]{Roelfsema2018PASA}
{Roelfsema} P.~R.,  et~al., 2018, \mn@doi [\pasa] {10.1017/pasa.2018.15}, \href
  {https://ui.adsabs.harvard.edu/abs/2018PASA...35...30R} {35, e030}

\bibitem[\protect\citeauthoryear{{Sazonov}, {Ostriker}  \& {Sunyaev}}{{Sazonov}
  et~al.}{2004}]{Sazonov:2004}
{Sazonov} S.~Y.,  {Ostriker} J.~P.,   {Sunyaev} R.~A.,  2004, \mn@doi [\mnras]
  {10.1111/j.1365-2966.2004.07184.x}, \href
  {https://ui.adsabs.harvard.edu/abs/2004MNRAS.347..144S} {347, 144}

\bibitem[\protect\citeauthoryear{{Schartmann}, {Meisenheimer}, {Camenzind},
  {Wolf}  \& {Henning}}{{Schartmann} et~al.}{2005}]{Schartmann:2005}
{Schartmann} M.,  {Meisenheimer} K.,  {Camenzind} M.,  {Wolf} S.,   {Henning}
  T.,  2005, \mn@doi [\aap] {10.1051/0004-6361:20042363}, \href
  {http://adsabs.harvard.edu/abs/2005A\%26A...437..861S} {437, 861}

\bibitem[\protect\citeauthoryear{{Schawinski} et~al.,}{{Schawinski}
  et~al.}{2006}]{Schawinski2006Natur}
{Schawinski} K.,  et~al., 2006, \mn@doi [\nat] {10.1038/nature04934}, \href
  {https://ui.adsabs.harvard.edu/abs/2006Natur.442..888S} {442, 888}

\bibitem[\protect\citeauthoryear{{Schaye} et~al.,}{{Schaye}
  et~al.}{2015}]{Schaye2015MNRAS}
{Schaye} J.,  et~al., 2015, \mn@doi [\mnras] {10.1093/mnras/stu2058}, \href
  {https://ui.adsabs.harvard.edu/abs/2015MNRAS.446..521S} {446, 521}

\bibitem[\protect\citeauthoryear{{Schleicher}, {Palla}, {Ferrara}, {Galli}  \&
  {Latif}}{{Schleicher} et~al.}{2013}]{schleicher2013}
{Schleicher} D. R.~G.,  {Palla} F.,  {Ferrara} A.,  {Galli} D.,   {Latif} M.,
  2013, \mn@doi [\aap] {10.1051/0004-6361/201321949}, \href
  {https://ui.adsabs.harvard.edu/abs/2013A&A...558A..59S} {558, A59}

\bibitem[\protect\citeauthoryear{{Schneider}, {Bianchi}, {Valiante}, {Risaliti}
   \& {Salvadori}}{{Schneider} et~al.}{2015}]{Schneider:2015}
{Schneider} R.,  {Bianchi} S.,  {Valiante} R.,  {Risaliti} G.,   {Salvadori}
  S.,  2015, \mn@doi [Astronomy and Astrophysics]
  {10.1051/0004-6361/201526105}, \href
  {https://ui.adsabs.harvard.edu/abs/2015A&A...579A..60S} {579, A60}

\bibitem[\protect\citeauthoryear{{Shakura} \& {Sunyaev}}{{Shakura} \&
  {Sunyaev}}{1973}]{Shakura:1973}
{Shakura} N.~I.,  {Sunyaev} R.~A.,  1973, \aap, \href
  {https://ui.adsabs.harvard.edu/abs/1973A&A....24..337S} {500, 33}

\bibitem[\protect\citeauthoryear{{Shang}, {Bryan}  \& {Haiman}}{{Shang}
  et~al.}{2010}]{shang2010}
{Shang} C.,  {Bryan} G.~L.,   {Haiman} Z.,  2010, \mn@doi [\mnras]
  {10.1111/j.1365-2966.2009.15960.x}, \href
  {https://ui.adsabs.harvard.edu/abs/2010MNRAS.402.1249S} {402, 1249}

\bibitem[\protect\citeauthoryear{{Shen}, {Hopkins}, {Faucher-Gigu{\`e}re},
  {Alexander}, {Richards}, {Ross}  \& {Hickox}}{{Shen}
  et~al.}{2020}]{Shen:2020}
{Shen} X.,  {Hopkins} P.~F.,  {Faucher-Gigu{\`e}re} C.-A.,  {Alexander} D.~M.,
  {Richards} G.~T.,  {Ross} N.~P.,   {Hickox} R.~C.,  2020, \mn@doi [\mnras]
  {10.1093/mnras/staa1381}, \href
  {https://ui.adsabs.harvard.edu/abs/2020MNRAS.495.3252S} {495, 3252}

\bibitem[\protect\citeauthoryear{{Sijacki}, {Springel}, {Di Matteo}  \&
  {Hernquist}}{{Sijacki} et~al.}{2007}]{Sijacki2007MNRAS}
{Sijacki} D.,  {Springel} V.,  {Di Matteo} T.,   {Hernquist} L.,  2007, \mn@doi
  [\mnras] {10.1111/j.1365-2966.2007.12153.x}, \href
  {https://ui.adsabs.harvard.edu/abs/2007MNRAS.380..877S} {380, 877}

\bibitem[\protect\citeauthoryear{{Silk}}{{Silk}}{2005}]{Silk2005MNRAS}
{Silk} J.,  2005, \mn@doi [\mnras] {10.1111/j.1365-2966.2005.09672.x}, \href
  {https://ui.adsabs.harvard.edu/abs/2005MNRAS.364.1337S} {364, 1337}

\bibitem[\protect\citeauthoryear{{Silverman} et~al.,}{{Silverman}
  et~al.}{2020}]{Silverman:2020}
{Silverman} J.~D.,  et~al., 2020, \mn@doi [\apj] {10.3847/1538-4357/aba4a3},
  \href {https://ui.adsabs.harvard.edu/abs/2020ApJ...899..154S} {899, 154}

\bibitem[\protect\citeauthoryear{{Snyder}, {Hayward}, {Sajina}, {Jonsson},
  {Cox}, {Hernquist}, {Hopkins}  \& {Yan}}{{Snyder}
  et~al.}{2013}]{Snyder2013ApJ}
{Snyder} G.~F.,  {Hayward} C.~C.,  {Sajina} A.,  {Jonsson} P.,  {Cox} T.~J.,
  {Hernquist} L.,  {Hopkins} P.~F.,   {Yan} L.,  2013, \mn@doi [\apj]
  {10.1088/0004-637X/768/2/168}, \href
  {https://ui.adsabs.harvard.edu/abs/2013ApJ...768..168S} {768, 168}

\bibitem[\protect\citeauthoryear{{Sommovigo}, {Ferrara}, {Pallottini},
  {Carniani}, {Gallerani}  \& {Decataldo}}{{Sommovigo}
  et~al.}{2020}]{Sommovigo:2020}
{Sommovigo} L.,  {Ferrara} A.,  {Pallottini} A.,  {Carniani} S.,  {Gallerani}
  S.,   {Decataldo} D.,  2020, \mn@doi [\mnras] {10.1093/mnras/staa1959}, \href
  {https://ui.adsabs.harvard.edu/abs/2020MNRAS.497..956S} {497, 956}

\bibitem[\protect\citeauthoryear{{Spinoglio} et~al.,}{{Spinoglio}
  et~al.}{2017}]{Spinoglio:2017}
{Spinoglio} L.,  et~al., 2017, \mn@doi [\pasa] {10.1017/pasa.2017.48}, \href
  {https://ui.adsabs.harvard.edu/abs/2017PASA...34...57S} {34, e057}

\bibitem[\protect\citeauthoryear{{Springel}}{{Springel}}{2005}]{Springel:2005}
{Springel} V.,  2005, \mn@doi [\mnras] {10.1111/j.1365-2966.2005.09655.x},
  \href {https://ui.adsabs.harvard.edu/abs/2005MNRAS.364.1105S} {364, 1105}

\bibitem[\protect\citeauthoryear{{Springel} \& {Hernquist}}{{Springel} \&
  {Hernquist}}{2003}]{Springel:2003}
{Springel} V.,  {Hernquist} L.,  2003, \mn@doi [\mnras]
  {10.1046/j.1365-8711.2003.06206.x}, \href
  {https://ui.adsabs.harvard.edu/abs/2003MNRAS.339..289S} {339, 289}

\bibitem[\protect\citeauthoryear{{Springel}, {Di Matteo}  \&
  {Hernquist}}{{Springel} et~al.}{2005}]{Springel2005MNRAS}
{Springel} V.,  {Di Matteo} T.,   {Hernquist} L.,  2005, \mn@doi [\mnras]
  {10.1111/j.1365-2966.2005.09238.x}, \href
  {https://ui.adsabs.harvard.edu/abs/2005MNRAS.361..776S} {361, 776}

\bibitem[\protect\citeauthoryear{{Stalevski}, {Fritz}, {Baes}, {Nakos}  \&
  {Popovi{\'c}}}{{Stalevski} et~al.}{2012}]{Stalevski:2012}
{Stalevski} M.,  {Fritz} J.,  {Baes} M.,  {Nakos} T.,   {Popovi{\'c}} L.~C.,
  2012, \mn@doi [\mnras] {10.1111/j.1365-2966.2011.19775.x}, \href
  {http://adsabs.harvard.edu/abs/2012MNRAS.420.2756S} {420, 2756}

\bibitem[\protect\citeauthoryear{{Stalevski}, {Ricci}, {Ueda}, {Lira}, {Fritz}
  \& {Baes}}{{Stalevski} et~al.}{2016}]{Stalevski:2016}
{Stalevski} M.,  {Ricci} C.,  {Ueda} Y.,  {Lira} P.,  {Fritz} J.,   {Baes} M.,
  2016, \mn@doi [\mnras] {10.1093/mnras/stw444}, \href
  {https://ui.adsabs.harvard.edu/abs/2016MNRAS.458.2288S} {458, 2288}

\bibitem[\protect\citeauthoryear{{Stefan} et~al.,}{{Stefan}
  et~al.}{2015}]{Stefan:2015}
{Stefan} I.~I.,  et~al., 2015, \mn@doi [\mnras] {10.1093/mnras/stv1108}, \href
  {https://ui.adsabs.harvard.edu/abs/2015MNRAS.451.1713S} {451, 1713}

\bibitem[\protect\citeauthoryear{{Stratta}, {Gallerani}  \&
  {Maiolino}}{{Stratta} et~al.}{2011}]{Stratta:2011}
{Stratta} G.,  {Gallerani} S.,   {Maiolino} R.,  2011, \mn@doi [\aap]
  {10.1051/0004-6361/201016414}, \href
  {https://ui.adsabs.harvard.edu/abs/2011A&A...532A..45S} {532, A45}

\bibitem[\protect\citeauthoryear{{Tanaka} \& {Haiman}}{{Tanaka} \&
  {Haiman}}{2009}]{tanaka2009}
{Tanaka} T.,  {Haiman} Z.,  2009, \mn@doi [\apj]
  {10.1088/0004-637X/696/2/1798}, \href
  {https://ui.adsabs.harvard.edu/abs/2009ApJ...696.1798T} {696, 1798}

\bibitem[\protect\citeauthoryear{{Tazaki} \& {Ichikawa}}{{Tazaki} \&
  {Ichikawa}}{2020}]{Tazaki:2020Drift}
{Tazaki} R.,  {Ichikawa} K.,  2020, \mn@doi [\apj] {10.3847/1538-4357/ab72f6},
  \href {https://ui.adsabs.harvard.edu/abs/2020ApJ...892..149T} {892, 149}

\bibitem[\protect\citeauthoryear{{Tazaki}, {Ichikawa}  \& {Kokubo}}{{Tazaki}
  et~al.}{2020}]{Tazaki:2020Charging}
{Tazaki} R.,  {Ichikawa} K.,   {Kokubo} M.,  2020, \mn@doi [\apj]
  {10.3847/1538-4357/ab7822}, \href
  {https://ui.adsabs.harvard.edu/abs/2020ApJ...892...84T} {892, 84}

\bibitem[\protect\citeauthoryear{{Teyssier}, {Moore}, {Martizzi}, {Dubois}  \&
  {Mayer}}{{Teyssier} et~al.}{2011}]{Teyssier2011MNRAS}
{Teyssier} R.,  {Moore} B.,  {Martizzi} D.,  {Dubois} Y.,   {Mayer} L.,  2011,
  \mn@doi [\mnras] {10.1111/j.1365-2966.2011.18399.x}, \href
  {https://ui.adsabs.harvard.edu/abs/2011MNRAS.414..195T} {414, 195}

\bibitem[\protect\citeauthoryear{{Tielens}, {McKee}, {Seab}  \&
  {Hollenbach}}{{Tielens} et~al.}{1994}]{Tielens:1994}
{Tielens} A.~G.~G.~M.,  {McKee} C.~F.,  {Seab} C.~G.,   {Hollenbach} D.~J.,
  1994, \mn@doi [\apj] {10.1086/174488}, \href
  {https://ui.adsabs.harvard.edu/abs/1994ApJ...431..321T} {431, 321}

\bibitem[\protect\citeauthoryear{{Todini} \& {Ferrara}}{{Todini} \&
  {Ferrara}}{2001}]{Todini:2001}
{Todini} P.,  {Ferrara} A.,  2001, \mn@doi [\mnras]
  {10.1046/j.1365-8711.2001.04486.x}, \href
  {https://ui.adsabs.harvard.edu/abs/2001MNRAS.325..726T} {325, 726}

\bibitem[\protect\citeauthoryear{{Tornatore}, {Borgani}, {Dolag}  \&
  {Matteucci}}{{Tornatore} et~al.}{2007}]{Tornatore:2007}
{Tornatore} L.,  {Borgani} S.,  {Dolag} K.,   {Matteucci} F.,  2007, \mn@doi
  [\mnras] {10.1111/j.1365-2966.2007.12070.x}, \href
  {https://ui.adsabs.harvard.edu/abs/2007MNRAS.382.1050T} {382, 1050}

\bibitem[\protect\citeauthoryear{{Trayford} et~al.,}{{Trayford}
  et~al.}{2017}]{Trayford2017MNRAS}
{Trayford} J.~W.,  et~al., 2017, \mn@doi [\mnras] {10.1093/mnras/stx1051},
  \href {https://ui.adsabs.harvard.edu/abs/2017MNRAS.470..771T} {470, 771}

\bibitem[\protect\citeauthoryear{{Urry} \& {Padovani}}{{Urry} \&
  {Padovani}}{1995}]{Urry:1995}
{Urry} C.~M.,  {Padovani} P.,  1995, \mn@doi [\pasp] {10.1086/133630}, \href
  {https://ui.adsabs.harvard.edu/abs/1995PASP..107..803U} {107, 803}

\bibitem[\protect\citeauthoryear{{Valiante}, {Schneider}, {Bianchi}  \&
  {Andersen}}{{Valiante} et~al.}{2009}]{Valiante:2009}
{Valiante} R.,  {Schneider} R.,  {Bianchi} S.,   {Andersen} A.~C.,  2009,
  \mn@doi [\mnras] {10.1111/j.1365-2966.2009.15076.x}, \href
  {https://ui.adsabs.harvard.edu/abs/2009MNRAS.397.1661V} {397, 1661}

\bibitem[\protect\citeauthoryear{{Vallini}, {Ferrara}, {Pallottini}  \&
  {Gallerani}}{{Vallini} et~al.}{2017}]{Vallini:2017}
{Vallini} L.,  {Ferrara} A.,  {Pallottini} A.,   {Gallerani} S.,  2017, \mn@doi
  [\mnras] {10.1093/mnras/stx180}, \href
  {https://ui.adsabs.harvard.edu/abs/2017MNRAS.467.1300V} {467, 1300}

\bibitem[\protect\citeauthoryear{Van~Rossum \& Drake}{Van~Rossum \&
  Drake}{2009}]{python3}
Van~Rossum G.,  Drake F.~L.,  2009, Python 3 Reference Manual.
CreateSpace, Scotts Valley, CA

\bibitem[\protect\citeauthoryear{Van~Rossum \& de Boer}{Van~Rossum \&
  de~Boer}{1991}]{python2}
Van~Rossum G.,  de Boer J.,  1991, CWI Quarterly, 4, 283

\bibitem[\protect\citeauthoryear{Vanden~Berk et~al.,}{Vanden~Berk
  et~al.}{2001}]{VandenBerk:2001}
Vanden~Berk D.~E.,  et~al., 2001, \mn@doi [The Astronomical Journal]
  {10.1086/321167}, 122, 549–564

\bibitem[\protect\citeauthoryear{{Venanzi}, {H{\"o}nig}  \&
  {Williamson}}{{Venanzi} et~al.}{2020}]{venanzi:2020}
{Venanzi} M.,  {H{\"o}nig} S.,   {Williamson} D.,  2020, \mn@doi [\apj]
  {10.3847/1538-4357/aba89f}, \href
  {https://ui.adsabs.harvard.edu/abs/2020ApJ...900..174V} {900, 174}

\bibitem[\protect\citeauthoryear{{Venemans}, {McMahon}, {Warren},
  {Gonzalez-Solares}, {Hewett}, {Mortlock}, {Dye}  \& {Sharp}}{{Venemans}
  et~al.}{2007}]{Venemans2007UKIDSS}
{Venemans} B.~P.,  {McMahon} R.~G.,  {Warren} S.~J.,  {Gonzalez-Solares} E.~A.,
   {Hewett} P.~C.,  {Mortlock} D.~J.,  {Dye} S.,   {Sharp} R.~G.,  2007,
  \mn@doi [\mnras] {10.1111/j.1745-3933.2007.00290.x}, \href
  {https://ui.adsabs.harvard.edu/abs/2007MNRAS.376L..76V} {376, L76}

\bibitem[\protect\citeauthoryear{{Venemans} et~al.,}{{Venemans}
  et~al.}{2012}]{Venemans:2012ApJ}
{Venemans} B.~P.,  et~al., 2012, \mn@doi [\apjl] {10.1088/2041-8205/751/2/L25},
  \href {https://ui.adsabs.harvard.edu/abs/2012ApJ...751L..25V} {751, L25}

\bibitem[\protect\citeauthoryear{{Venemans} et~al.,}{{Venemans}
  et~al.}{2013}]{Venemans2013VIKING}
{Venemans} B.~P.,  et~al., 2013, \mn@doi [\apj] {10.1088/0004-637X/779/1/24},
  \href {https://ui.adsabs.harvard.edu/abs/2013ApJ...779...24V} {779, 24}

\bibitem[\protect\citeauthoryear{{Venemans} et~al.,}{{Venemans}
  et~al.}{2015}]{Venemans2015VIKING}
{Venemans} B.~P.,  et~al., 2015, \mn@doi [\mnras] {10.1093/mnras/stv1774},
  \href {https://ui.adsabs.harvard.edu/abs/2015MNRAS.453.2259V} {453, 2259}

\bibitem[\protect\citeauthoryear{{Venemans}, {Walter}, {Zschaechner},
  {Decarli}, {De Rosa}, {Findlay}, {McMahon}  \& {Sutherland}}{{Venemans}
  et~al.}{2016}]{Venemans2016ApJ}
{Venemans} B.~P.,  {Walter} F.,  {Zschaechner} L.,  {Decarli} R.,  {De Rosa}
  G.,  {Findlay} J.~R.,  {McMahon} R.~G.,   {Sutherland} W.~J.,  2016, \mn@doi
  [\apj] {10.3847/0004-637X/816/1/37}, \href
  {https://ui.adsabs.harvard.edu/abs/2016ApJ...816...37V} {816, 37}

\bibitem[\protect\citeauthoryear{{Venemans} et~al.,}{{Venemans}
  et~al.}{2017a}]{venemans:2017CO}
{Venemans} B.~P.,  et~al., 2017a, \mn@doi [\apj] {10.3847/1538-4357/aa81cb},
  \href {https://ui.adsabs.harvard.edu/abs/2017ApJ...845..154V} {845, 154}

\bibitem[\protect\citeauthoryear{{Venemans} et~al.,}{{Venemans}
  et~al.}{2017b}]{Venemans:2017ApJ}
{Venemans} B.~P.,  et~al., 2017b, \mn@doi [\apjl] {10.3847/2041-8213/aa943a},
  \href {https://ui.adsabs.harvard.edu/abs/2017ApJ...851L...8V} {851, L8}

\bibitem[\protect\citeauthoryear{{Venemans} et~al.,}{{Venemans}
  et~al.}{2018}]{Venemans:2018}
{Venemans} B.~P.,  et~al., 2018, \mn@doi [\apj] {10.3847/1538-4357/aadf35},
  \href {https://ui.adsabs.harvard.edu/abs/2018ApJ...866..159V} {866, 159}

\bibitem[\protect\citeauthoryear{{Vignali} et~al.,}{{Vignali}
  et~al.}{2018}]{Vignali2018}
{Vignali} C.,  et~al., 2018, \mn@doi [\mnras] {10.1093/mnras/sty682}, \href
  {https://ui.adsabs.harvard.edu/abs/2018MNRAS.477..780V} {477, 780}

\bibitem[\protect\citeauthoryear{{Virtanen} et~al.,}{{Virtanen}
  et~al.}{2020}]{scipy}
{Virtanen} P.,  et~al., 2020, \mn@doi [Nature Methods]
  {10.1038/s41592-019-0686-2}, \href
  {https://ui.adsabs.harvard.edu/abs/2020NatMe..17..261V} {17, 261}

\bibitem[\protect\citeauthoryear{{Vito}, {Gilli}, {Vignali}, {Comastri},
  {Brusa}, {Cappelluti}  \& {Iwasawa}}{{Vito} et~al.}{2014}]{Vito2014}
{Vito} F.,  {Gilli} R.,  {Vignali} C.,  {Comastri} A.,  {Brusa} M.,
  {Cappelluti} N.,   {Iwasawa} K.,  2014, \mn@doi [\mnras]
  {10.1093/mnras/stu2004}, \href
  {https://ui.adsabs.harvard.edu/abs/2014MNRAS.445.3557V} {445, 3557}

\bibitem[\protect\citeauthoryear{{Vito} et~al.,}{{Vito}
  et~al.}{2018}]{Vito2018_obscured}
{Vito} F.,  et~al., 2018, \mn@doi [\mnras] {10.1093/mnras/stx2486}, \href
  {https://ui.adsabs.harvard.edu/abs/2018MNRAS.473.2378V} {473, 2378}

\bibitem[\protect\citeauthoryear{{Vito} et~al.,}{{Vito}
  et~al.}{2019a}]{Vito:2019_cand}
{Vito} F.,  et~al., 2019a, \mn@doi [\aap] {10.1051/0004-6361/201935924}, \href
  {https://ui.adsabs.harvard.edu/abs/2019A&A...628L...6V} {628, L6}

\bibitem[\protect\citeauthoryear{{Vito} et~al.,}{{Vito}
  et~al.}{2019b}]{Vito:2019}
{Vito} F.,  et~al., 2019b, \mn@doi [\aap] {10.1051/0004-6361/201936217}, \href
  {https://ui.adsabs.harvard.edu/abs/2019A&A...630A.118V} {630, A118}

\bibitem[\protect\citeauthoryear{{Volonteri}, {Haardt}  \& {Madau}}{{Volonteri}
  et~al.}{2003}]{volonteri2003}
{Volonteri} M.,  {Haardt} F.,   {Madau} P.,  2003, \mn@doi [\apj]
  {10.1086/344675}, \href
  {https://ui.adsabs.harvard.edu/abs/2003ApJ...582..559V} {582, 559}

\bibitem[\protect\citeauthoryear{{Wada}, {Schartmann}  \& {Meijerink}}{{Wada}
  et~al.}{2016}]{Wada2016ApJ}
{Wada} K.,  {Schartmann} M.,   {Meijerink} R.,  2016, \mn@doi [\apjl]
  {10.3847/2041-8205/828/2/L19}, \href
  {https://ui.adsabs.harvard.edu/abs/2016ApJ...828L..19W} {828, L19}

\bibitem[\protect\citeauthoryear{{Walter} et~al.,}{{Walter}
  et~al.}{2003}]{Walter:2003}
{Walter} F.,  et~al., 2003, \mn@doi [\nat] {10.1038/nature01821}, \href
  {https://ui.adsabs.harvard.edu/abs/2003Natur.424..406W} {424, 406}

\bibitem[\protect\citeauthoryear{{Walter}, {Riechers}, {Cox}, {Neri},
  {Carilli}, {Bertoldi}, {Weiss}  \& {Maiolino}}{{Walter}
  et~al.}{2009}]{Walter2009Natur}
{Walter} F.,  {Riechers} D.,  {Cox} P.,  {Neri} R.,  {Carilli} C.,  {Bertoldi}
  F.,  {Weiss} A.,   {Maiolino} R.,  2009, \mn@doi [\nat]
  {10.1038/nature07681}, \href
  {https://ui.adsabs.harvard.edu/abs/2009Natur.457..699W} {457, 699}

\bibitem[\protect\citeauthoryear{{Wang} et~al.,}{{Wang}
  et~al.}{2013}]{Wang2013ApJ}
{Wang} R.,  et~al., 2013, \mn@doi [\apj] {10.1088/0004-637X/773/1/44}, \href
  {https://ui.adsabs.harvard.edu/abs/2013ApJ...773...44W} {773, 44}

\bibitem[\protect\citeauthoryear{{Wang} et~al.,}{{Wang}
  et~al.}{2016}]{Wang:2016ApJ}
{Wang} R.,  et~al., 2016, \mn@doi [\apj] {10.3847/0004-637X/830/1/53}, \href
  {https://ui.adsabs.harvard.edu/abs/2016ApJ...830...53W} {830, 53}

\bibitem[\protect\citeauthoryear{{Wang} et~al.,}{{Wang}
  et~al.}{2018}]{Wang2018ApJ}
{Wang} F.,  et~al., 2018, \mn@doi [\apjl] {10.3847/2041-8213/aaf1d2}, \href
  {https://ui.adsabs.harvard.edu/abs/2018ApJ...869L...9W} {869, L9}

\bibitem[\protect\citeauthoryear{{Wang} et~al.,}{{Wang}
  et~al.}{2019}]{Wang2019ApJ}
{Wang} F.,  et~al., 2019, \mn@doi [\apj] {10.3847/1538-4357/ab2be5}, \href
  {https://ui.adsabs.harvard.edu/abs/2019ApJ...884...30W} {884, 30}

\bibitem[\protect\citeauthoryear{{Wang} et~al.,}{{Wang}
  et~al.}{2021}]{Wang2021ApJ}
{Wang} F.,  et~al., 2021, \mn@doi [\apjl] {10.3847/2041-8213/abd8c6}, \href
  {https://ui.adsabs.harvard.edu/abs/2021ApJ...907L...1W} {907, L1}

\bibitem[\protect\citeauthoryear{{Weinberger} et~al.,}{{Weinberger}
  et~al.}{2018}]{Weinberger2018MNRAS}
{Weinberger} R.,  et~al., 2018, \mn@doi [\mnras] {10.1093/mnras/sty1733}, \href
  {https://ui.adsabs.harvard.edu/abs/2018MNRAS.479.4056W} {479, 4056}

\bibitem[\protect\citeauthoryear{{Weingartner} \& {Draine}}{{Weingartner} \&
  {Draine}}{2001}]{Weingartner:2001}
{Weingartner} J.~C.,  {Draine} B.~T.,  2001, \mn@doi [\apj] {10.1086/318651},
  \href {https://ui.adsabs.harvard.edu/abs/2001ApJ...548..296W} {548, 296}

\bibitem[\protect\citeauthoryear{{Werner} et~al.,}{{Werner}
  et~al.}{2004}]{Werner2004ApJS}
{Werner} M.~W.,  et~al., 2004, \mn@doi [\apjs] {10.1086/422992}, \href
  {https://ui.adsabs.harvard.edu/abs/2004ApJS..154....1W} {154, 1}

\bibitem[\protect\citeauthoryear{{Wiedner} et~al.,}{{Wiedner}
  et~al.}{2020}]{Wiedner2020arXiv}
{Wiedner} M.~C.,  et~al., 2020, arXiv e-prints, \href
  {https://ui.adsabs.harvard.edu/abs/2020arXiv201202731W} {p. arXiv:2012.02731}

\bibitem[\protect\citeauthoryear{{Wiersma}, {Schaye}  \& {Smith}}{{Wiersma}
  et~al.}{2009}]{Wiersma2009MNRAS}
{Wiersma} R. P.~C.,  {Schaye} J.,   {Smith} B.~D.,  2009, \mn@doi [\mnras]
  {10.1111/j.1365-2966.2008.14191.x}, \href
  {https://ui.adsabs.harvard.edu/abs/2009MNRAS.393...99W} {393, 99}

\bibitem[\protect\citeauthoryear{{Willott}, {McLure}  \& {Jarvis}}{{Willott}
  et~al.}{2003}]{Willott2003ApJ}
{Willott} C.~J.,  {McLure} R.~J.,   {Jarvis} M.~J.,  2003, \mn@doi [\apjl]
  {10.1086/375126}, \href
  {https://ui.adsabs.harvard.edu/abs/2003ApJ...587L..15W} {587, L15}

\bibitem[\protect\citeauthoryear{{Willott} et~al.,}{{Willott}
  et~al.}{2007}]{Willott2007CFHQS}
{Willott} C.~J.,  et~al., 2007, \mn@doi [\aj] {10.1086/522962}, \href
  {https://ui.adsabs.harvard.edu/abs/2007AJ....134.2435W} {134, 2435}

\bibitem[\protect\citeauthoryear{{Willott} et~al.,}{{Willott}
  et~al.}{2010}]{Willott2010AJ}
{Willott} C.~J.,  et~al., 2010, \mn@doi [\aj] {10.1088/0004-6256/139/3/906},
  \href {https://ui.adsabs.harvard.edu/abs/2010AJ....139..906W} {139, 906}

\bibitem[\protect\citeauthoryear{{Willott}, {Bergeron}  \& {Omont}}{{Willott}
  et~al.}{2017}]{Willott:2017ApJ}
{Willott} C.~J.,  {Bergeron} J.,   {Omont} A.,  2017, \mn@doi [\apj]
  {10.3847/1538-4357/aa921b}, \href
  {https://ui.adsabs.harvard.edu/abs/2017ApJ...850..108W} {850, 108}

\bibitem[\protect\citeauthoryear{{Wiseman}, {Schady}, {Bolmer}, {Kr{\"u}hler},
  {Yates}, {Greiner}  \& {Fynbo}}{{Wiseman} et~al.}{2017}]{Wiseman:2017}
{Wiseman} P.,  {Schady} P.,  {Bolmer} J.,  {Kr{\"u}hler} T.,  {Yates} R.~M.,
  {Greiner} J.,   {Fynbo} J.~P.~U.,  2017, \mn@doi [\aap]
  {10.1051/0004-6361/201629228}, \href
  {https://ui.adsabs.harvard.edu/abs/2017A&A...599A..24W} {599, A24}

\bibitem[\protect\citeauthoryear{{Wu} et~al.,}{{Wu} et~al.}{2015}]{Wu2015Natur}
{Wu} X.-B.,  et~al., 2015, \mn@doi [\nat] {10.1038/nature14241}, \href
  {https://ui.adsabs.harvard.edu/abs/2015Natur.518..512W} {518, 512}

\bibitem[\protect\citeauthoryear{{Wyithe} \& {Bolton}}{{Wyithe} \&
  {Bolton}}{2011}]{Wyithe:2011}
{Wyithe} J. S.~B.,  {Bolton} J.~S.,  2011, \mn@doi [\mnras]
  {10.1111/j.1365-2966.2010.18030.x}, \href
  {https://ui.adsabs.harvard.edu/abs/2011MNRAS.412.1926W} {412, 1926}

\bibitem[\protect\citeauthoryear{{Xu}, {Sun}  \& {Xue}}{{Xu}
  et~al.}{2020}]{Xu2020ApJ}
{Xu} J.,  {Sun} M.,   {Xue} Y.,  2020, \mn@doi [\apj]
  {10.3847/1538-4357/ab811a}, \href
  {https://ui.adsabs.harvard.edu/abs/2020ApJ...894...21X} {894, 21}

\bibitem[\protect\citeauthoryear{{Xue} et~al.,}{{Xue} et~al.}{2011}]{xue2011}
{Xue} Y.~Q.,  et~al., 2011, \mn@doi [\apjs] {10.1088/0067-0049/195/1/10}, \href
  {https://ui.adsabs.harvard.edu/abs/2011ApJS..195...10X} {195, 10}

\bibitem[\protect\citeauthoryear{{Younger}, {Hayward}, {Narayanan}, {Cox},
  {Hernquist}  \& {Jonsson}}{{Younger} et~al.}{2009}]{Younger2009MNRAS}
{Younger} J.~D.,  {Hayward} C.~C.,  {Narayanan} D.,  {Cox} T.~J.,  {Hernquist}
  L.,   {Jonsson} P.,  2009, \mn@doi [\mnras]
  {10.1111/j.1745-3933.2009.00663.x}, \href
  {https://ui.adsabs.harvard.edu/abs/2009MNRAS.396L..66Y} {396, L66}

\bibitem[\protect\citeauthoryear{{Zafar}, {Watson}, {Fynbo}, {Malesani},
  {Jakobsson}  \& {de Ugarte Postigo}}{{Zafar} et~al.}{2011}]{Zafar2011}
{Zafar} T.,  {Watson} D.,  {Fynbo} J.~P.~U.,  {Malesani} D.,  {Jakobsson} P.,
  {de Ugarte Postigo} A.,  2011, \mn@doi [\aap] {10.1051/0004-6361/201116663},
  \href {https://ui.adsabs.harvard.edu/abs/2011A&A...532A.143Z} {532, A143}

\bibitem[\protect\citeauthoryear{{Zafar} et~al.,}{{Zafar}
  et~al.}{2018}]{Zafar2018MNRASx}
{Zafar} T.,  et~al., 2018, \mn@doi [\mnras] {10.1093/mnras/sty1876}, \href
  {https://ui.adsabs.harvard.edu/abs/2018MNRAS.480..108Z} {480, 108}

\bibitem[\protect\citeauthoryear{{Zinn}, {Middelberg}, {Norris}  \&
  {Dettmar}}{{Zinn} et~al.}{2013}]{Zinn2013ApJ}
{Zinn} P.~C.,  {Middelberg} E.,  {Norris} R.~P.,   {Dettmar} R.~J.,  2013,
  \mn@doi [\apj] {10.1088/0004-637X/774/1/66}, \href
  {https://ui.adsabs.harvard.edu/abs/2013ApJ...774...66Z} {774, 66}

\bibitem[\protect\citeauthoryear{{Zubovas}, {Nayakshin}, {King}  \&
  {Wilkinson}}{{Zubovas} et~al.}{2013}]{Zubovas2013MNRAS}
{Zubovas} K.,  {Nayakshin} S.,  {King} A.,   {Wilkinson} M.,  2013, \mn@doi
  [\mnras] {10.1093/mnras/stt952}, \href
  {https://ui.adsabs.harvard.edu/abs/2013MNRAS.433.3079Z} {433, 3079}

\bibitem[\protect\citeauthoryear{van~der Walt, Colbert  \& Varoquaux}{van~der
  Walt et~al.}{2011}]{numpy}
van~der Walt S.,  Colbert S.~C.,   Varoquaux G.,  2011, \mn@doi [Computing in
  Science Engineering] {10.1109/MCSE.2011.37}, 13, 22

\makeatother
\end{thebibliography}
